\documentclass[structabstract]{aa}  
\usepackage{graphicx,natbib,amsmath}
\usepackage{txfonts}

\def\kps{~{\rm kms}^{-1}}
\def\deg2{{\rm deg}^2}
\def\degm2{\rm deg^{-2}}
\def\arc2{~{\rm arcmin}^2}

\begin{document}
   \title{The WIRCAM Deep Infrared Cluster Survey I: Groups and Clusters at $z\gtrsim1.1$
   \thanks{Based on observations obtained with MegaPrime/MegaCam, a joint project of CFHT and CEA/DAPNIA, at the Canada-France-Hawaii Telescope (CFHT) which is operated by the National Research  Council (NRC) of Canada, the Institut National des Science de l'Univers of the Centre National de la Recherche Scientifique (CNRS) of France, and the University of Hawaii. This work is based in part on data products produced at TERAPIX and  the Canadian Astronomy Data Centre as part of the Canada-France-Hawaii Telescope Legacy Survey, a collaborative project of NRC and CNRS.}
}
   \titlerunning{The WIRCAM Deep Infrared Cluster Survey I}

   \author{R. M. Bielby\inst{1}, A. Finoguenov\inst{2}, M. Tanaka\inst{3,4}, H. J. McCracken\inst{1}, E. Daddi\inst{5}, P. Hudelot\inst{1}, O. Ilbert\inst{6}, J. P. Kneib\inst{6}, O.~Le~F\`{e}vre\inst{6}, Y. Mellier\inst{1}, K. Nandra\inst{2,7}, P. Petitjean\inst{1}, R. Srianand\inst{8}, C. S. Stalin\inst{9}, C. J. Willott\inst{10}
          }
   \institute{Institut dÕAstrophysique de Paris, UMR7095 CNRS, Universit\'e  Pierre et Marie Curie, 98 bis Boulevard Arago, 75014 Paris, France \\
    \email{rmbielby@gmail.com}
       \and
              Max Planck Institut f\"ur Extraterrestrische Physik, Giessenbachstrasse, 85478 Garching, Germany      
       \and
              European Southern Observatory, Karl-Schwarzschild-Str 2, 85748, Garching, Germany
       \and
              Institute for the Physics and Mathematics of the Universe, University of Tokyo, Kashiwa 2778582, Japan
       \and
		Service d'Astrophysique, CEA/Saclay, 91191 Gif-sur-Yvette, France
       \and
              Laboratoire d'Astrophysique de Marseille, Universit\'e Aix-Marseille, 38 rue Fr\'ed\'eric Joliot-Curie, 13388 Marseille, France
	\and
 		Astrophysics Group, Blackett Laboratory, Imperial College London, Prince Consort Road, London SW7 2BW, UK
	\and
              Inter University Center for Astronomy and Astrophysics, Post Bag 4, Ganesh Khind, Pune 411 007, India
	\and
		Indian Institute of Astrophysics, Koramangala, Bangalore 560 034, India
       \and
             	Herzberg Institute of Astrophysics, National Research Council, 5071 West Saanich Road, Victoria, BC V9E 2E7, Canada
             }
   \authorrunning{R. M. Bielby et al}
   \date{}

  \abstract
   {}
   {We use a combination of CFHTLS deep optical data, WIRCam Deep Survey (WIRDS) near-infrared data and XMM-Newton survey data to identify $z\gtrsim1.1$ clusters in the CFHTLS D1 and D4 fields. Counterparts to such clusters can not be identified without deep near-infrared data and as such the total of $\approx1\deg2$ of $J$, $H$ and $K_s$ band imaging provided by WIRDS is an indispensable tool in such work.}
   {Using public XMM X-ray data, we identify extended X-ray sources in the two fields. The resulting catalogue of extended X-ray sources was then analyzed for optical/near-infrared counterparts, using a red-sequence algorithm applied to the deep optical and near-infrared data. Redshifts of candidate groups and clusters were estimated using the median photometric redshifts of detected counterparts and where available these were combined with spectroscopic data (from VVDS Deep and Ultra-Deep and using AAT AAOmega data). Additionally, we surveyed X-ray point sources for potential group systems at the limit of our detection range in the X-ray data. A catalogue of $z>1.1$ cluster candidates in the two fields has been compiled and cluster masses, radii and temperatures have been estimated using the scaling relations.}
   {The catalogue of group and cluster candidates consists of 15 $z\gtrsim1.1$ objects. We find several massive clusters ($M\gtrsim10^{14}\rm{M_\odot}$) and a number of lower mass clusters/groups. Three of the detections are previously published extended X-ray sources. Of note is JKSC 041 (previously detected via Chandra X-ray data and reported as a $z=1.9$ cluster based on UKIDSS infrared imaging) for which we identify a number of structures at redshifts of $z=0.8$, $z=0.96$, $z=1.13$ and $z=1.49$ (but see no evidence of a structure at $z=1.9$). We also make an independent detection of the massive cluster, XMMXCS J2215.9-1738, for which we estimate a redshift of $z=1.37$ (compared to the spectroscopically confirmed redshift of $z=1.45$). We use the $z\gtrsim1.1$ catalogue to compare the cluster number counts in these fields with models based on WMAP 7-year cosmology and find that the models slightly over-predict the observations, whilst at $z>1.5$ we do not detect any clusters. We note that cluster number counts at $z\gtrsim1.1$ are highly sensitive to the cosmological model, however a significant reduction in present statistical (due to available survey area) and systematic (due to cluster scaling relations) uncertainties is required in order to confidently constrain cosmological parameters using cluster number counts at high redshift.}
   {}
   \keywords{methods: data analysis Ð surveys Ð galaxies: clusters: general Ð cosmology: large-scale structure of Universe}
   \maketitle
%

\section{Introduction}

Development of large-scale structure in the Universe is a sensitive
cosmological tool, which indepedently confirms the success of the
$\Lambda$CDM cosmological paradigm of inflationary theory (e.g.
\citealt{komatsu09,vikhlinin09}). Recently, attention of theoretical
cosmologists has been focused on the possibility of constraining different
models of inflation through their predictions of departures from Gaussianity
in the primordial curvature fluctuations \citep{desjacques09}. In addition,
some coupling between dark matter and dark energy fluids are possible, which
can give testable changes in growth of large scale structure
\citep{2010MNRAS.403.1684B}.

Clusters of galaxies, which reside in the deepest gravitational potential
wells, present an excellent cosmological probe as a tracer of the the peaks
of structure formation. For example, \citet{allen08} used the cluster X-ray
mass fraction (i.e. the ratio of the baryonic mass density to the total mass
density) to place constraints on the cosmological mass density, $\Omega_m$,
whilst also providing an independent constraint on the cosmic acceleration.
Further to this, \citet{vikhlinin09} used the cluster mass function to
measure structure growth to constrain the evolution of cosmic acceleration,
measuring the cosmic acceleration parameter, $w_0$, with a result of
$w_0=-1.14\pm0.21$.

Recently, \citet{jee09} presented a weak lensing analysis of the
$z=1.4$ cluster, XMMU J2235.3-2557 \citep{mullis05}, measuring a
cluster mass of $\approx8.5\times10^{14}\rm{M_{\odot}}$. Based on WMAP 5yr
cosmology they predict the expected number of clusters of this mass at this
redshift to be $10^{-3}$ in their $11\deg2$ survey. It is suggested by the
authors that the discovery of such a rare cluster within the given survey
area is indicative of either a higher value of $\sigma_8$ or structure
growth from non-gaussian primordial density fluctuations. \citet{jiminez09}
investigate the effect of non-gaussianity on the number density of such
massive clusters at the redshift of XMMU J2235.3-2557 and conclude that mass
determination of high-redshift, massive clusters can offer a complementary
probe of primordial non-gaussianity. Further to this, the detection of the
massive X-ray selected cluster, XMMXCS J2215.9-1738 \citep{stanford06} at a redshift of $z=1.45$ has
been treated as another supportive argument to these findings
\citep{jiminez09}.

Uncertainty in the current measurements of $\sigma_8$ is also hinted at from studies of the thermal Sunyaev-Zel'dovich (tSZ) effect from clusters. A
number of papers have claimed lower than expected tSZ signals from clusters in Cosmic Microwave Background (CMB) maps (e.g.
\citealt{2006ApJ...648..176L,2007MNRAS.382.1196B,2010ApJ...719.1045L,2010MNRAS.402.1179D,2010arXiv1001.4538K}) and
in particular \citet{2010ApJ...719.1045L} claim this to be suggestive of a
lower value of $\sigma_8$. From their measurements using the South Pole
Telescope (SPT), they give constraints of $\sigma_8=0.773\pm0.025$ compared
to WMAP 5-year constraints of $\sigma_8=0.812\pm0.026$ from the
WMAP$+$BAO$+$SN combination (although from WMAP alone a more consistent
value of $\sigma_8=0.796\pm0.036$ is calculated). Although these differences
are only at the $\sim1\sigma$ level, they can have significant effects when
used in simulations of structure formation (in particular on the cluster
number counts). From both tSZ and cluster number studies, it is evident that
significant uncertainty still remains in the determination of $\sigma_8$,
whilst these uncertainties can inhibit our ability to constrain
non-gaussianity in the primordial density fluctuations.

Critical to applying cluster observations to constraining the cosmological framework are the X-ray scaling laws that may be used to relate cluster properties such as mass, temperature and X-ray luminosity. From a simple model of pure gravitational collapse (referred to as the self-similar model) for example, power-law relations are predicted for the $L_X-T$ and $L_X-M$ cluster relations \citep{1986MNRAS.222..323K}. Much work has gone into characterising these relations (e.g. \citealt{1998ApJ...503...77M,2002ApJ...567..716R,2006ApJ...640..691V,2008MNRAS.387L..28R, 2010ApJ...715..162C}), with calibrations having been performed up to redshifts of $z\approx1$ \citep{leauthaud10}.

At the present time there are three promising ways to find a distant galaxy
cluster -- via the Sunyaev-Zel'Dovich Effect, through its extended X-ray emission, using the red-sequence
cluster method with near-infrared data. Optical surveys are limited to
$z\sim1.1$ as the 4000\AA\ break moves into the infra-red and can no longer
be used as an identification tool in optical bands. For example,
\citet{olsen07} presented a catalogue of galaxy clusters identified in all
four CFHTLS Deep fields purely from the CFHTLS optical $u^*griz$ data (i.e.
with no prior X-ray detections). The catalogue provides 169 cluster
candidate detections with a maximum redshift of $z=1.2$. Similarly, the Bonn
Lensing, Optical and X-ray (BLOX) galaxy cluster survey \citep{dietrich07}
provides a catalogue of weak-lensing and X-ray selected clusters in a number
of fields covering a total area of $6.4\deg2$. In this case the
limiting redshift of the catalogue is $z\sim1.0$ as their imaging data is
limited to the $BVRI$ optical bands, leaving a number of cluster detections
in their catalogue with no available redshift estimate. The
weak-lensing/X-ray observations can not by themselves separate out the
high-redshift cluster candidates and their spectroscopic follow-up is a
tedious (though rewarding e.g. \citealt{mullis04}) task.

{The Palomar Distant Cluster Survey (PDCS; \citealt{postman96}) performed an optical selection of galaxy clusters over an area of $5.1\deg2$}. This catalogue covered a range in redshifts of $0.2<z<1.2$ identifying 107 cluster candidates, but with no X-ray data provided no evidence of the potential wells that these potential clusters resided in. More recently, \citet{finoguenov07} presented cluster detections across the Cosmic Evolution Survey (COSMOS) field. Based on initial X-ray detections from XMM data, they identify 72 clusters across an area of $2\deg2$ using optical $+$ NIR data to determine photometric redshift estimates. Their full sample reaches a redshift limit of $z\sim1.3$. \citet{eisenhardt08} used the IRAC Shallow Survey \citep{eisenhardt04} to identify 106 $z>1$ galaxy cluster candidates over an area of $7.25\deg2$. Subsequent spectroscopic observations of cluster members confirmed 12 of these candidates with further observations to follow. They determine initial cluster redshift estimates using photometric redshifts based on deep optical data from the NOAO Deep Wide-Field Survey (NDWFS; \citealt{ndwfs}), NIR data from FLAMEX \citep{elston06} and IRAC Shallow Survey [3.6] and [4.5] imaging. With this combination they successfully identify clusters up to redshifts of $z\sim1.5$, with the success of the identification of a $z=1.41$ cluster \citep{stanford05}, at the time the highest redshift spectroscopically confirmed galaxy cluster. At $z>1.5$ however, they found their survey limited by the depth constraints of the optical data available, despite the capability of their IR data. This was followed by the discovery of the $z=1.45$ cluster XMMXCS J2215.9-1738 (located in the CFHTLS D4/LBQS 2212-1759 field) by \citet{stanford06}, which was initially detected in the X-ray with XMM data, with $I$ and $K_s$ band imaging and with Keck DEIMOS spectroscopic follow-up. More recently, \citet{2009A&A...507..147A} presented a cluster candidate at an estimated redshift of $z=1.9$ for an extended X-ray source, based on red-sequence analysis using deep NIR data from the UKIRT Deep Infrared Sky Survey. However, this remains unconfirmed by spectroscopic observations at the present time. The highest confirmed redshift cluster presently stands as the $z=1.62$ in the Subaru/XMM-Newton Deep Field (SXDF) discovered independently by \citet{2010ApJ...716.1503P} and \citet{2010ApJ...716L.152T}. Both present red-sequence analysis and spectroscopic follow-up showing that the red-sequence at bright magnitudes is in place at least to $z=1.6$. Recently the Spitzer Adaptation of the Red-Sequence Cluster Survey (SpARCS \citealt{2009ApJ...698.1934M,2009ApJ...698.1943W,2010ApJ...711.1185D}) has used the red-sequence method to identify $z>1$ clusters in six fields covering a total area of $\approx50\deg2$. A number of these have now been successfully confirmed with spectroscopic follow-up. 

Deep NIR surveys are required to identify clusters at $z\gtrsim1$, but are only available on a limited area of the sky. A number of such surveys have X-ray coverage sufficient for high-redshift cluster search \citep{2010MNRAS.403.2063F} and in this paper we present an overview of a promising combination between one of the best deep NIR data-sets available to date, the WIRCam Deep Survey (WIRDS; \citealt{bielby10c}), with the CFHTLS Deep optical data and X-ray data provided by XMM-Newton. Subsequent papers are planned, which will present the low-redshift detections and look at the properties of cluster members in the cluster sample. In section~\ref{sec:data}, we describe the data-sets used in this survey. Section~\ref{sec:method} describes the methods used for cluster detection and redshift estimation, section~\ref{sec:results} presents our high-z ($z\gtrsim1.1$) cluster candidates and section~\ref{sec:analysis} presents the analysis of the high-z catalogue data-set. We conclude in section~\ref{sec:conclusions}.

Unless otherwise stated we use a flat $\Lambda$CDM cosmology with $\Omega_m=0.25$ and $H_0=72\kps\rm{Mpc}^{-1}$, whilst magnitudes are given in the AB system.

\begin{table*}
\begin{minipage}[t]{\textwidth}
\renewcommand{\footnoterule}{}  
\caption{X-ray Observations.}             
\label{table:xrayobs}
\centering          
\begin{tabular}{lllllllll}     
\hline
\hline
Name & Obs. ID & R.A. & Dec. & PI & Filter\footnote{Note that in all observations the same filters were used for the pn, MOS1 and MOS2.} &\multicolumn{3}{c}{Net Exp (s)} \\
         &            &\multicolumn{2}{c}{(J2000)}&     &         & pn & MOS1 & MOS2\\
\hline
\hline
            XMM\_LSS\_2&  0037980201&    36.00000&    -3.83333&   M. Pierre&   Thin1&  7780.1& 11825.6& 12733.5\\
            XMM\_LSS\_3&  0037980301&    36.33334&    -3.83333&   M. Pierre&   Thin1&  7762.5& 13126.6& 13128.8\\
            XMM\_LSS\_4&  0037980401&    36.66666&    -3.83333&   M. Pierre&   Thin1&  3029.7&  5858.6&  6011.8\\
            XMM\_LSS\_5&  0037980501&    37.00000&    -3.83333&   M. Pierre&   Thin1& 10294.4& 14390.3& 14906.3\\
             XMDSOM\_1&  0109520101&    35.83333&    -4.16667&         K. Mason&   Thin1& 18571.5& 24756.3& 24768.6\\
             XMDSOM\_2&  0109520201&    36.83334&    -4.83333&         K. Mason&   Thin1& 17389.9& 24087.9& 24150.9\\
             XMDSOM\_3&  0109520301&    36.50000&    -4.83333&         K. Mason&   Thin1& 15097.4& 20780.1& 21299.1\\
             XMDSOM\_5&  0109520501&    35.83333&    -4.83333&         K. Mason&   Thin1& 17300.5& 23176.2& 23186.8\\
             XMDSOM\_6&  0109520601&    35.66667&    -4.50000&         K. Mason&   Thin1& 15336.5& 21098.1& 21768.6\\
            XMDSSSC\_1&  0111110101&    37.00000&    -5.16667&      M. Watson&   Thin1& 12051.0& 20382.5& 19159.7\\
            XMDSSSC\_2&  0111110201&    36.66666&    -5.16667&      M. Watson&   Thin1&  2142.7&  5907.8&  7307.9\\
            XMDSSSC\_3&  0111110301&    36.33334&    -5.16667&      M. Watson&   Thin1& 16788.3& 21222.0& 20939.6\\
            XMDSSSC\_4&  0111110401&    36.00000&    -5.16667&      M. Watson&   Thin1& 18321.3& 26235.8& 26616.6\\
            XMDSSSC\_2&  0111110701&    36.66666&    -5.16667&      M. Watson&   Thin1&  5857.6& 11226.3& 10820.7\\
                MLS-1&  0112680101&    36.83334&    -4.16667&       M. Turner&   Thin1& 20412.9& 23226.3& 23846.5\\
                MLS-2&  0112680201&    36.50000&    -4.16667&       M. Turner&   Thin1&  6390.5&  8159.4&  8148.5\\
                MLS-3&  0112680301&    36.16667&    -4.16667&       M. Turner&   Thin1& 16841.3& 21034.0& 20375.4\\
                MLS-5&  0112680401&    37.00000&    -4.50000&       M. Turner&   Thin1& 18032.3& 22559.3& 22823.7\\
                MLS-8&  0112680501&    36.00000&    -4.50000&       M. Turner&   Thin1& 14100.4& 20720.2& 20728.6\\
                MLS-7&  0112681001&    36.33334&    -4.50000&       M. Turner&   Thin1& 18199.3& 20108.7& 20330.6\\
                MLS-6&  0112681301&    36.66666&    -4.50000&       M. Turner&   Thin1&  9589.7& 15886.8& 15903.5\\
  XLSSJ022404.0-04132&  0210490101&    36.00834&    -4.18861&      L. Jones&  Medium& 61458.3& 73743.5& 71108.0\\
            XMM-LSS\_4&  0404960101&    36.66666&    -3.83333&   M. Pierre&   Thin1&  3752.0& 15728.5& 15341.6\\
             XMDSOM\_4&  0404960501&    36.16667&    -4.83333&   M. Pierre&   Thin1&  7480.8&  9590.4& 10312.9\\
\hline
        LBQS2212-1759&  0106660101&   333.88196&   -17.73492&         J. Clavel&   Thin1& 48139.6& 53831.9& 54507.2\\
        LBQS2212-1759&  0106660201&   333.88196&   -17.73492&         J. Clavel&   Thin1& 30559.4& 44450.2& 44544.6\\
        LBQS2212-1759&  0106660501&   333.88196&   -17.73492&         J. Clavel&   Thin1&  4391.1&  7675.9&  7775.8\\
        LBQS2212-1759&  0106660601&   333.88196&   -17.73492&         J. Clavel&   Thin1& 69304.4& 85376.6& 86686.9\\
          CFHTLS\_D4\_1&  0505460101&   334.22229&   -17.92925&       K. Nandra&  Medium& 15579.8& 24137.6& 24067.8\\
          CFHTLS\_D4\_2&  0505460201&   333.49088&   -17.87772&       K. Nandra&  Medium& 16975.0& 24725.6& 22828.1\\
          CFHTLS\_D4\_3&  0505460301&   334.14038&   -17.45453&       K. Nandra&  Medium& 21628.4& 27582.6& 27287.2\\
          CFHTLS\_D4\_4&  0505460401&   333.64246&   -17.44842&       K. Nandra&  Medium&  4293.1& 13787.9& 18142.2\\
          CFHTLS\_D4\_5&  0505460501&   333.85226&   -18.09394&       K. Nandra&  Medium& 11959.0& 19149.9& 19352.9\\
\hline
\hline
\end{tabular}
\end{minipage}
\end{table*}

\section{Data}
\label{sec:data}
\subsection{X-ray Data}

A description of the XMM-Newton observatory is given by \citet{jansen01}. In
this paper we use the data collected by the European Photon Imaging Cameras
(EPIC): the {\it pn}-CCD camera \citet{struder01} and the MOS-CCD
cameras \citep{turner01}.

XMM-Newton observations of the CFHTLS Deep fields are goals of different
surveys. The D1 field has been covered within the XMM-LSS survey (PI M.
Pierre).  The D2 field is located within the COSMOS field and has the
deepest X-ray coverage to date (PI G. Hasinger). The D3 field is also one of
DEEP2 fields and the X-ray data has been taken there as a part of AEGIS
survey.  The XMM observations of D4 field were obtained through a proposal
of the AEGIS team (PI K. Nandra).

To date most of these data have been publicly available through the ESA XMM-Newton Science Archive (XSA\footnote{http://xmm.esac.esa.int/xsa/}) and we have processed the X-ray data for all CFHTLS Deep fields. The coverage of the D3 by XMM is the smallest of all ($0.2\deg2$), since most AEGIS data has been obtained using Chandra and that program is still ongoing. The results of our survey for COSMOS has already been used for the identification of X-ray clusters in \citet{finoguenov07} and \citet{finoguenovinprep}, as well as cluster science exploration \citep{giodini09, leauthaud10}. As the COSMOS field is treated in depth in these other papers and the coverage of the D3 field is comparatively small, we focus here on the D1 and D4 fields.

The XMM coverage of both D1 and D4 are comparable with the entire area of these two fields being observed to10-20 ksec depth, thus presenting a homogeneous dataset. All X-ray observations that we have used in this work are listed in Table~\ref{table:xrayobs}.



The initial data processing has been done using the XMMSAS version 7.1 \citep{watson01, kirsch04SPIE, saxton055yrs}. Upon creating the calibrated event files, we perform a more conservative removal of time intervals affected by solar flares, following the procedure described in
\citet{zhang04}.  In order to increase our capability of detecting extended, low surface brightness features, we have applied the four-stage background subtraction of \citet{finoguenov07}. We also check for a high background that can be present in the MOS chips \citep{kuntzsnowden08}, identifying and removing from further analysis the following hot chips: MOS1 chip 4 in ODF 0404960101 and MOS2 chip 5 in ODFs 021049001, 0404960101, 0404960501. The resulting countrate-to-flux conversion in the 0.5--2 keV band excluding the lines is
$1.59\times10^{-12}$ for {\it pn} and $5.41\times10^{-12}$ for each MOS
detector, calculated for the source spectrum, corresponding to the APEC
\citep{smithr01} model for a collisional plasma of 2 keV temperature, 1/3
solar abundance and a redshift of 0.2. We note that in reconstructing the
properties of the identified groups and clusters of galaxies as well as in the
modelling of survey z-dependent characteristics, we implement the exact
corrections, based on the source spectral shape (as defined by the expected
temperature of the emission) and the measured redshift of the system.

After the background has been estimated for each observation and each
instrument separately, we produce the final mosaic of cleaned images and
correct it for the mosaic of the exposure maps in which we account for
differences in sensitivity between pn \& MOS detectors.

\subsection{Spectroscopic Redshifts}
\label{sec:specz}

In the CFHTLS D1 field, we use spectroscopic redshifts from the VVDS Deep \citep{2005A&A...439..845L} and Ultra-Deep \citep{cucciatiinprep} spectroscopic samples. The VVDS Deep sample is available publicly and consists of 8981 spectroscopically observed objects over an area of $1.2\deg2$ in the CFHTLS D1 field. It consists of a magnitude limited sample with a limit of $i<24$ and samples a redshift range of $0\leq z\leq5$. The Ultra-Deep sample consists of $\sim1500$ spectra over an area of $\approx0.14\deg2$ and covers a magnitude range of $22.5<I_{AB}<24.75$. Both of the VVDS spectroscopic catalogues attribute each object a flag based on the identification. These range from 1 to 4 with 1 being most unreliable and 4 being most reliable. In addition a flag 9 is given to objects identified based on a single emission line.

Spectroscopic redshift data in the D4 field were obtained using the AAOmega instrument at the Anglo-Australian Observatory (AAO) as part of a program to provide optical spectroscopy of X-ray point-sources in the CFHTLS \citep{2010MNRAS.401..294S}. Observations were taken on the dates 25th -- 27th September 2006 (program ID: 06B/027, PI: C. S. Stalin) and 11th -- 13th September 2007 (program ID: 07B/026, PI: P. Petitjean) with central coordinates of 22:15:30 and -17:44:00. The field of view of the AAT with the AAOmega instrument is defined by a diameter of $2^{\circ}$. The observations were made with individual exposure times of 1800s during the 2006 run and 1680s in the 2007 run. In order to maximize the wavelength coverage observations were made using two different set-ups, the first with the 580V grism (with a central wavelength of 4790\AA) and the second with the 385R grism (with a central wavelength of 7200\AA). Both grisms have a MOS resolution of $R=1300$ with a dispersion of 1.033\AA/pixel and 1.568\AA/pixel for the 580V and 385R grisms respectively.

Reductions were made using the AAO 2DFDR software to perform flat-fielding and arc-lamp wavelength calibration. Redshift identification was then performed on all objects individually by eye and using galaxy templates to determine galaxy redshifts. In total from the CFHTLS D4 AAOmega spectroscopic data, we have redshifts for 492 objects. 139 of these are QSOs, 52 are stars and 301 are galaxies, all at magnitudes of $i<22.5$. 

\subsection{CFHTLS Deep Optical Data}

All optical data used in this work was taken on the Canada-France-Hawaii Telescope (CFHT) as part of the CFHT Legacy Survey (CFHTLS). Specifically, we utilize the T0006 release of the Deep survey in the CFHT D1 and D4 fields\footnote{http://terapix.iap.fr/cplt/T0006-doc.pdf}. The CFHTLS Deep data consists of imaging with $u^*griz$ filters. $50\%$ point-source completeness limits for this data are given in table~\ref{table:photdepth}.

\subsection{WIRDS Near-Infrared Data}

In addition to the optical data from the CFHTLS, we have used near-infrared $J$, $H$ and $K_s$ deep imaging from the WIRCam InfraRed Deep Survey (WIRDS, \citealt{bielby10c}). WIRDS gives coverages of $\sim0.6\deg2$ and $\sim0.4\deg2$ in the CFHTLS D1 and D4 fields respectively. Again, $50\%$ point-source completeness limits are given in table~\ref{table:photdepth}.

\begin{table}
\caption{Photometric depths in the CFHTLS optical and WIRDS NIR imaging.}             
\label{table:photdepth}
\centering          
\begin{tabular}{lllllllll}     
\hline
&$u$&$g$&$r$&$i$&$z$& $J$ & $H$ & $K_s$ \\
&\multicolumn{8}{c}{($50\%$ point-source completeness)} \\
\hline
D1   &27.0 & 26.8 & 26.3 & 25.3 & 25.5 & 24.7 & 24.7 & 24.7 \\
D4   &26.5 & 26.7 & 26.3 & 25.1 & 25.4 & 25.1 & 24.6 & 24.6 \\
\hline
\end{tabular}
\end{table}

We construct matched catalogues for both fields incorporating all eight available imaging bands. The WIRDS and CFHTLS images are all provided with the same pixel-scale of $0.186''$ and the same area in each field, allowing catalogues to be simply extracted using SExtractor in dual-image mode. We therefore produce two catalogues for each field, the first using a $gri$ $\chi^2$ image as the detection image and the second using the $K_s$ image as the detection image. The $gri$ $\chi^2$ images were in turn produced using SWARP in $\chi^2$ combination mode using the CFHTLS $g$, $r$ and $i$ band images in each field. The $\chi^2$ mode produces an optimized combination of the input images where the output is effectively the probability of a given pixel being part of the sky-distribution, based on a reduced $\chi^2$ with the background distribution \citep{szalay99}. We note that all images have approximately the same seeing quality of $\approx0.6''-0.7''$ and as such we do not perform any smoothing of the images to match FWHMs before running SExtractor. Photometric calibration of the infrared observations was performed by matching to the 2MASS photometric data in these fields.

\subsection{Photometric Redshifts}

\begin{figure}
\centering
\includegraphics[width=9cm]{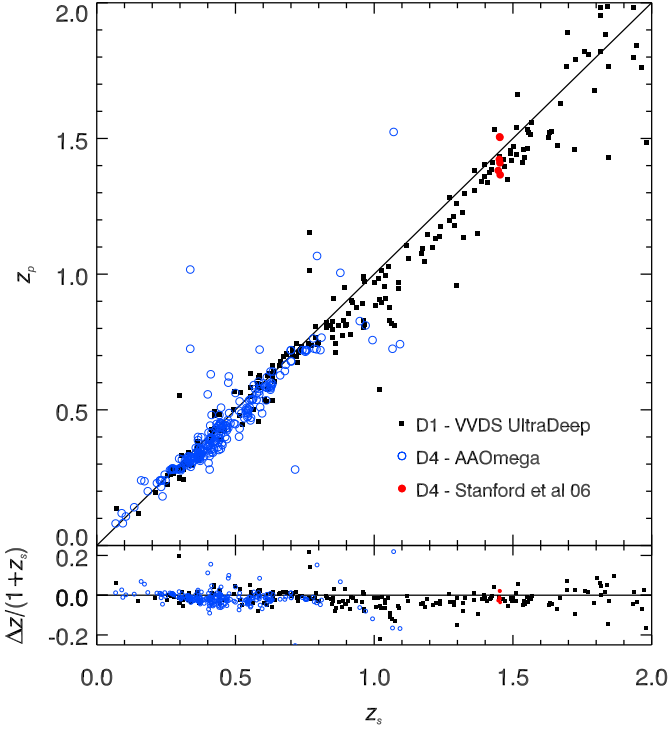}
\caption{Comparison of the spectroscopic and photometric redshifts in the CFHTLS D1 and D4 fields. All photometric redshifts are taken form our 8-band WIRDS photometric redshift catalogues. Spectroscopic redshifts are from the VVDS Ultra-Deep ($23<i<24.75$) sample for the D1 data (black squares) and the AAOmega spectra described in section~\ref{sec:specz} (open blue circles) and 5 spectroscopic redshifts from \citet{stanford06} (filled red circles) for the D4 data.}
\label{fig:specvphot}
\end{figure}

We derived photometric redshifts using the Le Phare code \citep{arnouts02,ilbert06} with a $\chi^2$ template-fitting method applied to the CFHTLS/WIRDS 8-band photometry ($u^*grizJHK_s$). Templates from \citet{2007ApJ...663...81P} were used in combination with the additional blue-galaxy templates of \citet{2009ApJ...690.1236I}. The photo-z were estimated using the median of the probability distribution function (PDFz) rather than the minimum of the $\chi^2$ distribution. We show the comparison of the photometric redshifts to available spectroscopic redshifts in the two fields (see section~\ref{sec:specz} for details of the spectroscopic data) in Fig.~\ref{fig:specvphot}. The black squares show the comparison between our photometric redshifts and the VVDS Ultra-Deep data in the CFHTLS D1 field, whilst the blue circles show the comparison with AAOmega spectroscopic redshifts in the D4 field and the red filled circles show the results for galaxies identified in the $z=1.45$ cluster of \citet{stanford06}. We find accuracies of the photometric redshifts of $\sigma_{\Delta z/(1+z)}=0.032$ with a failure rate of 2.5\% when comparing the photometric redshifts to the VVDS Deep data in the D1 field. Comparing with the VVDS Ultra-Deep spectroscopic redshifts, we find $\sigma_{\Delta z/(1+z)}=0.037$ with a failure rate of 1\%. These figures are derived across the redshift range $0<z<2$ and using only flag 3 and 4 objects from the spectroscopic datasets. We note that we see a slight systematic under-prediction of redshifts at $z\gtrsim1$. As discussed, the PSF in the optical and infrared images are relatively consistent and we have not performed any PSF matching. However, the minor differences in the PSF are the likely cause of the observed systematic offset. Although it will cause us to marginally underestimate cluster redshifts (by $\Delta z\sim0.05$), it should not affect our ability to identify galaxy over-densities, whilst the cluster redshifts may be more precisely identified using spectroscopic follow-up.

\section{Cluster Detection}
\label{sec:method}
\subsection{Extended X-ray Source Detections}

We use the prescription of \citet{2010MNRAS.403.2063F} for extended source detection, which consists of removing the PSF model for each detected point source from the data before applying the extended source search algorithm. We include an additional step in background refinement performed by \citet{2010MNRAS.403.2063F}, in which we repeat the background estimate steps for each detector and pointing. This refines the definition of the area used for the background evaluation, which is obtained through the analysis of the final mosaics. In particular, the area free from expected contamination of sources is refined based on the PSF model image, in which we excise zones where the effect of point sources leads to a statistical overestimate of the background (despite no effect being detected individually).

We use the 4$\sigma$ detection threshold in the wavelet analysis using the $32\arcsec$ and $64\arcsec$ scales applied to point-source free maps. In the following we will report direct flux estimates, made within ellipses determined by detection of the signal at 90\% confidence level. When the signal strength on flux measurement is below $4\sigma$, simulations show that contamination from unresolved point sources can be high \citep{2008AstL...34..367B}.

\begin{figure*}
\centering
\includegraphics[width=9cm]{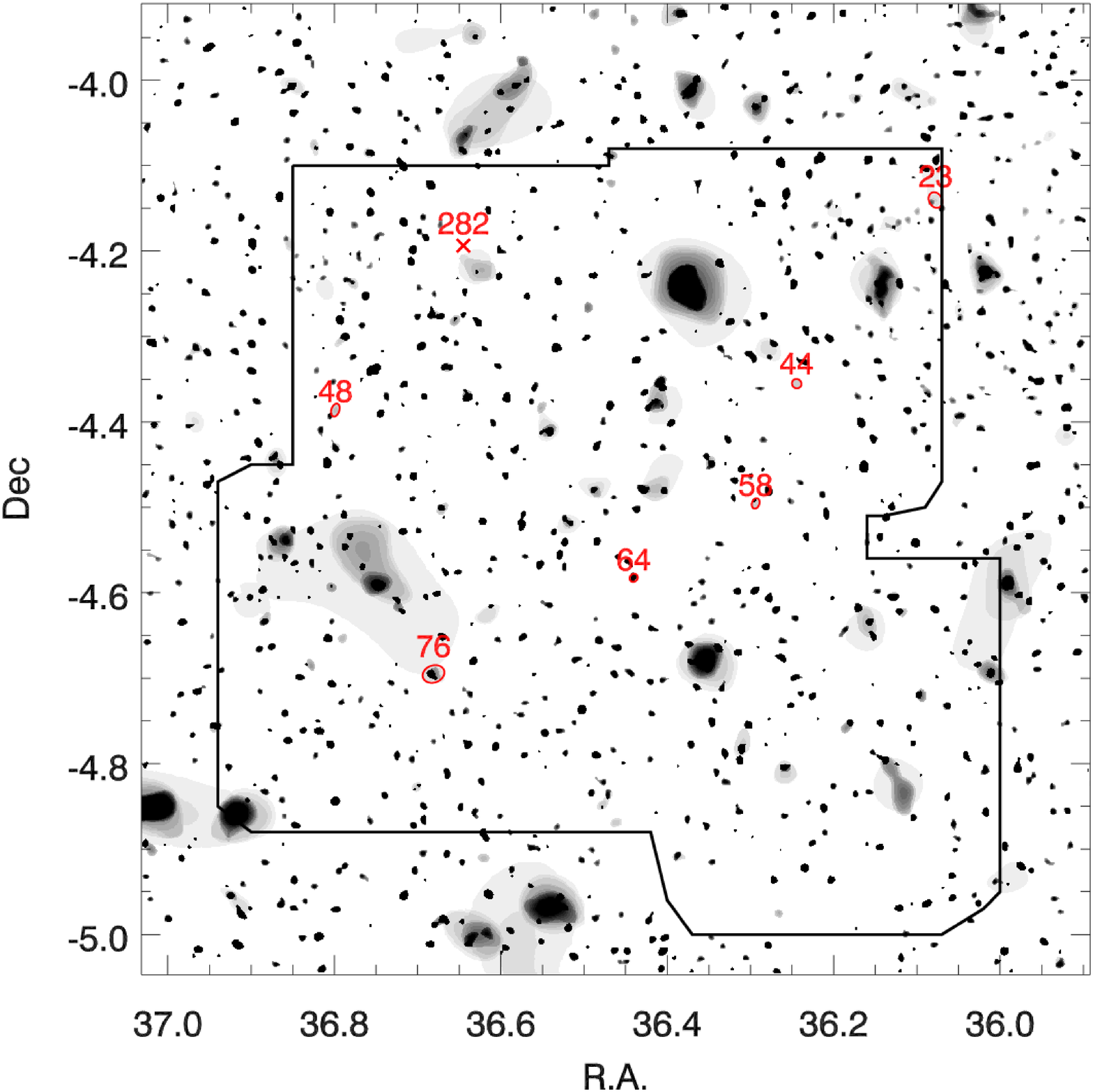}
\includegraphics[width=9cm]{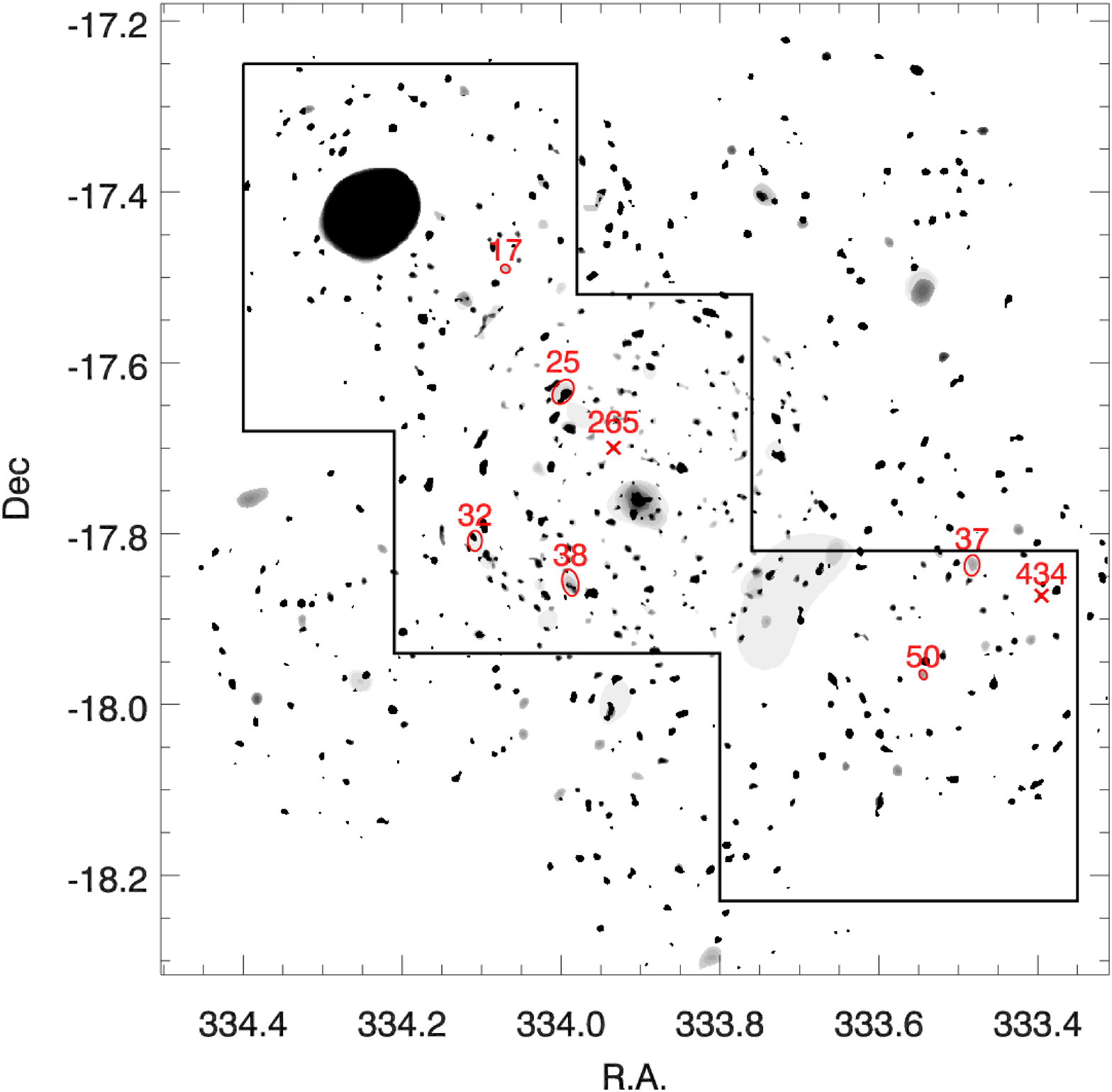}
   \caption{X-ray contour maps of the CFHTLS D1 (left) and D4 (right) fields. The grey-scale shows the smoothed X-ray flux. Detected extended X-ray cluster candidates are ringed and numbered, whilst point-source X-ray cluster candidates are marked by red crosses and numbered field. Both maps cover an area of $1^\circ\times1^\circ$, the entirety of which is covered by the CFHTLS Deep optical $u^*griz$ data. The solid lines denote the extent of our near-infrared $JHK_s$ imaging data.}
      \label{fig:xray-maps}
\end{figure*}

\subsection{Cluster Red-Sequence}

In order to identify optical counterparts of the extended X-ray sources, we have made use of the red-sequence technique used in \citet{2010MNRAS.403.2063F}, which is itself a refinement of the photo-z concentration technique used by \citet{finoguenov07}.

The method is based on the modeling of the red-sequence, which is performed using the \citet{bruzualcharlot03} population synthesis code, with a passive evolution model of a single stellar population (SSP) and assuming the Chabrier initial mass function \citep{2003PASP..115..763C} and no dust-extinction. The slope of the red-sequence is then reproduced by fitting the SSP models (formed at $z_f=5$) to the Coma cluster red-sequence with a range of metallicities. We assume that the slope of the red-sequence is entirely due to the mass-metalicity relation, as suggested by both observations and theoretical work \citep{kodamaarimoto97,stanford98}.

For a given X-ray source, we fit galaxy colours and magnitudes from the $K_s$ selected catalogues at the cluster location to the red-sequence model described above over the redshift range $0<z<2$. At each redshift step within this range, galaxies within an area of radius 0.5Mpc of the centre of the X-ray detection and within $\Delta z_{phot}<0.2$ of the given redshift step are extracted and the significance of an overdensity of red galaxies around the model red-sequence is estimated. This significance is determined with a weighting applied such that galaxies closer to the X-ray centre are given a higher weighting. All significant red-sequence detections are then recorded. The weighting, $w(z)$, is given by:

\begin{equation}
w(z)=\sum_i{exp\left[-\left(\frac{c_{i}-c_m(z,m_i)}{\sigma_{i,c}}\right)^2-\left(\frac{m_{i}-m_{m}^{\star}(z)}{\sigma_{mag}}\right)^2-\left(\frac{r_i}{\sigma_r}\right)^2\right]}
\label{eq:rs_weighting}
\end{equation}

\noindent where $c_i$ and $m_i$ are the observed colour and magnitude of the $i$th galaxy. $\sigma_{i,c}$ is the error on the observed colour, $c_m(z,m_i)$ is the model red-sequence colour at the observed magnitude, $m_i$, whilst $m_{m}^{\star}(z)$ is the characteristic magnitude based on the model and $\sigma_{mag}$ is a smoothing parameter. Finally $r_i$ is the distance from the X-ray centre and $\sigma_r$ is a spatial smoothing factor. The significance of the red-sequence measure is estimated based on this weighting calculation relative to the median red-sequence signal measured at random positions and redshifts within the field. No explicit magnitude cut is placed on the galaxy sample for this analysis, however the presence of the errors on the colour and the magnitude in the denominators means that faint sources with high magnitude errors are given low weightings.

The three different terms in equation~\ref{eq:rs_weighting} perform weighting on colour, magnitude and spatial clustering respectively. Appropriate colour-magnitude scales are chosen based on the redshift range being probed. These are as follows:

\begin{itemize}
\item$0<z<0.5$: $(b-i)$ colour, $i$ magnitude
\item$0.5<z<1$: $(r-z)$ colour, $z$ magnitude
\item$1<z<1.5$: $(i-J)$ colour, $J$ magnitude
\item$1.5<z<2$: $(z-K_s)$ colour, $K_s$ magnitude
\end{itemize}

As stated, we focus solely on the high redshift sources identified by the red-sequence analysis, as it is at $z\gtrsim1.1$ that the deep near-infrared data provides significant improvements over what is possible using optical data alone. For example, \citet{olsen07} perform a cluster search in the CFHTLS fields using the CFHTLS deep optical data and are limited to identifying redshifts of counterparts for clusters at $z<1.1$ only.

\begin{figure}
\centering
\includegraphics[width=8cm]{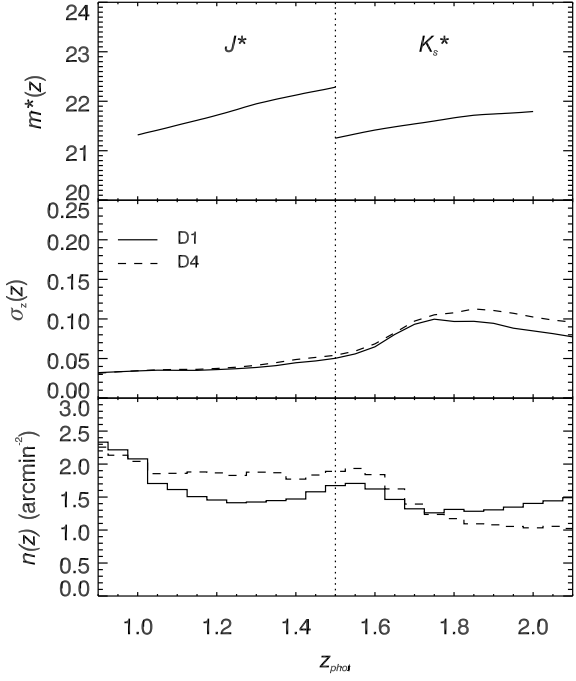}
   \caption{Top panel: The characteristic magnitude as a function of redshift used in equation~\ref{eq:rs_weighting}. Note that the $J$ band magnitudes are used in the redshift range $1<z<1.5$ and the $K_s$ band magnitudes are used at $1.5<z<2.0$. Middle panel: Median photometric redshift errors of galaxies in the magnitude range $m_m^*(z)\pm1$ as a function of redshift for the D1 (solid line) and D4 (dashed line) fields. Bottom panel: The photometric redshift distribution of galaxies in the $K_s$ selected catalogue with magnitudes of $m=m_m^*(z)\pm1$.}
      \label{fig:mstar}
\end{figure}

The success of applying the red-sequence method to identifying cluster members at $z>1.1$ is reliant on a number of factors. The first factor is the sensitivity of the data to the presence of the red-sequence. We show the characteristic magnitude as a function of redshift in the top panel of Fig.~\ref{fig:mstar}. In the range $1<z<1.5$, this is $J$ band magnitude is used, whilst at $1.5<z<2.0$, the $K_s$ band magnitude is used. Additionally, we show the median photometric redshift errors as a function of redshift for all galaxies within $m_m^*\pm1$ in both the D1 (solid line) and D4 (dashed line) fields (middle panel) and the $n(z)$ distribution of this population again for both the D1 (solid line) and D4 (dashed line) datasets (bottom panel). The increase in the photometric redshift errors will act to smear the redshift distribution. However, for the $m=m_m^*\pm1$ population potted, the median photometric redshift error remains well below the redshift selection window ($z_{rs}\pm0.2$) we apply prior to the red-sequence analysis.

Furthermore, the method is reliant on the red-sequence being in place in clusters in the survey volume. At the present time, the red-sequence has been confirmed to be in place at least to $z=1.62$ based on the observations of \citet{2010ApJ...716L.152T}. However, observations of clusters at $z>1.5$ are somewhat limited and the prevalence of the red-sequence and it's dominance over star-forming galaxies in clusters at these redshifts is far from certain. A fraction of groups and clusters at such redshifts may be dominated by star-forming galaxies, in which case the success rate of the red-sequence analysis would be reduced.

We note that the model red sequence shows an offset with respect to the observed red sequence at a given redshift. This is due to the systematic uncertainty in the models and also due to small photometric zero point errors in the observed data.  We do not calibrate the offset with spectroscopically confirmed clusters as it is
difficult due to the lack of spectroscopic redshifts particularly at high-$z$. The cluster redshifts are estimated using the median redshifts of the galaxies identified as cluster members via the above method. 

For all of the detections, we perform visual checks of the red-sequence identifications and assign a flag based on the quality of the identification. These were assigned based on the following criteria:

\begin{enumerate}
\item Good X-ray centre, reliable redshift estimate.
\item Unable to use X-ray centre, reliable redshift estimate.
\item Spectroscopic redshift confirmation required.
\item Potential projection effects.
\item Redshift identification potentially unreliable.
\item Identification ok, but not a group/cluster.
\item No identification possible.
\item Detection out of IR data area.
\end{enumerate}

Note that as we only present a subset of the cluster identifications in this paper, not all the flag values are utilized in this paper, however we list the full range of possible flags for consistency with the full catalogue, which includes the $z\lesssim1.1$ detections.

\subsection{Cluster Properties \& Scaling Relations}

From the cluster apperture flux measurements and redshift estimates we
derive the total cluster fluxes, luminosities and masses. The XMM
X-ray observations will not be sensitive to the outer faint regions of the
cluster X-ray emission, particularly so given the distant nature of the
clusters presented here. Deriving an accurate estimate of the total flux
therefore relies on two possible methods, re-observation to gain greater
depth or modeling of the X-ray profile. In this work we extrapolate the
total flux for each cluster from the observed flux, $F_d$, by iteratively
fitting a $\beta$-model to the observed X-ray emission, from which we
estimate the correction factor, $C_\beta(z,T)$. Thus the total flux is given
by:

\begin{equation}
F(<r_{500})=C_{\beta}(z,T)F_d
\end{equation}

\noindent where $r_{500}$ is the radius enclosing the matter density
$500\times$ the critical density, which corresponds to the observed
steepening in the surface brightness profiles of clusters
\citep{2006ApJ...640..691V}. \citet{2002ApJ...567..716R}, amongst others,
have shown that the bulk of the cluster flux is contained within this
radius. The $\beta$-model \citep{1976A&A....49..137C} is given by: 

\begin{equation}
S(r)=S(0)\left[1 + \left(\frac{r}{r_{core}}\right)^2\right]^{-3\beta+1/2}
\end{equation}

\noindent where $S(0)$ is the central flux, $r$ is the cluster radius and we
use the parameters $\beta$ and $r_{core}$ given by the scaling relations: 

\begin{equation}
\beta=0.4(kT/1\text{keV})^{1/3}
\end{equation}

\noindent and

\begin{equation}
r_{core}=0.07(kT/1\text{keV})^{0.63}r_{500}
\end{equation}

We note that although the $\beta$-model is considered a good description of
the cluster shape, characterization of the outskirts remains uncertain,
whilst groups show a larger scatter in their X-ray profiles. Further, given
the redshift range we are probing in this paper, it is important to note
that these scaling relations have been derived at redshifts of $z<1$.

In order to estimate the intrinsic cluster luminosity, we apply the
$K$-correction following the approach used by and described in
\citet{2002ApJ...567..716R}, \citet{2004A&A...425..367B} and
\citet{finoguenov07} among others. Therefore, applying the $K$-correction to
account for the temperature and redshift of individual clusters, the
intrinsic luminosity is given by:

\begin{equation}
L_{0.1-2.4\rm{keV}} = 4\pi d^2_LK(z,T)C_\beta(z,T)F_d
\end{equation}

\noindent where $d_L$ is the luminosity distance to the cluster.

From the luminosity, we estimate both the mass and temperatures for a given cluster based on derived scaling relations \citep{1998ApJ...503...77M}. The cluster temperature is therefore given by:

\begin{equation}
kT=0.2+6\times10^{[\rm{log}_{10}(L_{0.1-2\rm{keV}}/E_z)-44.45]/2.1}
\end{equation}

\noindent where $E_z$ is given by:

\begin{equation}
E_z=\sqrt{\Omega_M(1+z)^3+\Omega_\Lambda}
\end{equation}

The radius measure, $r_{500}$, is then related to the cluster temperature, $kT$, by:

\begin{equation}
r_{500}=0.391\textrm{Mpc}~(kT/\rm{keV})^{0.63}/E_z
\end{equation}

We solve for the above inter-related parameters iteratively for each cluster starting from the observed X-ray signal.

The above procedure of measuring the cluster luminosity has been calibrated
against weak lensing mass measurement in \citet{leauthaud10} within
$z<1$, yielding

\begin{equation}
M_{200} E_z= 10^{13.7} M_\odot \left(1.07\pm0.15\right) \left({L_X E^{-1}_z \over 10^{42.7} \rm{ergs}.\rm{s}^{-1}}\right)^{0.64\pm0.03}
\end{equation}

This is consistent with local measurements of the mass-luminosity relations, within the adopted correction for expected evolution with redshift. In this work we perform the mass estimates of the clusters by extrapolating this relation to higher redshifts.


\section{Cluster Candidates at $z\gtrsim1.1$}
\label{sec:results}

\begin{table*}
\begin{minipage}[t]{\textwidth}
\caption{WIRD Cluster Survey $z\gtrsim1.1$ cluster candidates.}             
\label{table:rs_overview}      
\centering
\renewcommand{\footnoterule}{}  
\begin{tabular}{lllllllllll}     
\hline
\hline                      
ID         &Field & Cat no.\footnote{Running ID number in full extended X-ray source catalogue}& R.A.    & Dec.      & $z_{rs}$ & $\sigma_{rs}$ & $z_{phot}$ & $z_{spec}$\footnote{Redshifts in italics are tentative based on a single spectroscopic redshift, spectroscopic redshifts of galaxies not selected in the red-sequence analysis or the redshift of a QSO.} & Flag\\
 & & & \multicolumn{2}{c}{(J2000)} & & & & & \\
\hline
\hline               
WIRDXC J0224.3-0408 &D1&23 & 36.0777 & -4.1403 & $1.26^{+0.13}_{-0.08}$ & 9.1 & 1.45 & ---    & 3 \\[0.08cm]
WIRDXC J0225.0-0421 &D1&44 & 36.2442 & -4.3550 & $1.24^{+0.26}_{-0.05}$ & 6.5 & 1.39 & ---    & 3 \\[0.08cm]
WIRDXC J0227.2-0423 &D1&48 & 36.7988 & -4.3861 & $1.24^{+0.04}_{-0.09}$ & 2.9 & 1.24 & \em{1.32} & 5 \\[0.08cm]
WIRDXC J0225.2-0429 &D1&58 & 36.2938 & -4.4955 & $1.46^{+0.04}_{-0.08}$ & 4.0 & 1.46 & ---    & 5 \\[0.08cm]
WIRDXC J0225.8-0434 &D1&64 & 36.4403 & -4.5826 & $1.14^{+0.04}_{-0.01}$ & 2.7 & 1.14 & \em{1.10} & 5 \\[0.08cm]
JKCS 041 (a)               &D1&76 & 36.6808 & -4.6951 & $1.07^{+0.06}_{-0.06}$ & 5.4 & 1.10 & 1.13 & 1 \\[0.08cm]
JKCS 041 (b)               &D1&76 & 36.6808 & -4.6951 & $1.49^{+0.28}_{-0.06}$ & 6.0 & 1.49 & ---   & 3 \\[0.08cm]
\hline
WIRDXC J2216.3-1729 &D4&17 &334.0701&-17.4900 & $1.46^{+0.04}_{-0.01}$ & 2.1 & 1.46 & ---    & 5 \\[0.08cm]
XMMXCS J2215.9-1738 &D4&25 &333.9975&-17.6340 & $1.37^{+0.09}_{-0.19}$ &17.9& 1.37 & 1.45 & 1 \\[0.08cm]
WIRDXC J2216.4-1748 &D4&32 &334.1084&-17.8084 & $1.41^{+0.09}_{-0.08}$ & 3.1 &1.40 & \em{1.42} & 3 \\[0.08cm]
WIRDXC J2213.9-1750 &D4&37 &333.4828&-17.8372 & $1.16^{+0.11}_{-0.06}$ & 5.4 & 1.16 & ---    & 3 \\[0.08cm]
BLOX J2215.9-1751.6  &D4&38 &333.9883&-17.8574 & $1.11^{+0.12}_{-0.06}$ & 8.5 & 1.17 & ---    & 3 \\[0.08cm]
WIRDXC J2214.2-1757 &D4&50 &333.5445&-17.9652 & $1.20^{+0.11}_{-0.04}$ & 2.5 & 1.28 & ---    & 5 \\[0.08cm]
\hline
\hline     
\end{tabular}
\end{minipage}
\end{table*}

Our XMM-Newton maps of both fields (CFHTLS D1 and D4) are shown in Fig.~\ref{fig:xray-maps}, with the $z\geq1.1$ cluster candidates denoted by the red ellipses and numbers (the numbers corresponding directly to the IDs given in table~\ref{table:det_overview}). The extent of the WIRDS data is shown by the solid outlines and corresponds to 5 WIRCam pointings in the D1 field and 3 in the D4 field. In total we detect 62 extended X-ray sources within the D1 field and 40 within the D4 field. Typically all sources are based on detections of $>30$ counts. In the D1, we find that 46 out of the 62 detections appear to be associated with galaxy over-densities identified using the red-sequence analysis, with 40 being at $z<1.1$ and 6 being at $z\geq1.1$. In the D4, the number of X-ray sources associated with a red-sequence detection is 28 out of the 40, with 22 being at $z<1.1$ and 6 being at $z\geq1.1$. The X-ray detection found to not be associated with galaxy over-densities are most likely the result of source confusion of close pairs. This is an issue that we are actively working to resolve, whilst our requirement of dual detection of both an extended X-ray emission and a clustered red-sequence should minimize the effect of false detections in our final catalogue. As discussed, in this paper we focus on those sources with red-sequence identifications at $z\geq1.1$ as it is the identification of this population that the deep near-infrared data facilitates.

In table~\ref{table:rs_overview} we present the locations of the extended X-ray detections and the results of the red-sequence analysis. We list the red-sequence redshift estimate, $z_{rs}$ (column 6) with the significance of the result, $\sigma_{rs}$ (column 7). The median photometric redshift of the galaxies selected via the red-sequence analysis, $z_{phot}$, is given in column 8. Spectroscopic redshift identifications are given in column 9, where redshift in italics represent tentative identifications based on either a QSO that may be identified with the cluster, non-red-sequence galaxies at the estimated cluster redshift or just a single red-sequence identified galaxy having a spectroscopic redshift. Normal text denotes multiple corroborating spectroscopic identifications of red-sequence galaxies.

For each candidate we present the spatial distribution of red-sequence selected galaxies, the magnitude-colour diagram used to identify the red-sequence, the red-sequence 'signal' as a function of redshift and the redshift histogram within the analysis radius. We also provide for each cluster a $2.5\arcmin\times2.5\arcmin$ (or larger for the more extended detections) $giK_s$ colour image constructed using the \texttt{STIFF} software \citep{stiff}. We use the $g$-band image to provide the blue channel, the $i$-band image for the green channel and the $K_s$-band image to provide the red channel and tailor the gamma correction and lower and upper brightness limits in order to optimize the images for publication. X-ray contours have been overlayed on the colour images in each case, whilst galaxies selected as cluster members via the red-sequence method are highlighted (white arrows) and, where available, spectroscopic redshifts are denoted (white circles).

We provide the details of our cluster analysis based on the estimated redshifts and X-ray data in table~\ref{table:det_overview}. All candidates listed here were detected in the X-ray data and are estimated to be at $z\geq1.1$ based on the red-sequence analysis described above in conjunction with photometric redshifts. Cluster IDs are given (column 1). The best estimated redshift (photometric or spectroscopic) is given in column 2. For each cluster we give $r_{200}$ (i.e. the radius enclosing a matter density $200\times$ the critical density, column 3), the iteratively determined total flux, $F(r<r_{500})$ (column 4), and mass, $M_{200}$ (column 5), the X-ray luminosity, $L_x$ (column 6) and the cluster temperature, $T_X$ (column 7). Finally the detection reliability flag is given in column 10. Note that for the calculations of the cluster properties, we have used a flat cosmology with $\Omega_m=0.25$ and $h=0.72$ in order to be consistent with the \citet{leauthaud10}.

All the following candidates were detected using the available X-ray data and analyzed using the full 8-band photometry available in the optical from CFHTLS-Deep and the near-infrared from WIRDS. We have checked all our candidates against known clusters using the NASA Extragalactic Database (NED). In cases where a candidate has already been identified, with or without a redshift estimate, we compare our results with the previous detection.

\begin{table*}
\begin{minipage}[t]{\textwidth}
\caption{WIRD Cluster Survey $z\gtrsim1.1$ cluster candidates.}             
\label{table:det_overview}      
\centering
\renewcommand{\footnoterule}{}  
\begin{tabular}{l l l l l l l l}     
\hline
\hline                      
ID                       & z\footnote{Best estimate cluster redshift - photometric redshift unless multiple corroborating spectroscopic redshifts are available.}   &$r_{200}$ & $F(<r_{500})$ & $M_{200}$ & $L_{x}(0.1$-$2.4\rm{keV})$& $T_{X}$  & Flag \\ 
                          &      &($^{\circ}$) & ($10^{-15}\rm{ergs/s/cm}^2$)    &        ($10^{14}\rm{M_{\sun}})$  & $(10^{43}\rm{ergs/s})$ &  (keV)     &           \\
\hline
\hline               
WIRDXC J0224.3-0408 & 1.45 & 0.0169 & $1.92\pm0.31$            & $0.85\pm0.08$                  & $6.05\pm0.99$             &$2.18\pm0.15$        & 3       \\[0.04cm]
WIRDXC J0225.0-0421 & 1.39 & 0.0171 & $1.44\pm0.55$            &$0.73\pm0.16$                   & $4.61\pm2.06$             &$1.85\pm0.28$              & 3       \\  [0.04cm] 
WIRDXC J0227.2-0423 & 1.24 & 0.0189  & $2.43\pm0.62$           &$0.91\pm0.13$                   & $4.96\pm1.26$             &$2.10\pm0.22$                & 5       \\   [0.04cm]
WIRDXC J0225.2-0429 & 1.46 & 0.0186  &  $3.02\pm0.67$          &   $1.03\pm0.13$                & $7.49\pm1.67$             & $2.43\pm0.22$               & 5       \\   [0.04cm]
WIRDXC J0225.8-0434 & 1.14 & 0.0213  & $4.27\pm0.84$           & $1.15\pm0.13$                  & $6.48\pm1.27$             &$2.41\pm0.20$    & 5       \\[0.04cm]
JKCS 041\footnote{{We detect multiple structures with the red-sequence analysis for JKCS 041. However, we are unable to separate the X-ray emissions from the different structures and therefore give upper limits to the X-ray properties based on taking the entire X-ray emission for each.}}
                                & 1.13 & 0.0259  & $<11.02\pm1.59$          &$<1.78\pm0.15$                  & $<11.68\pm1.68$            &$3.20\pm0.24$             & 1        \\[0.04cm]
JKCS 041                   & 1.49 & 0.0224  & $<10.42\pm1.50$          &$<1.88\pm0.16$                  & $<21.42\pm3.08$            &$3.85\pm0.24$             & 3        \\[0.04cm]
\hline
WIRDXC J2216.3-1729 &1.46  & 0.0162     &  $1.19\pm0.46$  &  $0.68\pm0.15$          & $3.72\pm1.45$              &  $1.80\pm0.27$                & 5 \\   [0.04cm]
XMMXCS J2215.9-1738 &1.45  & 0.0229    & $10.80\pm0.45$  &$1.91\pm0.05$            & $20.50\pm0.85$              &$3.82\pm0.07$                  & 1 \\   [0.04cm]
WIRDXC J2216.4-1748 &1.40  & 0.0203     & $4.59\pm0.49$   & $1.25\pm0.08$           & $9.43\pm1.01$                &$2.73\pm0.13$                 & 3 \\   [0.04cm]
WIRDXC J2213.9-1750 &1.16  & 0.0213     & $3.65\pm0.74$   & $1.03\pm0.12$           & $4.91\pm1.00$                &$2.17\pm0.18$       & 3 \\   [0.04cm]
BLOX J2215.9-1751.6  &1.17  & 0.0224     & $5.13\pm0.39$   & $1.22\pm0.06$           & $6.64\pm0.50$                      &$2.47\pm0.08$    & 3 \\   [0.04cm]
WIRDXC J2214.2-1757 &1.28  & 0.0207     &$4.02\pm 0.95$   & $1.14\pm0.15$            &$6.83\pm1.62$              &$2.44\pm0.24$&5 \\   [0.04cm]
\hline
\hline     
\end{tabular}
\end{minipage}
\end{table*}

\subsection{CFHTLS D1 Extended Sources}

In the D1 field we detect 6 $z\gtrsim1.1$ extended X-ray cluster candidates using the combination of X-ray and NIR data. The X-ray data covers the entire $1\deg2$ of the CFHTLS Deep field, whilst the NIR data covers $\sim0.6\deg2$. Our $z\gtrsim1.1$ detections are necessarily limited to the area covered by the NIR imaging. The candidate locations are shown over-layed on the smoothed X-ray data in Fig.~\ref{fig:xray-maps}. Coordinates and cluster properties are presented in table~\ref{table:det_overview}. In the following sub-sections we present each of the cluster candidates identified in the CFHTLS D1 field.

\subsubsection{WIRDXC J0224.3-0408 (D1-23)}

\begin{figure}
\centering
\includegraphics[height=8.5cm,angle=270.]{15135f04a.ps}
\includegraphics[width=9cm]{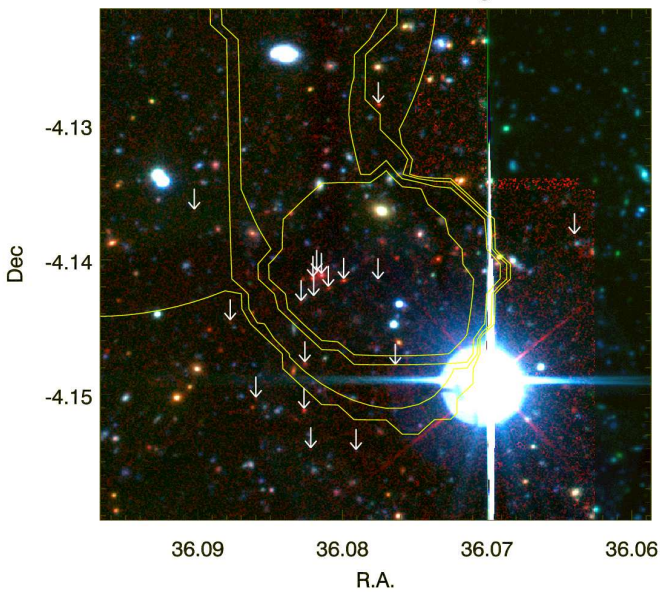}
\caption{Top: Spatial distribution, colour-magnitude plot and redshift distribution/significance from the red sequence analysis. The red points in the spatial distribution and colour-magnitude plot show the photometrically identified red-sequence galaxies. Blue points show galaxies not selected as being part of the red-sequence at the estimated cluster redshift and grey points show galaxies within the cluster radius at other redshifts. The third panel shows the redshift distribution of galaxies within the cluster radius (divided by the overall field redshift distribution - field black histogram), whilst the significance of the red-sequence detections as a function of redshift is given by the solid red line. The dashed horizontal lines give the significance scale. Bottom: Colour image combining CFHTLS $g$, $i$ and WIRDS $K_s$ band imaging of cluster candidate WIRDXC J0224.3-0408. The yellow contours show the X-ray emission. White arrows denote galaxies selected via the red-sequence analysis. The area covered by the image is $2.3\arcmin\times2.3\arcmin$.}
\label{fig:colim-WIRDXC_J0224.3-0408}
\end{figure}

In the first high redshift candidate in the D1 field, we find a clearly extended X-ray signal (yellow contours in Fig.~\ref{fig:colim-WIRDXC_J0224.3-0408}) with a extended flux signal of $6.2\sigma$. The region of the X-ray detection is somewhat obscured in the optical/NIR data due to a bright foreground star. We see a strong signal in the red-sequence analysis, however the signal is very broad and covers the range $z\approx1.1-1.6$. Two peaks in the analysis are visible, the first at $z=1.26^{+0.13}_{-0.08}$, with a significance of $9.1\sigma$ and the second at $z=1.53^{+0.20}_{-0.04}$, with a significance of $6.4\sigma$. The broadness of the two peaks in the red-sequence analysis suggest that they may be caused by the same structure at a single redshift (we note that the bright star close to this candidate may be affecting the photometry). Taking the photometric redshift distribution, we observe a large peak at high redshift with $z=1.45$. Given the large uncertainties on the red-sequence results and the general consensus with the photometric redshift peak, we take the median redshift of this peak (i.e. $z=1.45$) as the cluster redshift. There are no apparent red-sequence signals below $z\sim1.1$.

Based on this redshift estimate, we predict a cluster mass of $M_{200}=0.85\times10^{14}\rm{M_{\odot}}$ and an X-ray luminosity of$L_x=6.1\times10^{43}$ergs/s.

\subsubsection{WIRDXC J0225.0-0421 (D1-44)}

\begin{figure}
\centering
\includegraphics[height=8.5cm,angle=270]{15135f05a.ps}
\includegraphics[width=9.0cm]{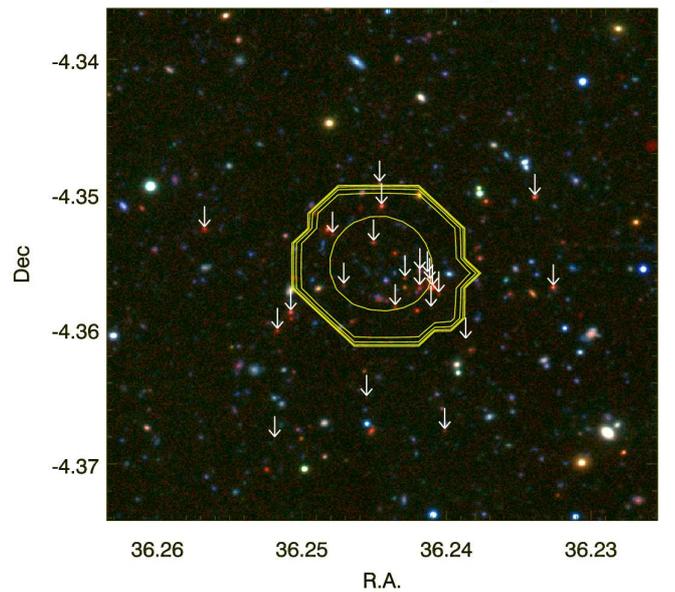}
\caption{As in Fig.~\ref{fig:colim-WIRDXC_J0224.3-0408}, but for candidate WIRDXC J0225.0-0421.}
\label{fig:colim-WIRDXC_J0225.0-0421}
\end{figure}

Candidate WIRDXC J0225.0-0421 is shown in Fig.~\ref{fig:colim-WIRDXC_J0225.0-0421}. The X-ray signal is measured with a significance of $\sim2.6\sigma$. Analyzing the red-sequence result, we see a broad peak from $z\sim1.2$ to $z\sim1.5$ with a significance of $\sim5\sigma$. The primary solution gives a redshift of $z=1.24^{+0.26}_{-0.05}$. This correlates with a peak in the redshift distribution and we find a median photometric redshift of the galaxies in this structure of $z=1.39$. Looking at the spatial distribution of the $z\approx1.39$ red-sequence galaxies, we find a strongly clustered group within the X-ray contours. A total of 13 galaxies are classified as red-sequence members at the cluster redshift within the observed X-ray contours. Again we see no significant signs of structures at redshifts of $z=1.1$ using either the NIR or optically selected catalogues.

Taking the $z=1.39$ solution, we estimate a cluster mass of $M_{200}=0.73\times10^{14}\rm{M_{\odot}}$, an X-ray luminosity of $L_x=4.6\times10^{43}$erg/s and a cluster radius of $r_{200}=0.017^{\circ}$.

\subsubsection{WIRDXC J0227.2-0423 (D1-48)}

\begin{figure}
\centering
\includegraphics[width=8.5cm]{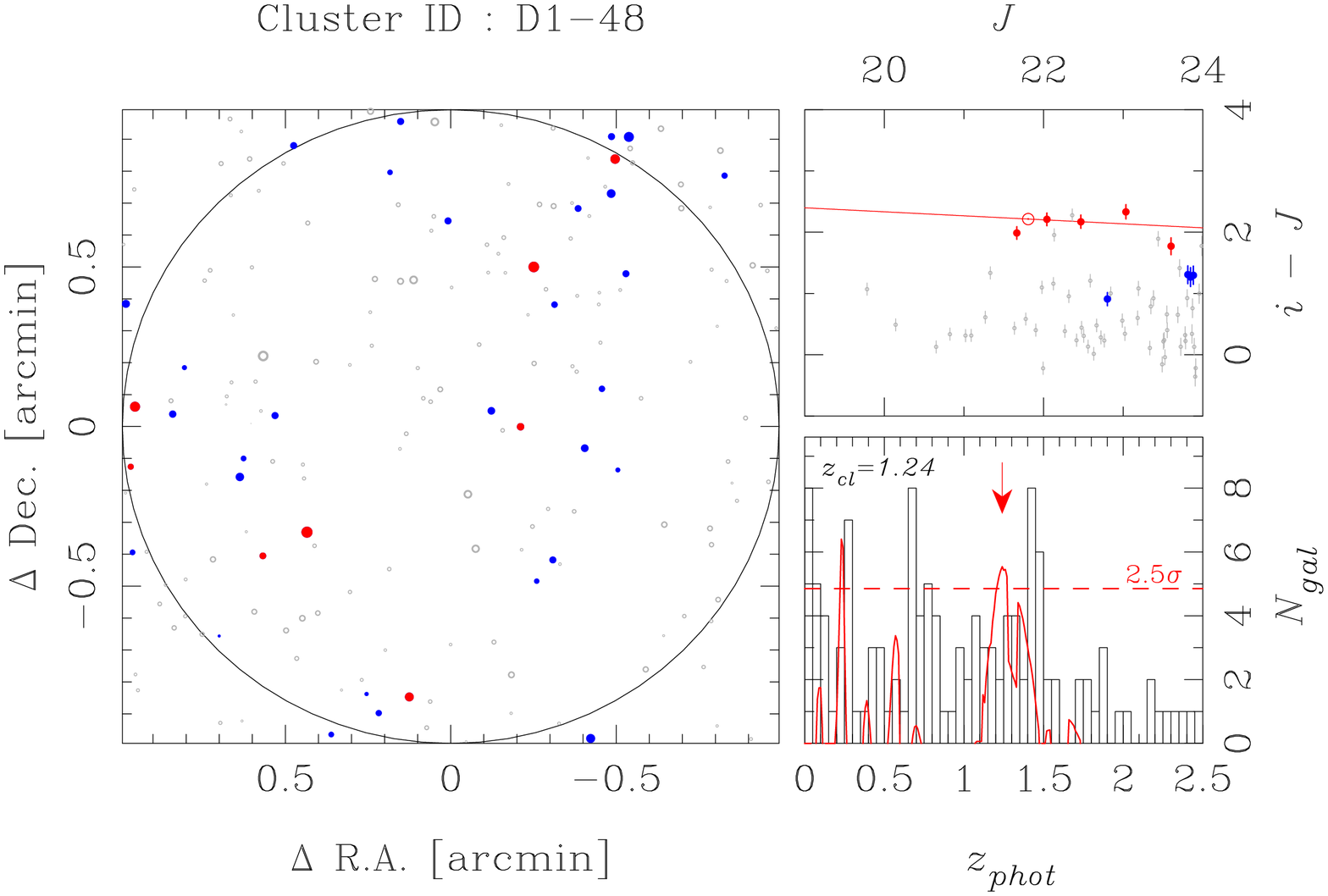}
\includegraphics[width=9.0cm]{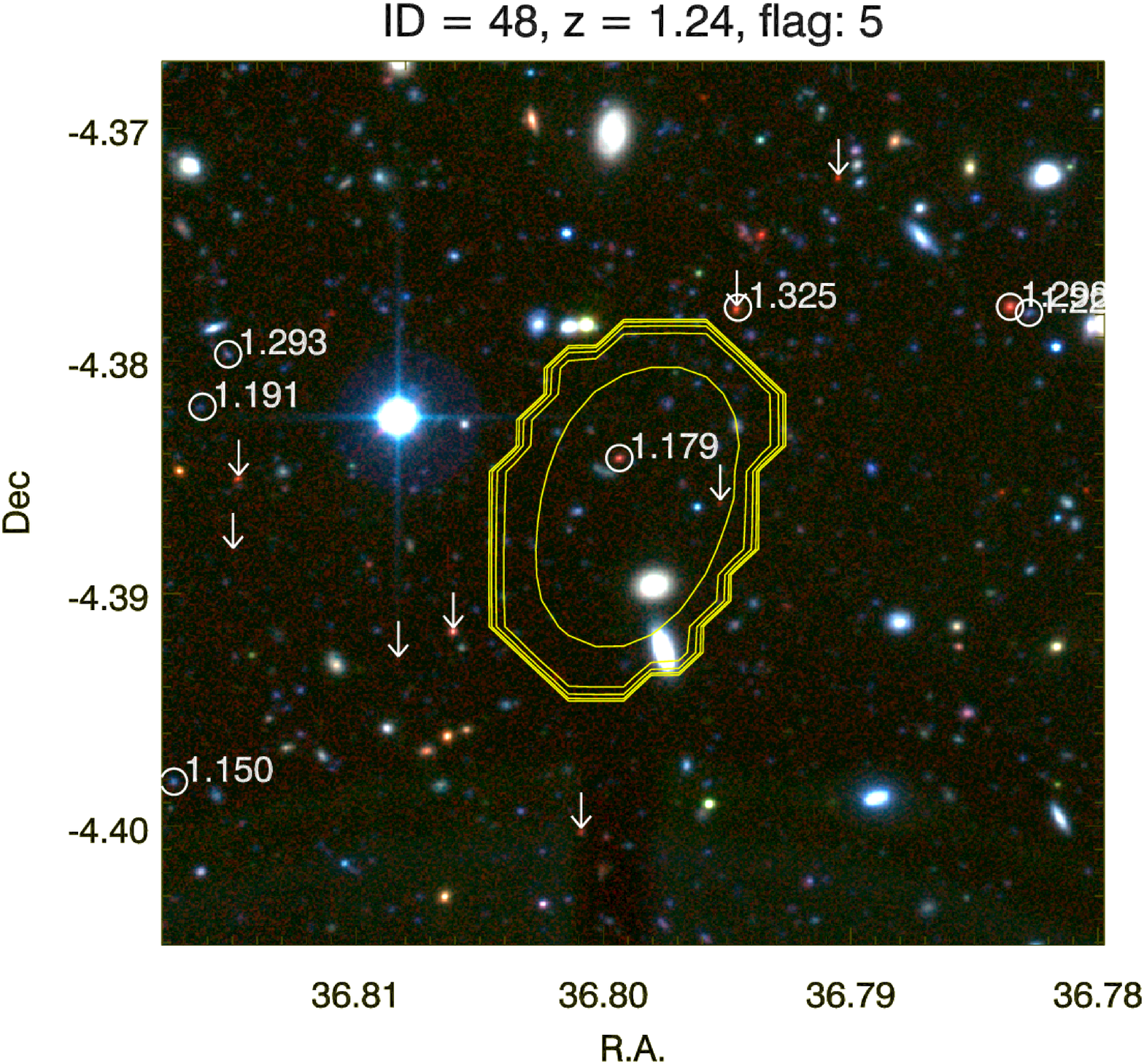}
\caption{As in Fig.~\ref{fig:colim-WIRDXC_J0224.3-0408}, but for candidate WIRDXC J0227.2-0423.}
\label{fig:colim-WIRDXC_J0227.2-0423}
\end{figure}

We find a $\sim4\sigma$ signal for this X-ray source, which is clearly extended (Fig.~\ref{fig:colim-WIRDXC_J0227.2-0423}). The red-sequence gives two possible solutions, one at $z=1.24^{+0.04}_{-0.09}$ with a significance of $2.0\sigma$ and the second at $z=0.23^{+0.03}_{-0.02}$ with a significance of $3.3\sigma$. 

A number of the $z\approx1.24$ red-sequence selected galaxies lie within the
X-ray signal, whilst several of the brightest lie to the North-West of the
X-ray detection. One of these has a spectroscopic confirmation of $z=1.325$
and lies close to the boundary of the X-ray signal. A further galaxy
selected as being a cluster member via the red-sequence method (and also
close to the edges of the X-ray detection) is found to have
spectroscopically measured redshift of $z=1.481$, somewhat higher than our
estimated cluster redshift. The full spectroscopic sample of galaxies within
$\sim1.2\arcmin$ of the X-ray centre is given in
table~\ref{table:spec_1_48}, ordered by their distance from the centre of
the X-ray detection, $\Delta r$. We also provide our photometric redshift
estimate ($z_{phot}$) for each object where available. The flag value
provided gives the confidence level of the spectroscopic redshift
measurement ($z_{spec}$) as described in section~\ref{sec:specz}.

\begin{table}
\caption{Spectroscopically confirmed galaxies around WIRDXC J0227.2-0423 close to the predicted cluster redshift.}             
\label{table:spec_1_48}      
\centering          
\begin{tabular}{lllllll}     
\hline
R.A.           &Dec.          & $\Delta r$ &$K_s$& $z_{phot}$&$z_{spec}$ & Flag\\
\multicolumn{2}{c}{(J2000)}& $(^{\circ})$& (AB) & & & \\
\hline
   36.7946 &   -4.3777 &    0.0094 &   21.09 &   1.40$^{+0.04}_{-0.04}$ &   1.325 &    2 \\[0.08cm]
   36.8162 &   -4.3820 &    0.0179 &   24.87 &   1.17$^{+0.06}_{-0.05}$ &   1.191 &    2 \\[0.08cm]
   36.7994 &   -4.3842 &    0.0020 &   21.14 &   1.53$^{+0.06}_{-0.05}$ &   1.179 &    1 \\[0.08cm]
   36.7836 &   -4.3777 &    0.0174 &   20.45 &   1.55$^{+0.07}_{-0.05}$ &   1.298 &    1 \\[0.08cm]
   36.8152 &   -4.3797 &    0.0176 &   23.42 &   1.22$^{+0.06}_{-0.05}$ &   1.293 &    3 \\[0.08cm]
   36.7828 &   -4.3779 &    0.0180 &   22.87 &   1.81$^{+0.06}_{-0.27}$ &   1.227 &   22 \\[0.08cm]
   36.8174 &   -4.3981 &    0.0221 &   22.99 & --- &   1.150 &    2 \\[0.08cm]
\hline
\end{tabular}
\end{table}

We estimate a cluster mass of $M_{200}=0.91\times10^{14}\rm{M_{\odot}}$, an X-ray luminosity of $L_x=5.0\times10^{43}$erg/s and a cluster radius of $r_{200}=0.019^{\circ}$ based on our $z=1.32$ cluster redshift estimate. The high sensitivity of X-rays towards low redshift groups means that the contribution to the X-ray detection from the z=0.23 component will be very important, even in the presence of a confirmed high-z cluster. In this case the $L_x$ and $M_{200}$ should be treated as upper limits.

\subsubsection{WIRDXC J0225.2-0429 (D1-58)}

\begin{figure}
\centering
\includegraphics[height=8.5cm,angle=270.]{15135f07a.ps}
\includegraphics[width=9.0cm]{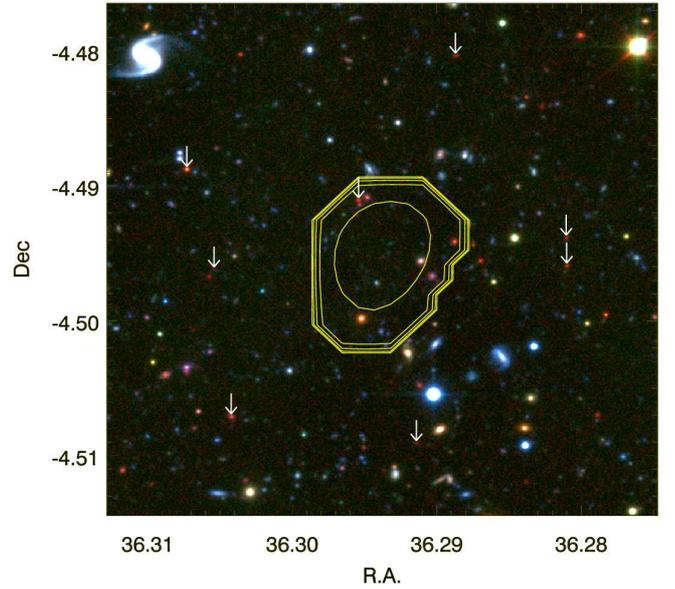}
\caption{As in Fig.~\ref{fig:colim-WIRDXC_J0224.3-0408}, but for candidate WIRDXC J0225.2-0429.}
\label{fig:colim-WIRDXC_J0225.2-0429}
\end{figure}

X-ray source 58 the D1 field is a flag 5 cluster candidate with an estimated
redshift of $z=1.46$. The red-sequence analysis results and thumbnail of the
candidate are shown in Fig.~\ref{fig:colim-WIRDXC_J0225.2-0429}. The red-sequence
detection at $z=1.46$ is relatively weak compared to other candidates, but
we see few signs of alternative red-sequence redshifts based on the
analysis. The X-ray signal is detected at a level of $\sim4\sigma$ and we
include this source in the high redshift catalogue to help in indicating the
upper limits on the numbers of high redshift clusters in this field.

Iteratively solving for the cluster properties using the scaling relations and the $z=1.46$ redshift solution, we find a cluster mass of $M_{200}=1.03\pm0.13\times10^{14}~M_{\odot}$, an X-ray luminosity of $L_x=7.49\pm2.43\times10^{43}~\rm{ergs}/\rm{s}$ and a radius of $r_{200}=0.0186^{\circ}$.

\subsubsection{WIRDXC J0225.8-0434 (D1-64)}

\begin{figure}
\centering
\includegraphics[width=8.5cm]{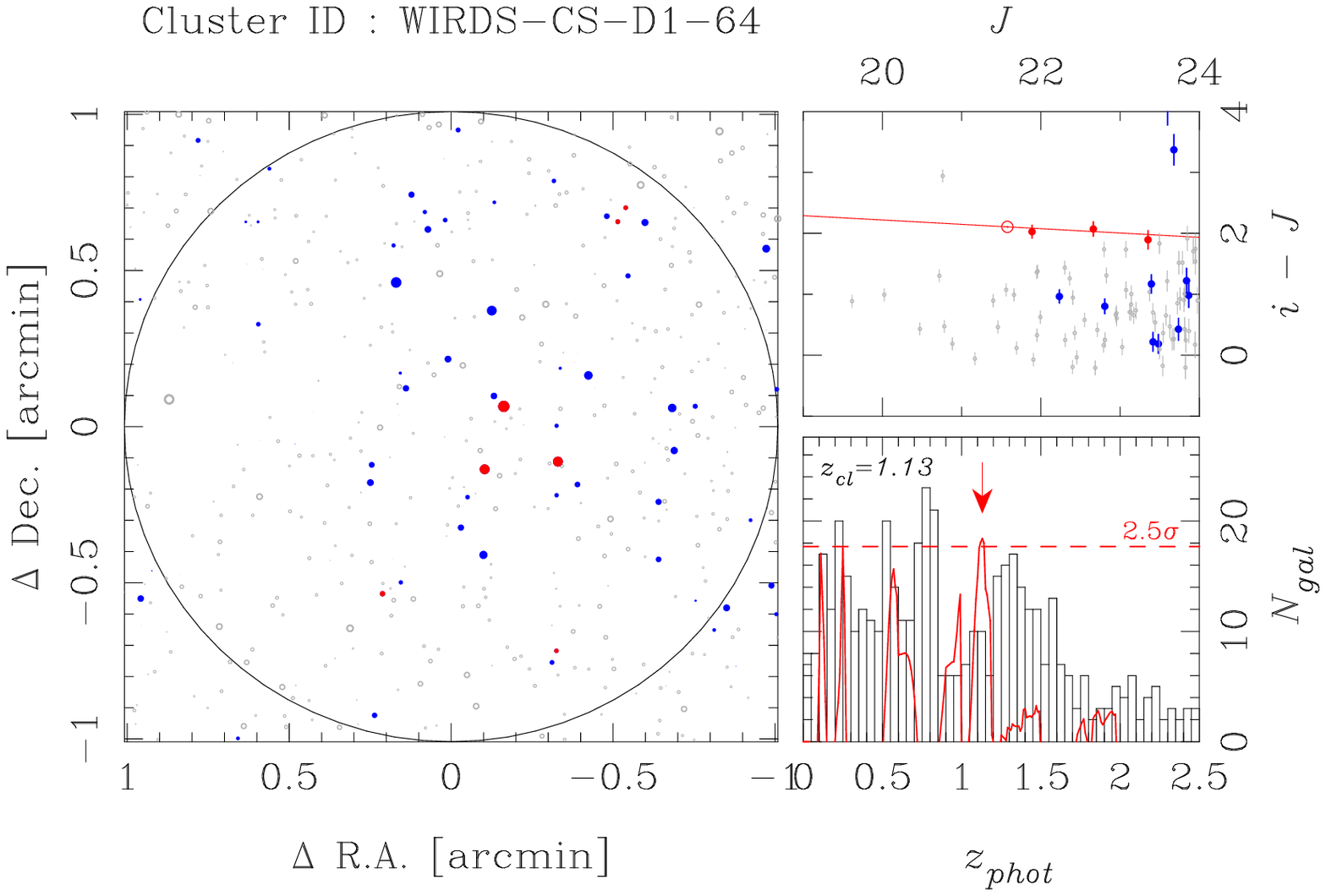}
\includegraphics[width=9.0cm]{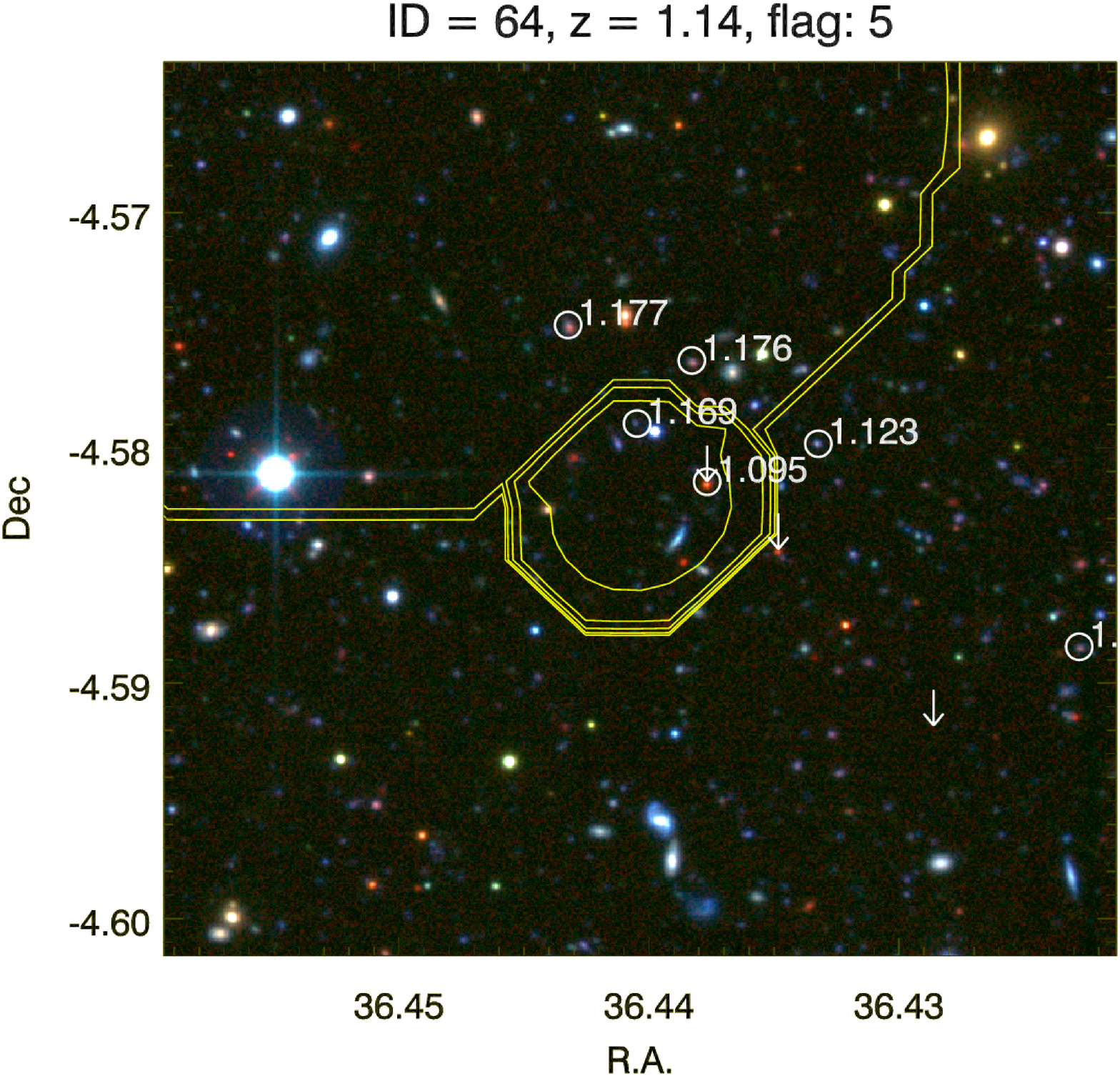}
\caption{As in Fig.~\ref{fig:colim-WIRDXC_J0224.3-0408}, but for candidate WIRDXC J0225.8-0434. Spectroscopic redshifts are shown in green in the lower panel.}
\label{fig:colim-WIRDXC_J0225.8-0434}
\end{figure}

The thumbnail for candidate WIRDXC J0225.8-0434 is shown in
Fig.~\ref{fig:colim-WIRDXC_J0225.8-0434}. The extended X-ray signal, detected at
a signal of $\sim5\sigma$, shows a relatively compact structure. The
red-sequence analysis estimates a cluster redshift of
$z=1.13^{+0.06}_{-0.06}$ with a confidence of $\sim2.6\sigma$.

We find three $z_{phot}\approx1.13$ red-sequence selected objects within the
X-ray contours, with a number of other sources at comparable photometric
redshifts close-by. A number of spectroscopically observed objects are
available around this object (see Fig.~\ref{table:spec_1_64}), two of which
lie close to the estimated cluster redshift: the first at $z_{spec}=1.169$
and the second at $z_{spec}=1.123$. We note that neither is selected as
being a cluster member via the red-sequence analysis.

\begin{table}
\caption{Spectroscopically confirmed objects around WIRDXC J0225.8-0434 within $\Delta z=0.1$ of the estimated cluster redshift.}             
\label{table:spec_1_64}      
\centering          
\begin{tabular}{lllllll}     
\hline
R.A.           &Dec.          & $\Delta r$ &$K_s$& $z_{phot}$&$z_{spec}$ & Flag\\
\multicolumn{2}{c}{(J2000)}& $(^{\circ})$& (AB) & & & \\
\hline
   36.4405 &   -4.5790 &    0.0036 &   23.81 &   1.06$^{+0.05}_{-0.06}$ &   1.169 &    3 \\[0.08cm]
   36.4332 &   -4.5798 &    0.0076 &   22.61 &   1.07$^{+0.03}_{-0.03}$ &   1.123 &    3 \\[0.08cm]
   36.4377 &   -4.5814 &    0.0029 &   20.69 &   1.05$^{+0.04}_{-0.04}$ &   1.095 &    2 \\[0.08cm]
   36.4383 &   -4.5763 &    0.0066 &   21.73 &   1.12$^{+0.03}_{-0.03}$ &   1.176 &    3 \\[0.08cm]
   36.4432 &   -4.5748 &    0.0083 &   21.05 &   1.23$^{+0.03}_{-0.03}$ &   1.177 &    3 \\[0.08cm]
   36.4228 &   -4.5884 &    0.0185 &   22.65 &   1.12$^{+0.04}_{-0.03}$ &   1.170 &    2 \\[0.08cm]
\hline                  
\end{tabular}
\end{table}

Taking the estimated cluster redshift of $z\sim1.13$, we estimate a cluster
mass of $M_{200}=1.15\times10^{14}\rm{M_{\odot}}$, an X-ray luminosity of
$L_x=6.5\times10^{43}$erg/s and a cluster radius of $r_{200}=0.021^{\circ}$.

\subsubsection{JKCS 041 (D1-76)}

\begin{figure}
\centering
\includegraphics[width=8.5cm]{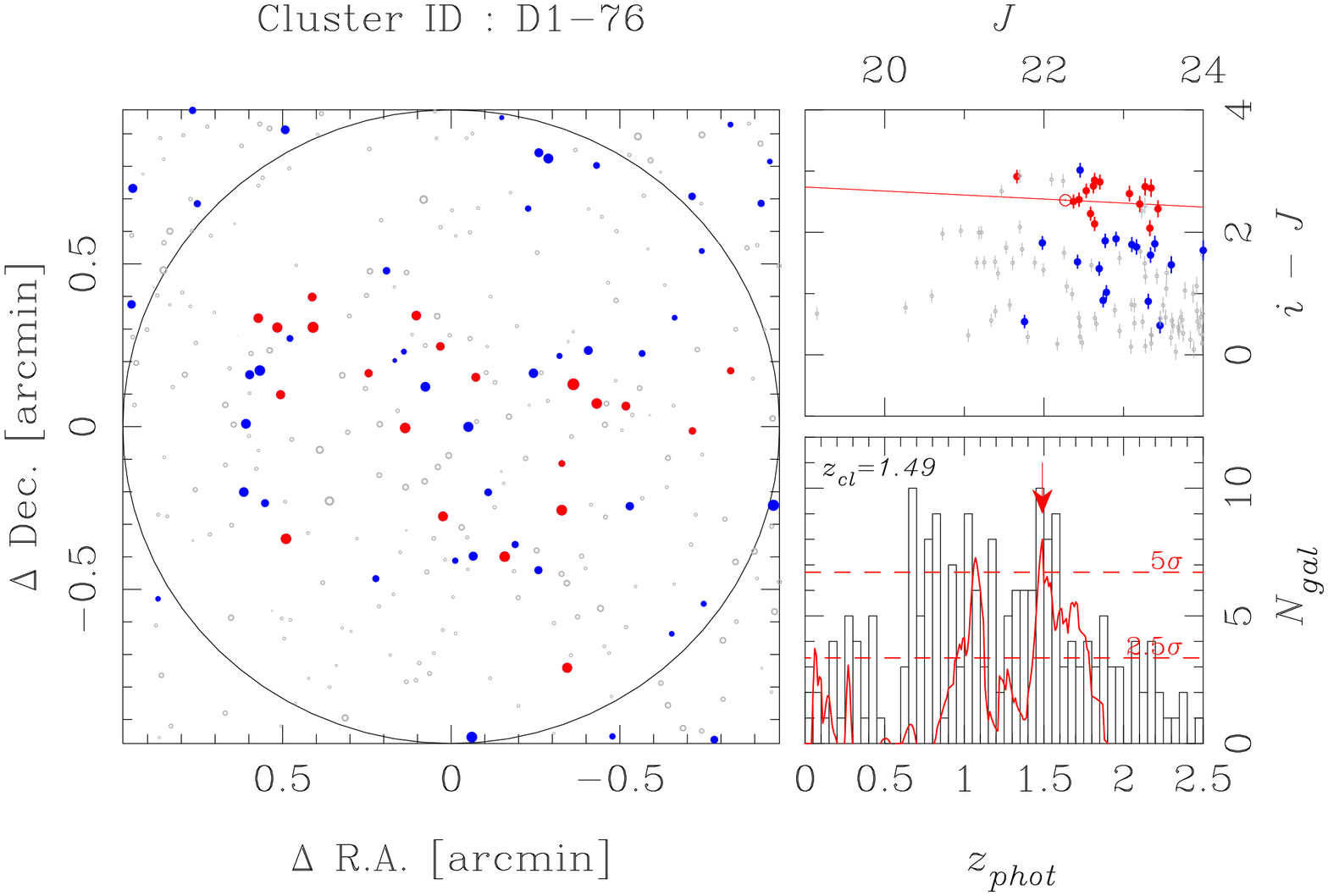}
\includegraphics[width=9.0cm]{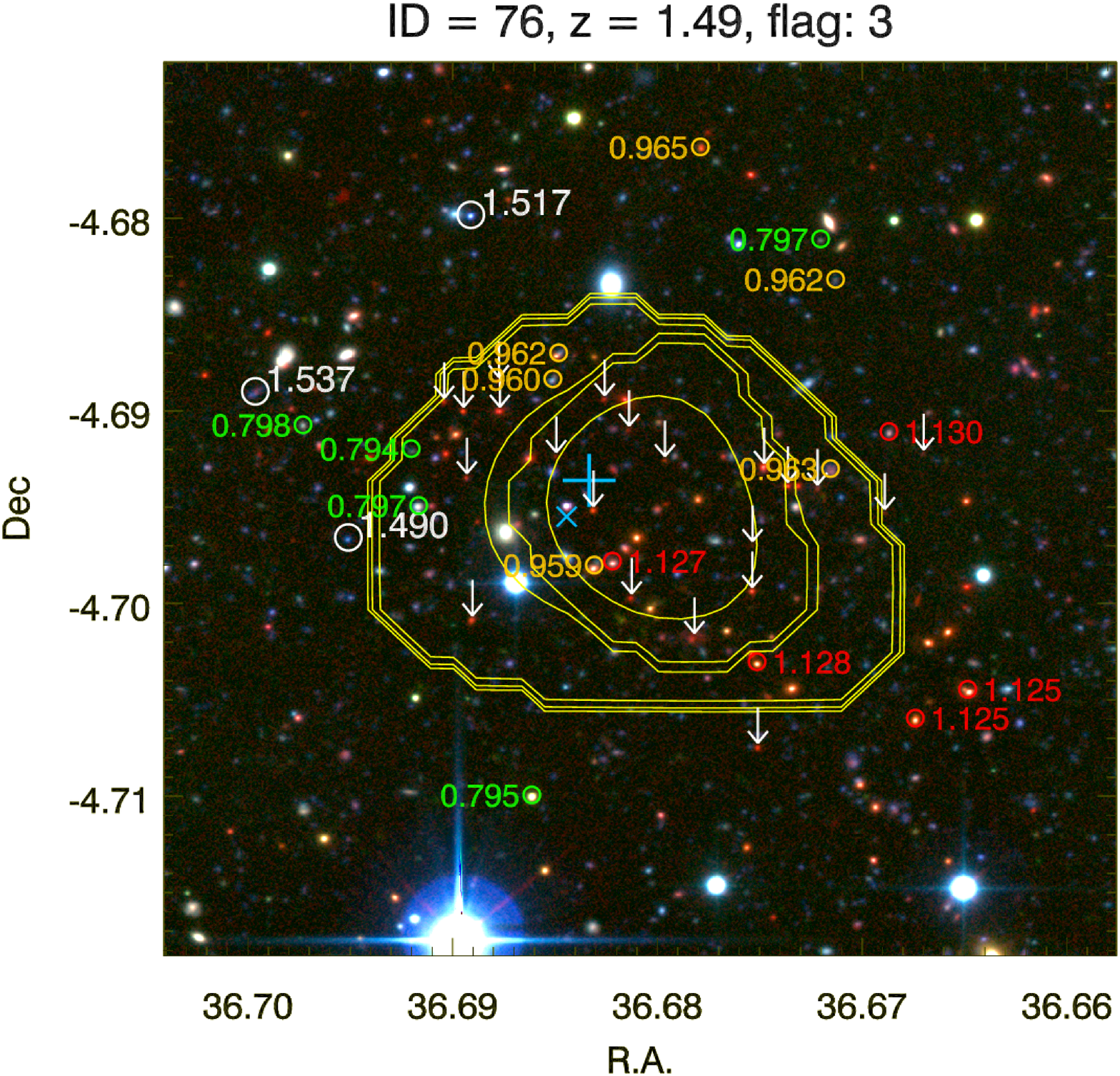}
\caption{As in Fig.~\ref{fig:colim-WIRDXC_J0224.3-0408}, but for candidate JKCS 041 and with an image size of $2.8\arcmin\times2.8\arcmin$. The coordinates of the cluster as given by \citet{2009A&A...507..147A} are shown by the blue cross.{The location of the filtered out point-source is given by the blue $\times$.}}
\label{fig:colim-JKCS_041}
\end{figure}

This candidate was reported by \citet{2009A&A...507..147A}. It is shown in Fig.~\ref{fig:colim-JKCS_041} and appears to be a complex combination of different structures along the line of sight. The X-ray detection is clearly extended with a detection of $10.4\pm1.5\times10^{-15}\text{ergs}/\text{s}/\text{cm}^2$, making it the second brightest object in our $z\gtrsim1.1$ cluster catalogue. {Within the extended X-ray emission we also detected a point-source signal in the XMM data, the location of which is given by the blue $\times$ in Fig.~\ref{fig:colim-JKCS_041}. This point-source was identified and filtered out by the point-source removal processing and was measured to have a flux of $5.6\times10^{-15}\text{ergs}/\text{s}/\text{cm}^2$.}

\begin{figure}
\centering
\includegraphics[width=9.0cm]{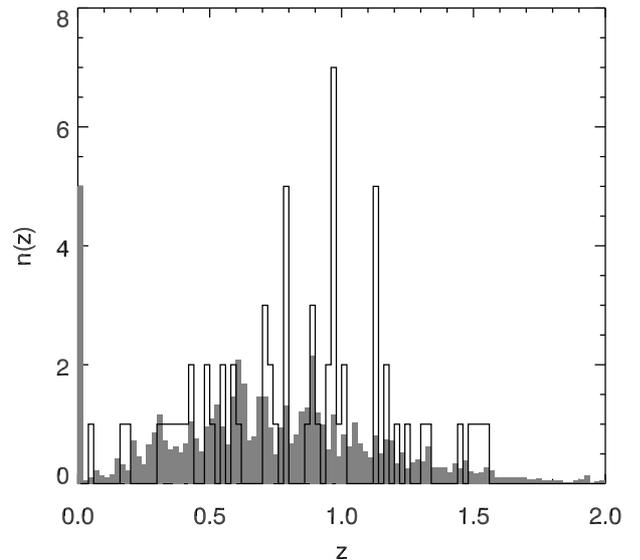}
\caption{Spectroscopic redshift distribution within $1.2\arcmin$ of the centre of the X-ray emission for candidate JKCS 041 (solid black line histogram). The redshift distribution of the entire sample is shown by the filled grey histogram.}
\label{fig:speczhist-JKCS 041}
\end{figure}

From the public VVDS Deep Survey, a large number of spectroscopic redshifts are available within the field. Analyzing the distribution of the spectroscopic identifications within $1.2\arcmin$ of the X-ray profile centre (black histogram in Fig.~\ref{fig:speczhist-JKCS 041}), three redshift peaks are evident at redshifts of $z=0.80$, $z=0.96$ and $z=1.13$ (the redshift distribution of the complete VVDS deep sample is given by the filled grey histogram in Fig.~\ref{fig:speczhist-JKCS 041}). We plot the galaxies identified at these redshifts in Fig.~\ref{fig:colim-JKCS_041}. Identified $z\approx0.8$ galaxies are given by green circles, $z\approx0.96$ galaxies by orange circles and $z\approx1.12$ galaxies by red circles.

Combining this data with the photometric redshift data and the red-sequence
analysis, we find firstly that the $z=0.80$ galaxies appear to be offset
from the X-ray emission, whilst we see no significant signal in the
red-sequence analysis. From the photometric redshift analysis, we find that
the photometric data for these galaxies are (where an accurate photometric
redshift has been calculated) best fit by starburst templates, corroborating
the lack of any red-sequence in these objects. Given the spatial
distribution of these galaxies and their blue colours, it is likely that
these are not associated with the bulk of the extended X-ray signal.

Taking the $z=0.96$ spectroscopic redshift peak, we find a small signal in
the red-sequence analysis, whilst the positions of the spectroscopically
confirmed $z=0.96$ galaxies are well correlated with the X-ray contours.
Again however, for the galaxies for which the photometric redshifts
corroborate the spectroscopic redshifts, the photometric data are best fit
by starburst templates. Again the blue colours of these galaxies suggest
that they may not be associated with the bulk of the X-ray producing gas,
although they remain a strong candidate for some sort of structure,
potentially a small group or filamentary structure along the line of sight.

Moving to the third solution, we find a strong signal from the red-sequence
analysis at $z=1.07\pm0.06$, which corroborates the $z=1.13$ estimate from
the spectroscopic redshift distribution. Taking the photometric fitting, the
photometry for two of the confirmed $z=1.13$ galaxies is best fit by
elliptical templates. Out of the three spectroscopically identified
solutions, we take the $z=1.13$ solution as the most plausible structure to
be associated with the bulk of the X-ray emission.

From the red-sequence analysis however, we find the strongest red-sequence solution at a redshift of $z=1.49^{+0.28}_{-0.06}$. Selected cluster members at this redshift are highlighted by white arrows in Fig.~\ref{fig:colim-JKCS_041} and are observed to align well with the X-ray contours. {Three spectroscopic redshifts at $z\approx1.5$ are available from the VVDS data, however these are not selected as part of the red-sequence analysis and are attributed with spectroscopic flags of only 1 and 2}. Despite the lack of spectroscopic confirmation, this apparent $z=1.49$ structure is a strong candidate as contributing source to the X-ray detection. 

\begin{table}
\caption{Spectroscopically confirmed objects around JKCS 041 within $\Delta z=0.1$ of the estimated cluster redshift.}             
\label{table:spec_1_76}      
\centering          
\begin{tabular}{lllllll}   
\hline
R.A.           &Dec.          & $\Delta r$ &$K_s$& $z_{phot}$&$z_{spec}$ & Flag\\
\multicolumn{2}{c}{(J2000)}& $(^{\circ})$& (AB) & & & \\
\hline
   36.6917 &   -4.6949 &    0.0108 &   20.4 &  0.76$^{+0.03}_{-0.03}$ &   0.797 &    4 \\[0.08cm]
   36.6920 &   -4.6920 &    0.0116 &   23.1 & 0.78$^{+0.04}_{-0.04}$ &   0.794 &    9 \\[0.08cm]
   36.6861 &   -4.7099 &    0.0158 &   19.6 &                                --- &   0.795 &    4 \\[0.08cm]
   36.6720 &   -4.6811 &    0.0166 &   23.0 &  0.75$^{+0.04}_{-0.04}$ &   0.797 &    2 \\[0.08cm]
   36.6973 &   -4.6907 &    0.0170 &   21.8 &  0.75$^{+0.03}_{-0.03}$ &   0.798 &    2 \\[0.08cm]
   36.6831 &   -4.6980 &    0.0037 &      --- &  2.10$^{+0.44}_{-0.71}$ &   0.959 &    3 \\[0.08cm]
   36.6851 &   -4.6883 &    0.0080 &   22.1 &  0.87$^{+0.08}_{-0.05}$ &   0.960 &    3 \\[0.08cm]
   36.6848 &   -4.6870 &    0.0090 &   20.4 &  0.99$^{+0.04}_{-0.03}$ &   0.962 &    2 \\[0.08cm]
   36.6715 &   -4.6930 &    0.0095 &   21.2 &  1.20$^{+0.03}_{-0.03}$ &   0.963 &    2 \\[0.08cm]
   36.6713 &   -4.6832 &    0.0153 &   23.0 &  0.94$^{+0.04}_{-0.06}$ &   0.962 &    2 \\[0.08cm]
   36.6779 &   -4.6763 &    0.0190 &   20.8 &  0.82$^{+0.04}_{-0.04}$ &   0.965 &    2 \\[0.08cm]
   36.6822 &   -4.6978 &    0.0030 &   24.8 &  0.95$^{+0.10}_{-0.11}$ &   1.127 &   22 \\[0.08cm]
   36.6751 &   -4.7030 &    0.0098 &   20.3 &  1.04$^{+0.03}_{-0.03}$ &   1.128 &    2 \\[0.08cm]
   36.6687 &   -4.6911 &    0.0128 &   21.6 &  1.08$^{+0.03}_{-0.03}$ &   1.130 &    3 \\[0.08cm]
   36.6674 &   -4.7059 &    0.0173 &   20.0 &  1.12$^{+0.04}_{-0.04}$ &   1.125 &    3 \\[0.08cm]
   36.6648 &   -4.7045 &    0.0185 &   19.6 &  0.96$^{+0.03}_{-0.03}$ &   1.125 &    3 \\[0.08cm]
   36.6951 &   -4.6966 &    0.0143 &   24.0 &  2.78$^{+0.04}_{-0.04}$ &   1.490 &    2 \\[0.08cm]
   36.6891 &   -4.6798 &    0.0174 &   22.4 &  1.56$^{+0.14}_{-0.07}$ &   1.517 &    2 \\[0.08cm]
   36.6996 &   -4.6890 &    0.0198 &   22.2 &   1.77$^{+0.13}_{-0.12}$ &   1.537 &    1 \\[0.08cm]
\hline                  
\end{tabular}
\end{table}

As stated, this X-ray source is at the same position as that reported by \citet{2009A&A...507..147A} as a $z=1.9$ cluster (the centre coordinates of the two detections are within $0.2\arcmin$ of each other). The \citet{2009A&A...507..147A} X-ray detection was made using observations from the Chandra X-ray telescope, making this XMM based observation an independent detection of this cluster. \citet{2009A&A...507..147A} performed a red-sequence analysis using UKIDSS DR3 deep $J$ and $K$ NIR data. The latest depths for these bands reported for the UKIDSS Seep Extragalactic Survey \citep{2007astro.ph..3037W} are $J(\rm{Vega})=22.2$ and $K(\rm{Vega})=20.9$ given $5\sigma$ detected point sources with $2\arcsec$ apertures ($\sim23.1$ and $\sim22.8$ respectively in the AB system). This compares to depths (in the AB system) from the WIRDS data of $J=24.3$, $H=24.1$ and $K_s=24.1$, again using $5\sigma$ detected point sources with $2\arcsec$ aperture.

Using the UKIDSS $JK$ data, \citet{2009A&A...507..147A} reported a cluster redshift of $z=1.9$ with a confidence of $6.5\sigma$. In conjunction with this detection, they also comment that at least one more cluster is detected along the line of sight. Based on the red-sequence analysis with the CFHTLS/WIRDS data, we find no significant detection of a red-sequence at $z=1.9$.

{We make estimates for cluster properties taking the $z=1.13$ and $z=1.49$ candidate structures. As discussed, the $z=1.13$ solution appears to be confirmed by spectroscopic observations, whilst the $z=1.49$ solution remains tentative and based on photometric/red-sequence observations. Taking these two solutions, and using the total extended flux measurement in each case, we place upper limits on the cluster mass for a cluster at $z=1.13$ of $M_{200}<1.78\pm0.15\times10^{14}\rm{M_{\sun}}$ and for a cluster at $z=1.49$ of $M_{200}<1.88\pm0.16\times10^{14}\rm{M_{\sun}}$. We note that based on this estimated upper limit for the mass of the possible $z=1.49$ structure that the spectroscopic redshifts of the three objects span too large a range to all be part of the same bound structure.}

\subsection{CFHTLS D4 Extended X-ray Sources}

We have a total of six $z\gtrsim1.1$ extended X-ray cluster candidates identified in the CFHTLS D4
field based on the initial X-ray detections in combination with the
red-sequence/photometric redshift analysis. The distribution of these
candidates are shown (circled in red) in Fig.~\ref{fig:xray-maps},
over-layed on the filtered X-ray map.

\subsubsection{WIRDXC J2216.3-1729 (D4-17)}

The X-ray detection for candidate WIRDXC J2216.3-1729 is detected at a level of $\sim2\sigma$. This is a relatively weak source, with a tentative red-sequence detection at $z=1.46$. This detection (at a level of $\sim2.5\sigma$) correlates with a peak in the photometric redshift distribution at the same redshift. Given the low detection signals involved, we attribute this candidate flag 5 and include it in the sample as part of placing upper limits on the numbers of clusters in the fields.

\begin{figure}
\centering
\includegraphics[height=8.5cm,angle=270.]{15135f11a.ps}
\includegraphics[width=9.0cm]{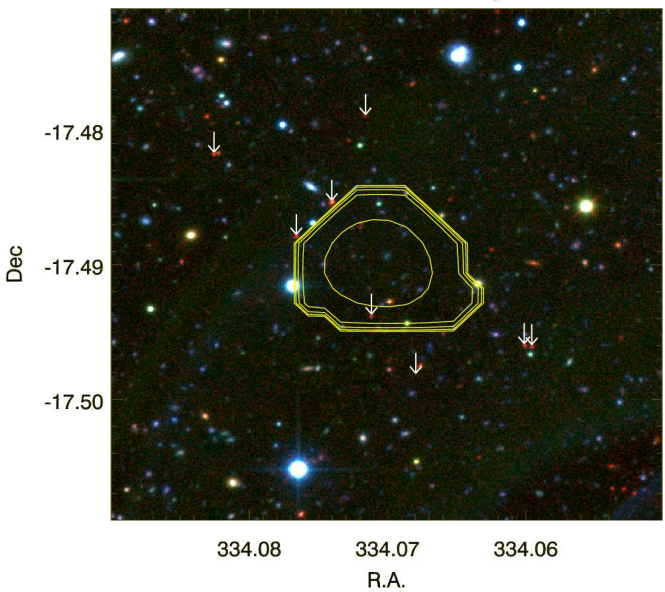}
\caption{As in Fig.~\ref{fig:colim-WIRDXC_J0224.3-0408}, but for candidate WIRDXC J2216.3-1729.}
\label{fig:colim-WIRDXC_J2216.3-1729}
\end{figure}

From the measured X-ray flux and the redshift estimate presented here, we find a cluster mass of $M_{200}=0.68\pm0.15\times10^{14}~\rm{M_{\sun}}$, an X-ray luminosity of $L_x=3.72\pm1.45\times10^{43}~\rm{ergs}/\rm{s}$ and a radius of $r_{200}=0.0162^{\circ}$.

\subsubsection{XMMXCS J2215.9+1738 (D4-25)}

\begin{figure}
\centering
\includegraphics[width=8.5cm]{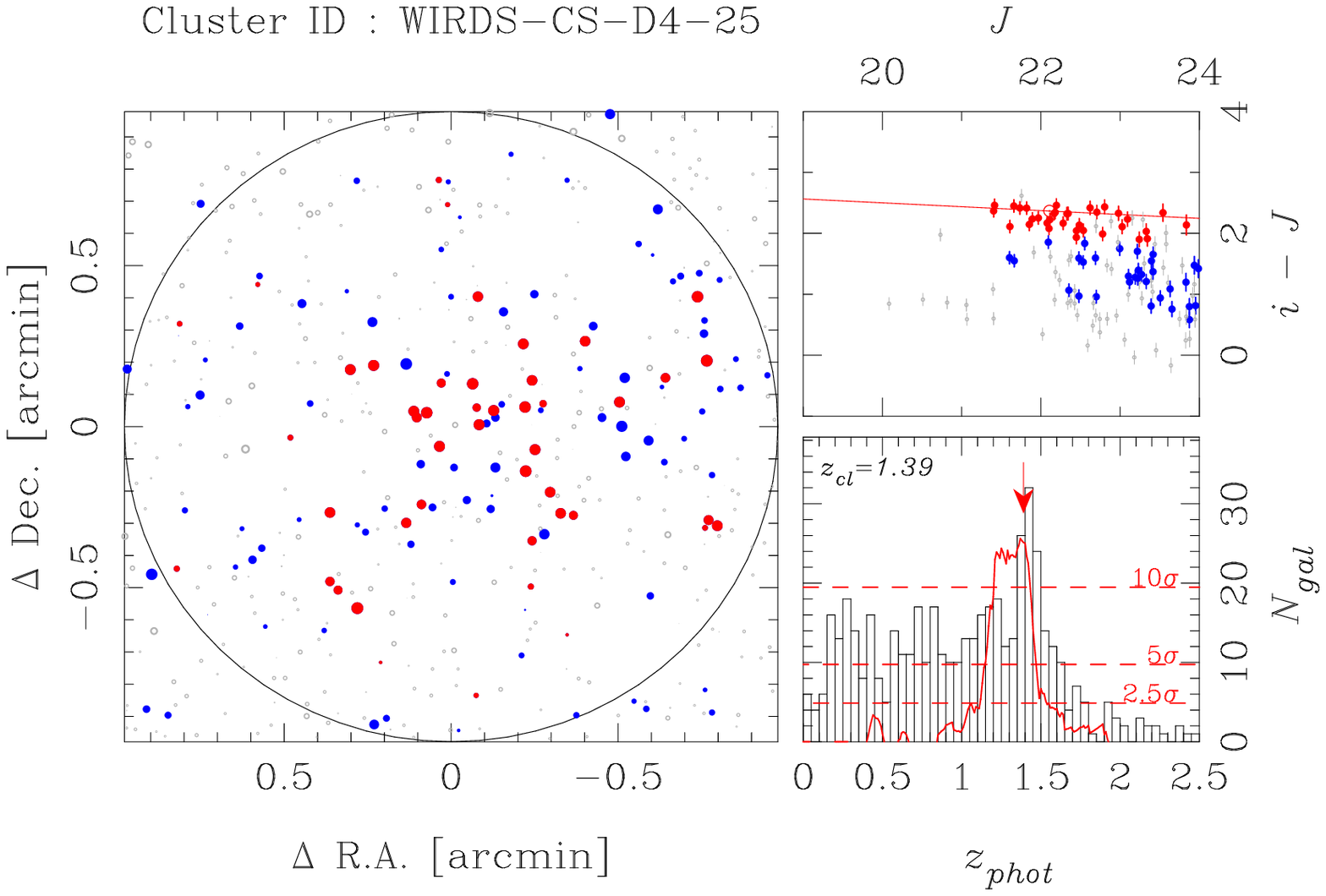}
\includegraphics[width=9.0cm]{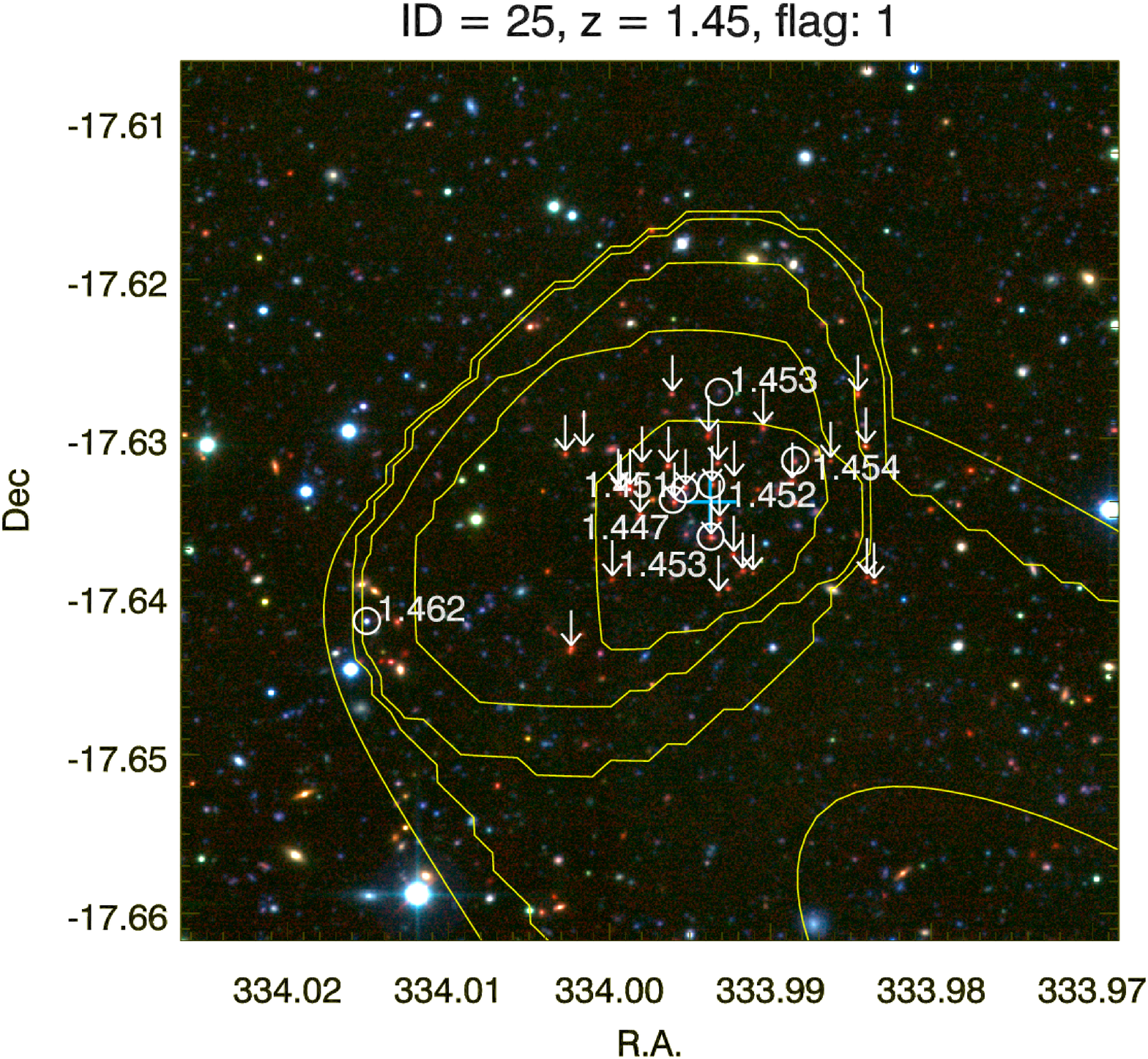}
\caption{As in Fig.~\ref{fig:colim-WIRDXC_J0224.3-0408}, but for candidate XMMXCS J2215.9-1738 and with an image size of $3.3\arcmin\times3.3\arcmin$. The white circles and numbers in the lower panel indicate objects with spectroscopic redshifts. The blue cross shows the coordinates of XMMXCS J2215.9-1738 as published by \citet{stanford06}.}
\label{fig:colim-XMMXCS_J2215.9-1738}
\end{figure}

This detection is the same as the spectroscopically confirmed $z=1.4$ cluster of \citet{stanford06}. From our imaging, we see a strong clustering of red-sequence galaxies both spatially (Fig.~\ref{fig:colim-XMMXCS_J2215.9-1738} lower panel) and in magnitude-colour space (Fig.~\ref{fig:colim-XMMXCS_J2215.9-1738} upper panels). The red-sequence analysis and photometric redshift distribution both indicate a cluster redshift of $z=1.37$. This is reassuringly close to the redshift measured by \citet{stanford06} of $z=1.45$ and confirms the reliability of our redshift estimation techniques. In Fig.~\ref{fig:colim-XMMXCS_J2215.9-1738}, we show those galaxies identified as cluster members by our red-sequence analysis. Four of these have been spectroscopically confirmed as cluster members by \citet{stanford06} and are circled and labeled in the figure. \citet{stanford06} present spectra for a further two cluster members that are not selected as cluster members as part of our analysis, whilst from the AAOmega spectroscopic data we observe a QSO at $z=1.462$ lying in what appear to be the cluster outskirts. We give all the available spectroscopic redshifts from \citet{stanford06} and AAOmega in table~\ref{table:spec_4_25}.

\begin{table}
\caption{Spectroscopically confirmed objects around XMMXCS J2215.9-1738 within $\Delta z=0.1$ of the estimated cluster redshift.}             
\label{table:spec_4_25}      
\centering
\begin{tabular}{lllllll}     
\hline
R.A.           &Dec.          & $\Delta r$ &$K_s$& $z_{phot}$&$z_{spec}$ & Class\\
\multicolumn{2}{c}{(J2000)}& $(^{\circ})$& (AB) & & & \\
\hline
  333.9961 &  -17.6339 &    0.0014 &   20.56 &   1.38$^{+0.04}_{-0.04}$ &   1.447 &     Gal \\[0.08cm]
  333.9953 &  -17.6332 &    0.0023 &   20.87 &   1.42$^{+0.04}_{-0.04}$ &   1.451 &     Gal \\[0.08cm]
  333.9937 &  -17.6329 &    0.0039 &   20.63 &   1.51$^{+0.04}_{-0.05}$ &   1.452 &     Gal \\[0.08cm]
  333.9937 &  -17.6362 &    0.0044 &   20.36 &   1.41$^{+0.03}_{-0.04}$ &   1.453 &     Gal \\[0.08cm]
  333.9932 &  -17.6271 &    0.0081 &   22.22 &   1.50$^{+0.06}_{-0.06}$ &   1.453 &     Gal \\[0.08cm]
  333.9885 &  -17.6314 &    0.0094 &   21.27 &   1.37$^{+0.10}_{-0.04}$ &   1.454 &     Gal \\[0.08cm]
  334.0151 &  -17.6415 &    0.0191 &   21.00 &   0.58$^{+0.05}_{-0.06}$ &   1.462 &     QSO \\[0.08cm]
\hline                  
\end{tabular}
\end{table}

Based on the spectroscopically measured cluster redshift of $z=1.45$, we estimate a cluster mass of $M_{200}=1.9\pm0.05\times10^{14}~\rm{M_{\sun}}$. Taking the velocity dispersion of $\sigma_v=580\pm140\kps$, measured by \citet{2007ApJ...670.1000H}, and combining this with the virial radius of $R_v=1.05~\rm{Mpc}$ estimated by the same authors, the total virial mass is estimated to be $M_{Tot}=1.5\pm0.6~\times~10^{14}~\rm{M_\odot}$, comparable to our estimate from the X-ray observation. The X-ray luminosity is estimated to be $L_x=20.50\pm0.85\times10^{43}~\rm{ergs}/\rm{s}$, with a cluster radius of $r_{200}=0.0229^{\circ}$.

\subsubsection{WIRDXC J2216.4-1748 (D4-32)}

\begin{figure}
\centering
\includegraphics[width=8.5cm]{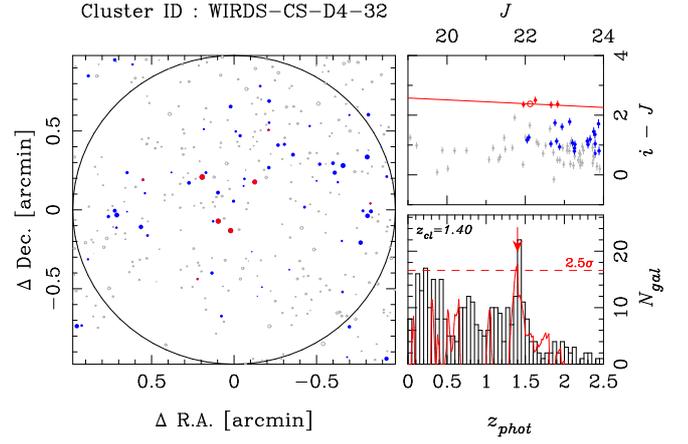}
\includegraphics[width=9.0cm]{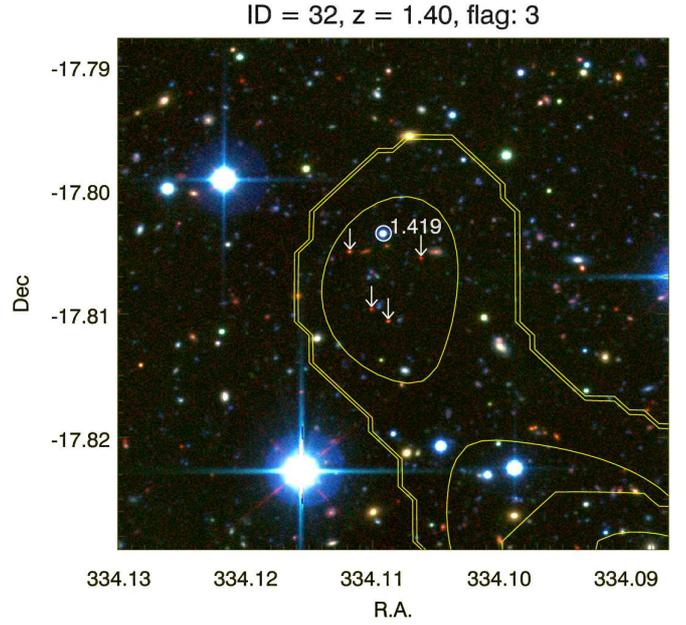}
\caption{As in Fig.~\ref{fig:colim-WIRDXC_J0224.3-0408}, but for candidate WIRDXC J2216.4-1748 and with an image size of $2.5\arcmin\times2.5\arcmin$.. The white circles and numbers in the lower panel indicate objects with spectroscopic redshifts.}
\label{fig:colim-WIRDXC_J2216.4-1748}
\end{figure}

WIRDXC J2216.4-1748 is a clear extended X-ray detection, with several red-sequence galaxies at $z=1.4$ found within the X-ray detection (Fig.~\ref{fig:colim-WIRDXC_J2216.4-1748}). The significance of the $z=1.4$ red-sequence detection is $2.7\sigma$, whilst the X-ray detection is measured with a signal of $\sim9\sigma$. Although the red-sequence analysis does not appear conclusive, we note that the detection coincides with a strong peak in the photometric redshift distribution and that no significant signal in the red-sequence analysis is seen at any other redshifts.

We note that, although we have no spectroscopic confirmation of selected cluster members, we observe a spectroscopically confirmed $z=1.42$ quasar (taken from the AAOmega data and given in table~\ref{table:spec_4_32}) within the extended X-ray detection, which we speculate may be part of this structure.

\begin{table}
\caption{Spectroscopically confirmed objects around WIRDXC J2216.4-1748 within $\Delta z=0.1$ of the estimated cluster redshift.}             
\label{table:spec_4_32}      
\centering          
\begin{tabular}{lllllll}     
\hline
R.A.           &Dec.          & $\Delta r$ &$K_s$& $z_{phot}$&$z_{spec}$ & Class\\
\multicolumn{2}{c}{(J2000)}& $(^{\circ})$& (AB) & & & \\
\hline
  334.1091 &  -17.8035 &    0.0050 &   19.41 & --- &   1.419 &     QSO \\
\hline                  
\end{tabular}
\end{table}

For this candidate, we find a cluster mass of $M_{200}=1.25\pm0.08\times10^{14}~\rm{M_{\sun}}$, an X-ray luminosity of $L_x=9.43\pm1.01\times10^{43}~\rm{ergs}/\rm{s}$ and a radius of $r_{200}=0.0203^{\circ}$.

\subsubsection{WIRDXC J2213.9-1750 (D4-37)}

\begin{figure}
\centering
\includegraphics[width=8.5cm]{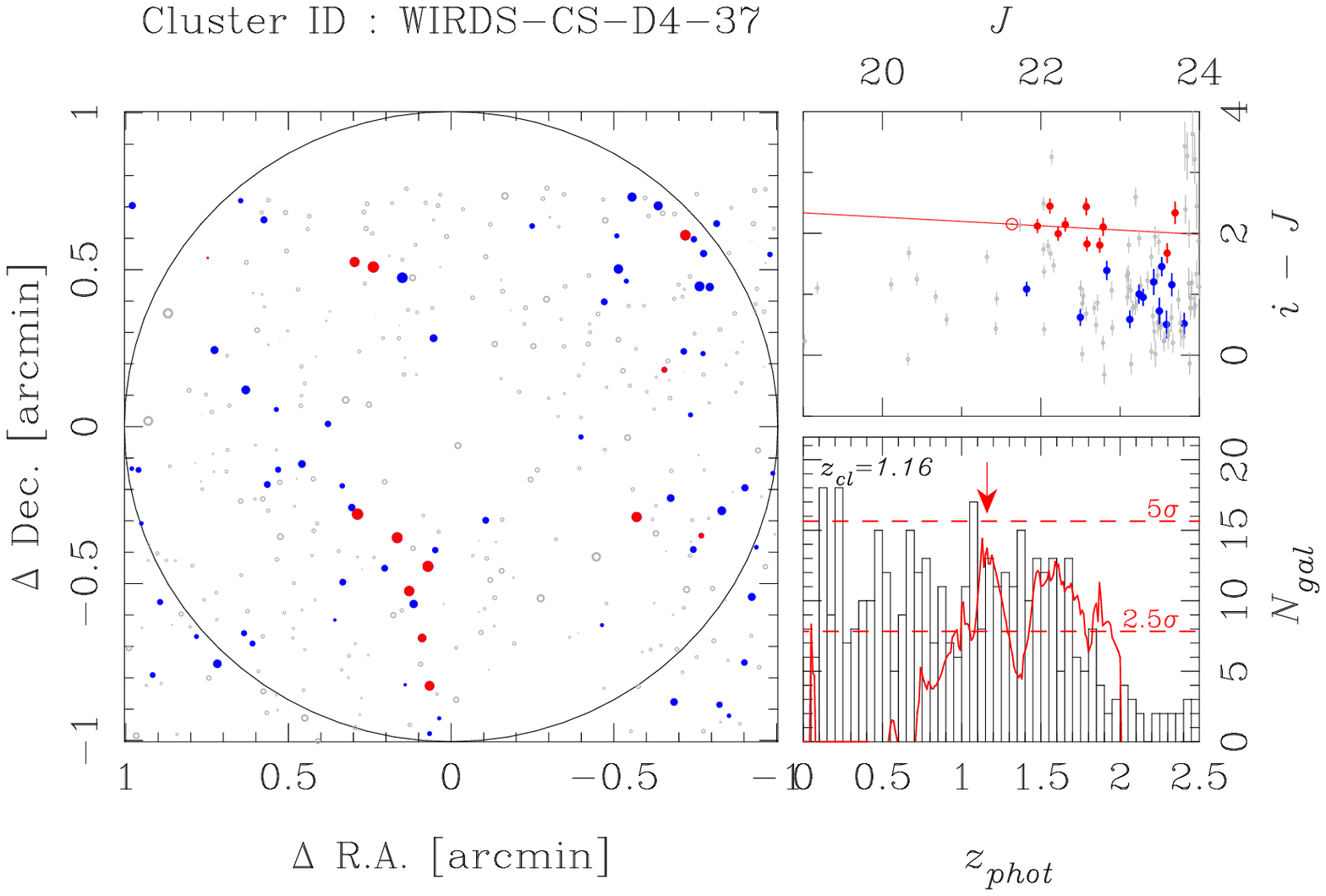}
\includegraphics[width=9.0cm]{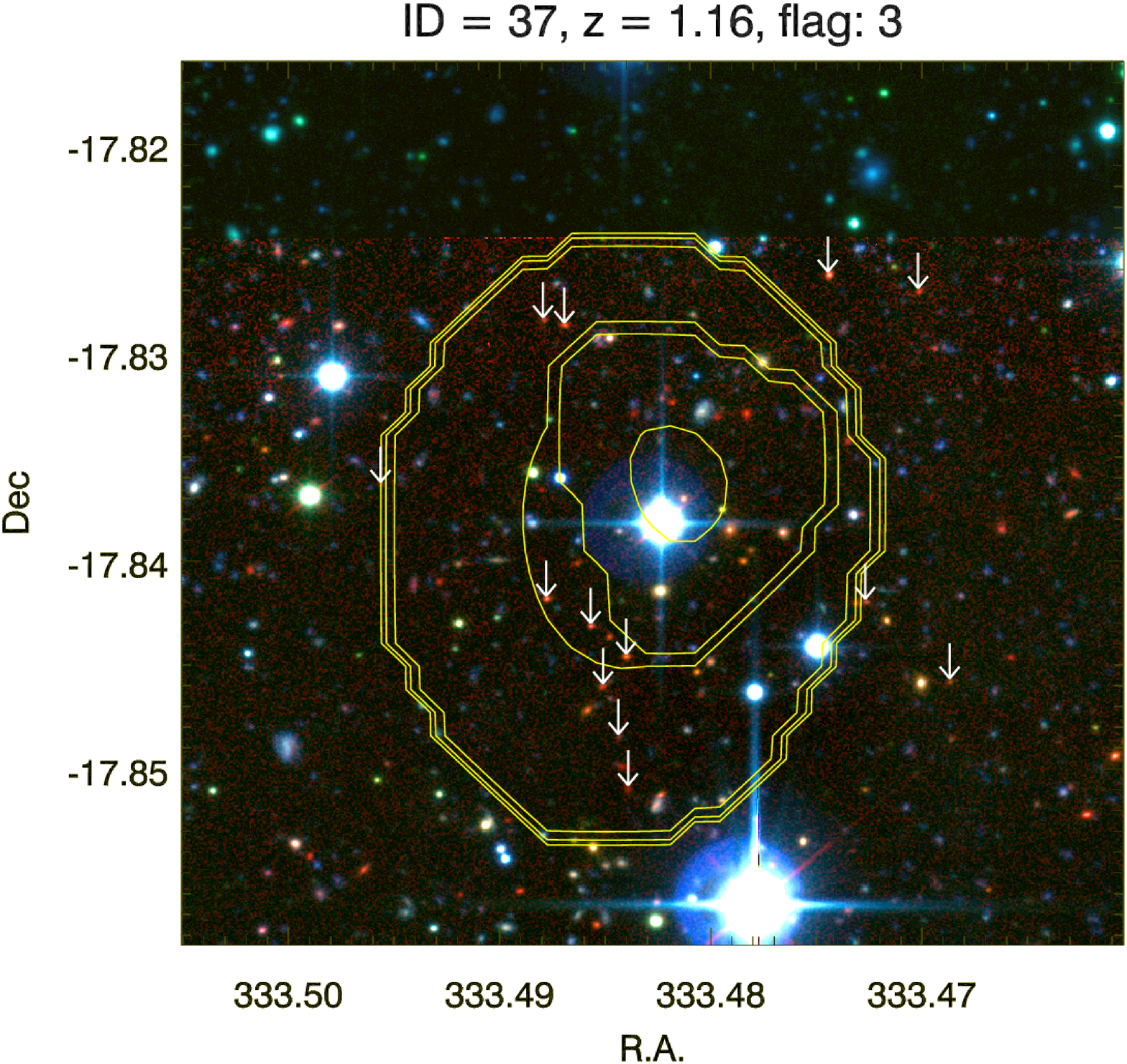}
\caption{As in Fig.~\ref{fig:colim-WIRDXC_J0224.3-0408}, but for candidate WIRDXC J2213.9-1750 and with an image size of $2.5\arcmin\times2.5\arcmin$.}
\label{fig:colim-WIRDXC_J2213.9-1750}
\end{figure}

The red-sequence analysis and $giK_s$ image with selected cluster members and spectroscopic identifications for cluster candidate WIRDXC J2213.9-1750 are shown in Fig.~\ref{fig:colim-WIRDXC_J2213.9-1750}. The X-ray detection for this source is measured at a signal of $\sim5\sigma$ and is clearly extended.

The detection lies close to the edge of our NIR data and thus have larger photometric errors than in the main regions of our data. In addition the central region is obscured by a bright star, hampering the red-sequence analysis. Despite these problems, we detect a strong peak in the red-sequence analysis at $z=1.16$ with a significance of $4.6\sigma$. The galaxies selected as cluster members are well clustered within the extended X-ray emission. 

We have only a single spectroscopic redshift from the AAOmega data in the region of this candidate, however it lies at a lower redshift and is not selected as a cluster member by our analysis. Based on a cluster redshift of $z=1.40$, we estimate a cluster mass of $M_{200}=1.0\times10^{14}\rm{M_{\odot}}$, an X-ray luminosity of $L_x=4.9\times10^{43}$erg/s and a cluster radius of $r_{200}=0.021^{\circ}$ 

\subsubsection{BLOX J2215.9-1751.6 (D4-38)}

\begin{figure}
\centering
\includegraphics[width=8.5cm]{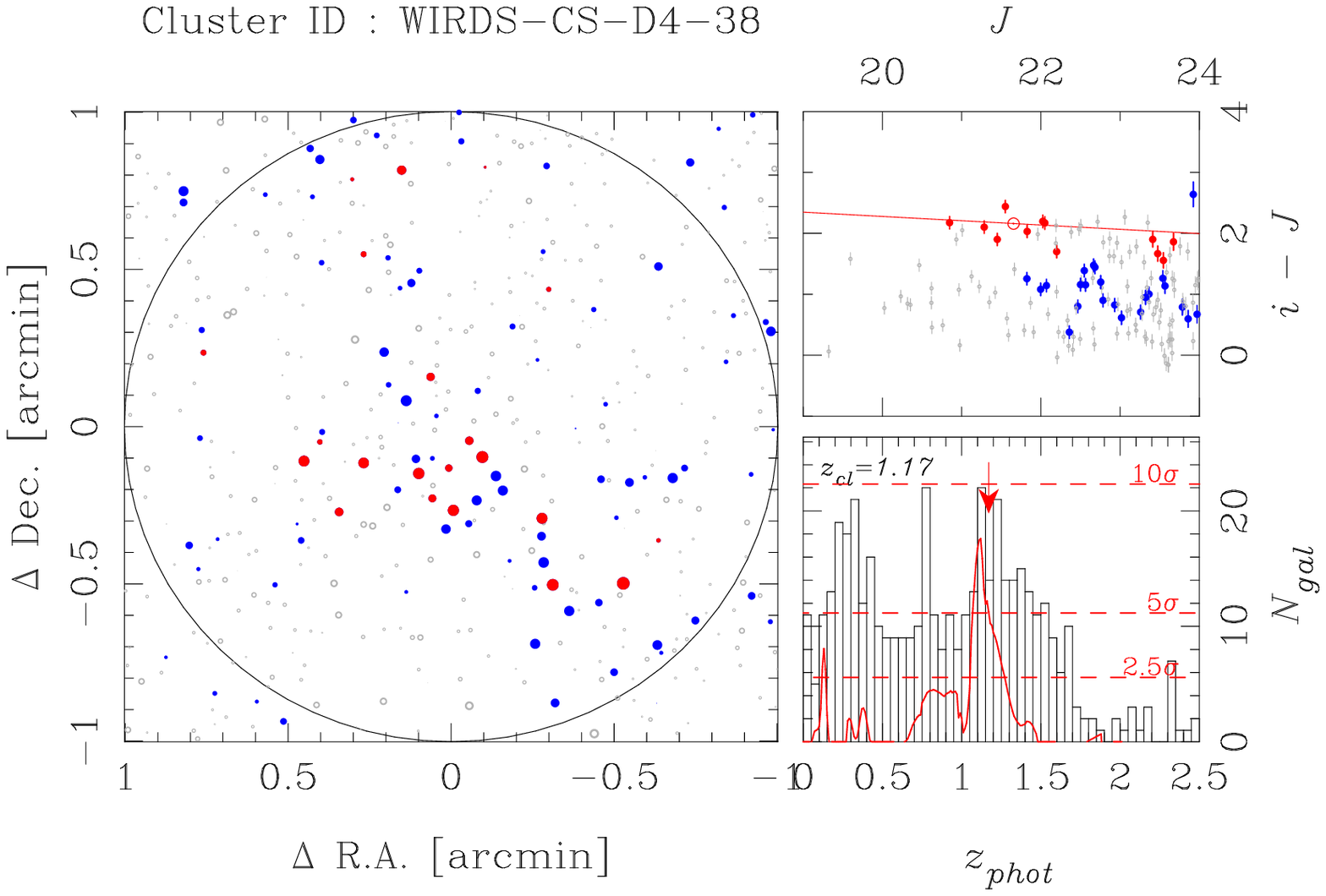}
\includegraphics[width=9.0cm]{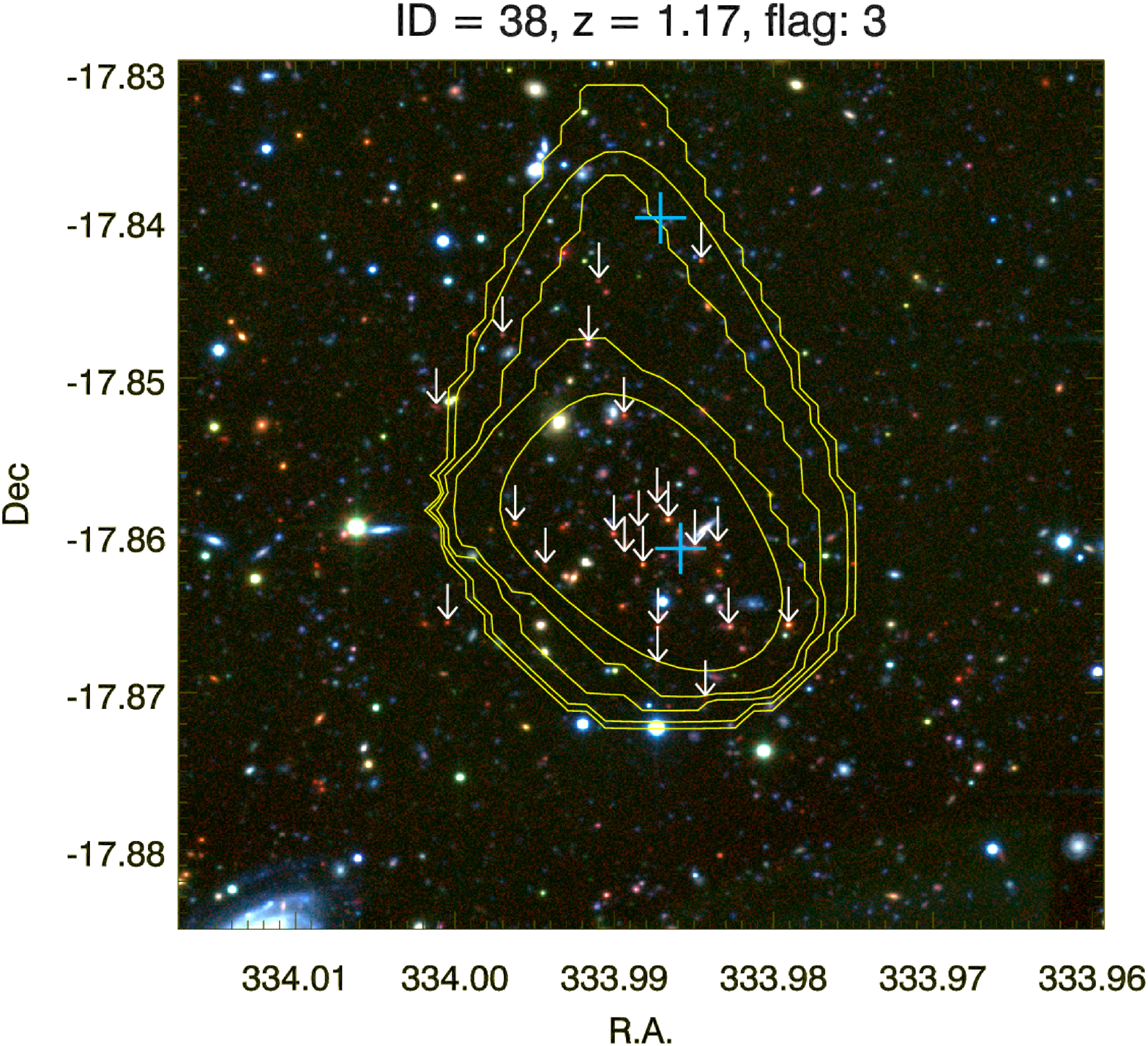}
\caption{As in Fig.~\ref{fig:colim-WIRDXC_J0224.3-0408}, but for candidate BLOX J2215.9-1751.6 and with an image size of $3.3\arcmin\times3.3\arcmin$. The blue crosses mark the central coordinates of the detections of BLOX J2215.9-1751.6 by \citet{dietrich07} and of CFHTLS-CL J221556-175023 (offset by $1.05\arcmin$ northwards of our detection) by \citet{olsen07}.}
\label{fig:colim-BLOX_J2215.9-1751.6}
\end{figure}

This candidate is also detected in the work of \citet{dietrich07} and is listed as BLOX J2215.9-1751.6. They find a reliable X-ray detection with a flux measurement of $5.7\pm0.6\times10^{-15}~\text{ergs}/\text{s}/\text{cm}^2$ and an angular diameter of the major axis of $16.8\arcsec$. This compares to our detection of $F(<r_{500})=5.1\pm0.59\times10^{-15}~\text{ergs}/\text{s}/\text{cm}^2$ and $r_{200}=0.0224^{\circ}$ ($\approx80\arcsec$). They are unable to measure a redshift for this source however as their analysis is limited to $BVRI$ imaging with no NIR data. The candidate may also be present in the cluster survey of \citet{olsen07}, who present a candidate (CFHTLS-CL J221556-175023) at a distance of $1.05\arcmin$ from our detected cluster core with a redshift estimate of $z=1.1$. The \citet{olsen07} analysis is limited to CFHTLS $u^*griz$ data and attribute the detection their lowest confidence grade (D). The locations of the X-ray detections for BLOX J2215.9-1751.6 and CFHTLS-CL J221556-175023 are given by the blue crosses in the $giK_s$ image in Fig.~\ref{fig:colim-BLOX_J2215.9-1751.6}. BLOX J2215.9-1751.6 aligns with the centre of our X-ray contours, whilst CFHTLS-CL J221556-175023 is offset to the North.

Using the optical plus NIR data, our red-sequence analysis (Fig.~\ref{fig:colim-BLOX_J2215.9-1751.6}) produces a significant peak in the probability distribution at $z=1.11$ (with a significance of $8.5\sigma$). This is reinforced by a strong peak in the photometric redshift distribution with a median redshift of $z=1.17$. Galaxies selected as part of the red-sequence at $z=1.17$ are highlighted with the white arrows in Fig.~\ref{fig:colim-BLOX_J2215.9-1751.6}. A number of the brightest of these galaxies are clearly clustered within the core of the X-ray extended signal and appear to follow the shape of the X-ray profile.

A small number of spectroscopic redshifts are available in the region around this cluster candidate, however none are available for any of the galaxies selected as cluster members via the red-sequence analysis.

Estimating the clustering properties from the given redshift, we find a cluster mass of $M_{200}=1.22\pm0.06\times10^{14}~\rm{M_{\sun}}$, an X-ray luminosity of $L_x=6.64\pm0.50\times10^{43}~\rm{ergs}/\rm{s}$ and a radius of $r_{200}=0.0224^{\circ}$.


\subsubsection{WIRDXC J2214.2-1757 (D4-50)}

\begin{figure}
\centering
\includegraphics[width=8.5cm]{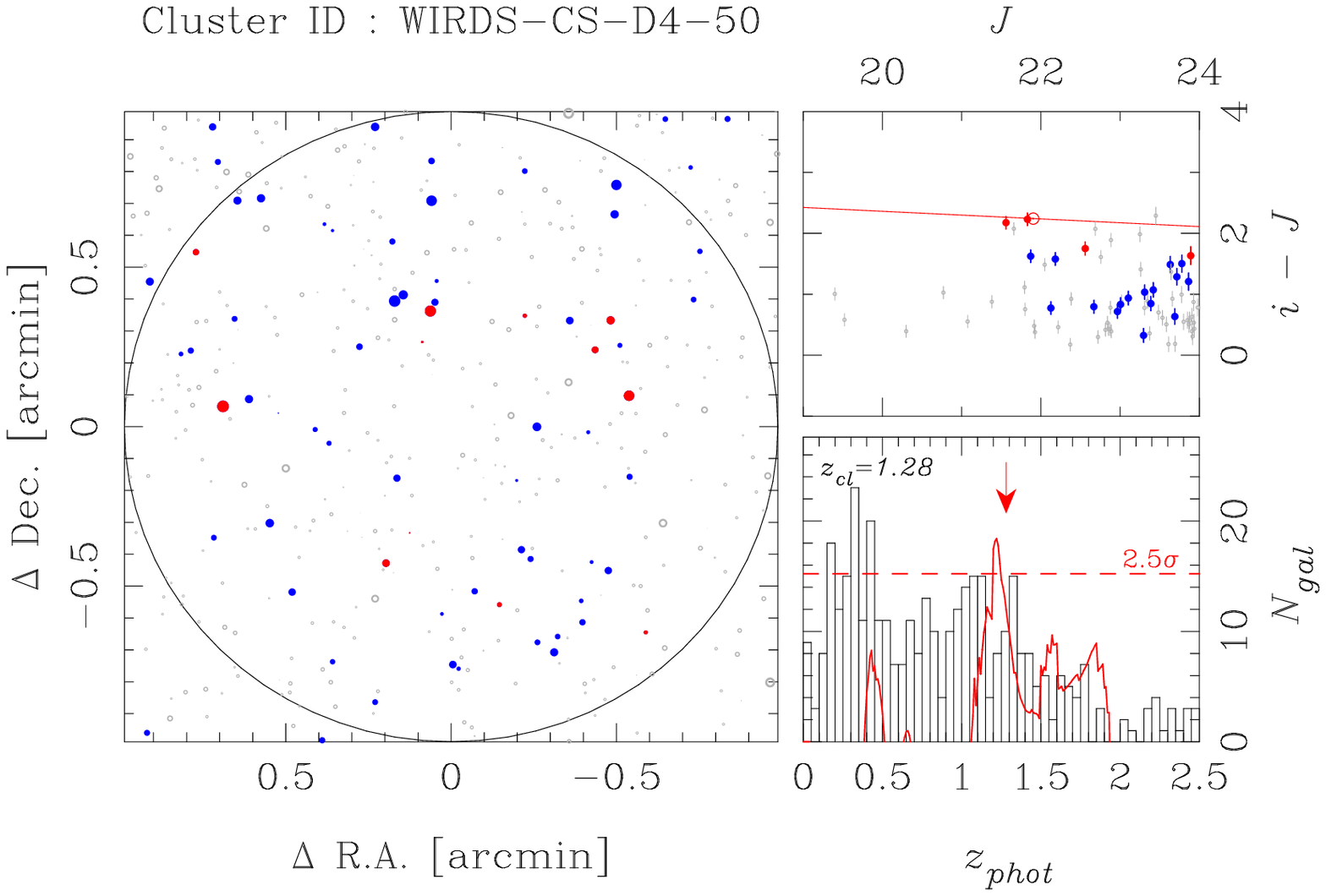}
\includegraphics[width=9.0cm]{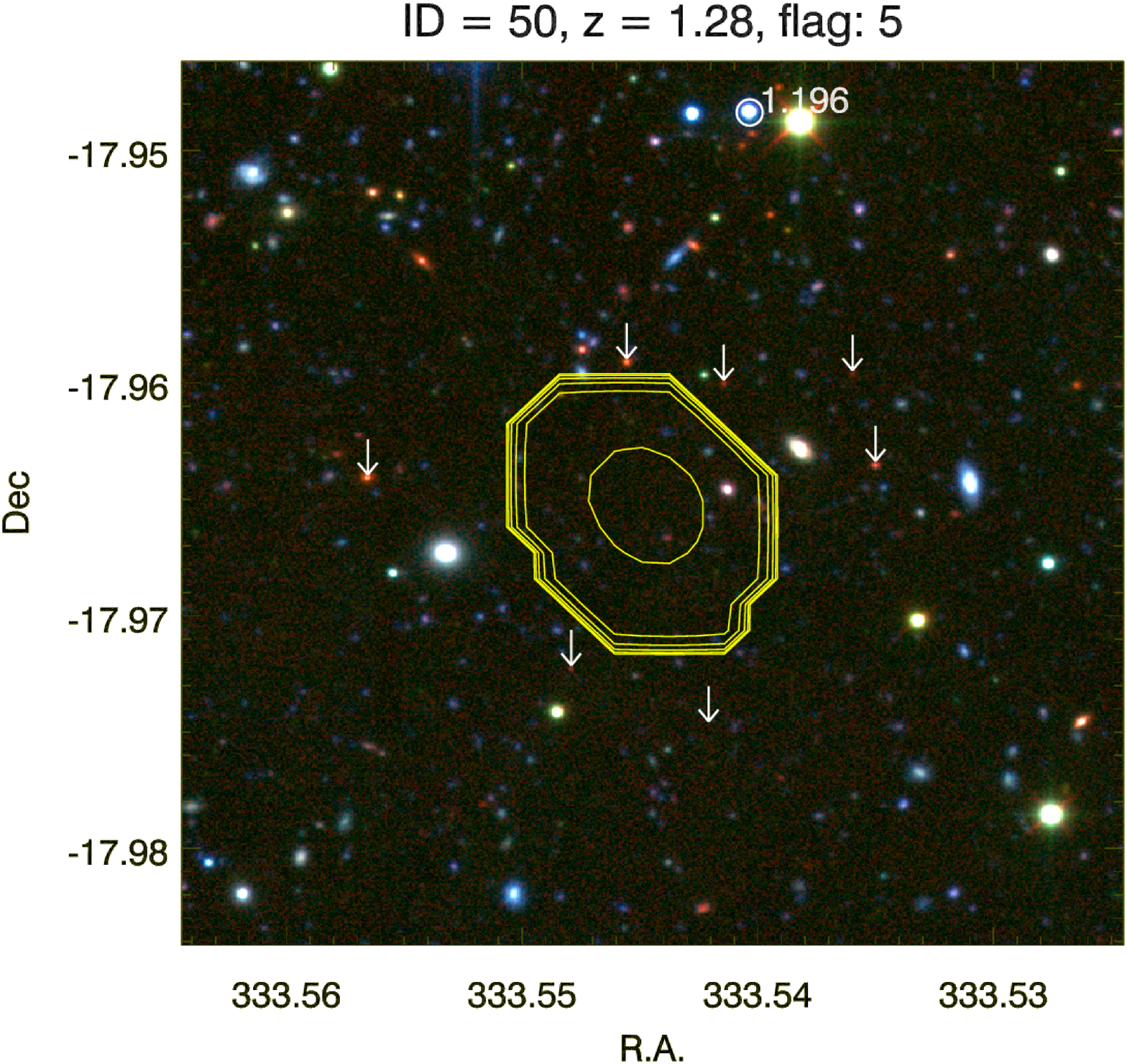}
\caption{As in Fig.~\ref{fig:colim-WIRDXC_J0224.3-0408}, but for candidate WIRDXC J2214.2-1757. The white circles and numbers in the lower panel indicate objects with spectroscopic redshifts.}
\label{fig:colim-WIRDXC_J2214.2-1757}
\end{figure}

The thumbnail for candidate WIRDXC J2214.2-1757 is shown in Fig.~\ref{fig:colim-WIRDXC_J2214.2-1757}. The extended X-ray signal, detected at a signal of $\sim4\sigma$, shows a relatively compact structure. The red-sequence analysis estimates a cluster redshift of $z=1.28$ with a confidence of $3\sigma$. Visually the red-sequence selected galaxies do not appear to be well spatially clustered.

We have no spectroscopic data for any of the galaxies in the X-ray emission region. However we note another QSO from the AAOmega spectroscopic data close to the X-ray emission and at a redshift of $z=1.196$. This is given in table~\ref{table:spec_4_50}.

\begin{table}
\caption{Spectroscopically confirmed objects around WIRDXC J2214.2-1757 within $\Delta z=0.1$ of the estimated cluster redshift.}             
\label{table:spec_4_50}      
\centering          
\begin{tabular}{lllllll}     
\hline
R.A.           &Dec.          & $\Delta r$ &$K_s$& $z_{phot}$&$z_{spec}$ & Class\\
\multicolumn{2}{c}{(J2000)}& $(^{\circ})$& (AB) & & & \\
\hline
  333.5403 &  -17.9484 &    0.0173 &   19.38 & --- &   1.196 &     QSO \\
\hline                  
\end{tabular}
\end{table}

Based on a redshift of 1.28, we estimate a cluster mass of $M_{200}=1.14\pm0.15\times10^{14}~\rm{M_{\sun}}$, an X-ray luminosity of $L_x=6.83\pm1.62\times10^{43}~\rm{ergs}/\rm{s}$ and a radius of $r_{200}=0.0207^{\circ}$.

\subsection{X-ray Unresolved Candidates?}

Thus far all the detected candidates have been part of the extended source
catalogue. As discussed, this excludes any sources that are close in size to
the PSF of the image within the constraints we have set. Especially as we
are concerned with high-redshift systems, this risks the possibility of
missing particularly compact/faint clusters. We have therefore reviewed the
point-source catalogue for the two fields, running the red-sequence
algorithm with the weighting tuned to select more compact spatial clustering
by setting $\sigma_r=300\text{kpc}$.

\begin{table*}
\begin{minipage}[t]{\textwidth}
\caption{WIRD Cluster Survey $z\gtrsim1.1$ cluster candidates.}             
\label{table:compact_overview}      
\centering
\renewcommand{\footnoterule}{}  
\begin{tabular}{l l l l l l l l l l}     
\hline
\hline                      
ID                       & R.A.    & Dec.                         & z\footnote{Cluster redshift estimated from red-sequence analysis or spectroscopic redshift if available.}   &$r_{200}$ & $F(<r_{500})$\footnote{Extrapolated cluster flux ($10^{-15}\rm{ergs/s/cm}^2$).} & $M_{200}$ & $L_{x}(0.1$-$2.4\rm{keV})$& $T_{X}$  & Flag \\ 
                           &\multicolumn{2}{c}{(J2000)}&   &($^{\circ}$) &              &   ($10^{14}\rm{M_{\sun}}$) &($10^{43}\rm{ergs/s}$) &  (keV)     &           \\
\hline
\hline               
WIRDCS-1-p282 &36.644929 &-4.1936979& 1.35& 0.0162 & $0.93\pm0.63$& $0.60\pm0.22$ & $2.60\pm1.76$ & $1.59\pm0.39$ & 5 \\   
WIRDCS-4-p265 &333.9341 &-17.699511& 1.26&   0.0165 & $0.87\pm0.20$& $0.56\pm0.08$ & $2.05\pm0.48$ &$1.47\pm0.13$ & 5 \\   
WIRDCS-4-p434 &333.3954 &-17.872525& 1.13&    0.0187&  $1.49\pm0.92$& $0.67\pm0.22$ & $2.27\pm1.40$& $1.57\pm0.35$ & 5 \\   
\hline
\hline     
\end{tabular}
\end{minipage}
\end{table*}

\begin{figure}
\centering
\includegraphics[width=8.5cm]{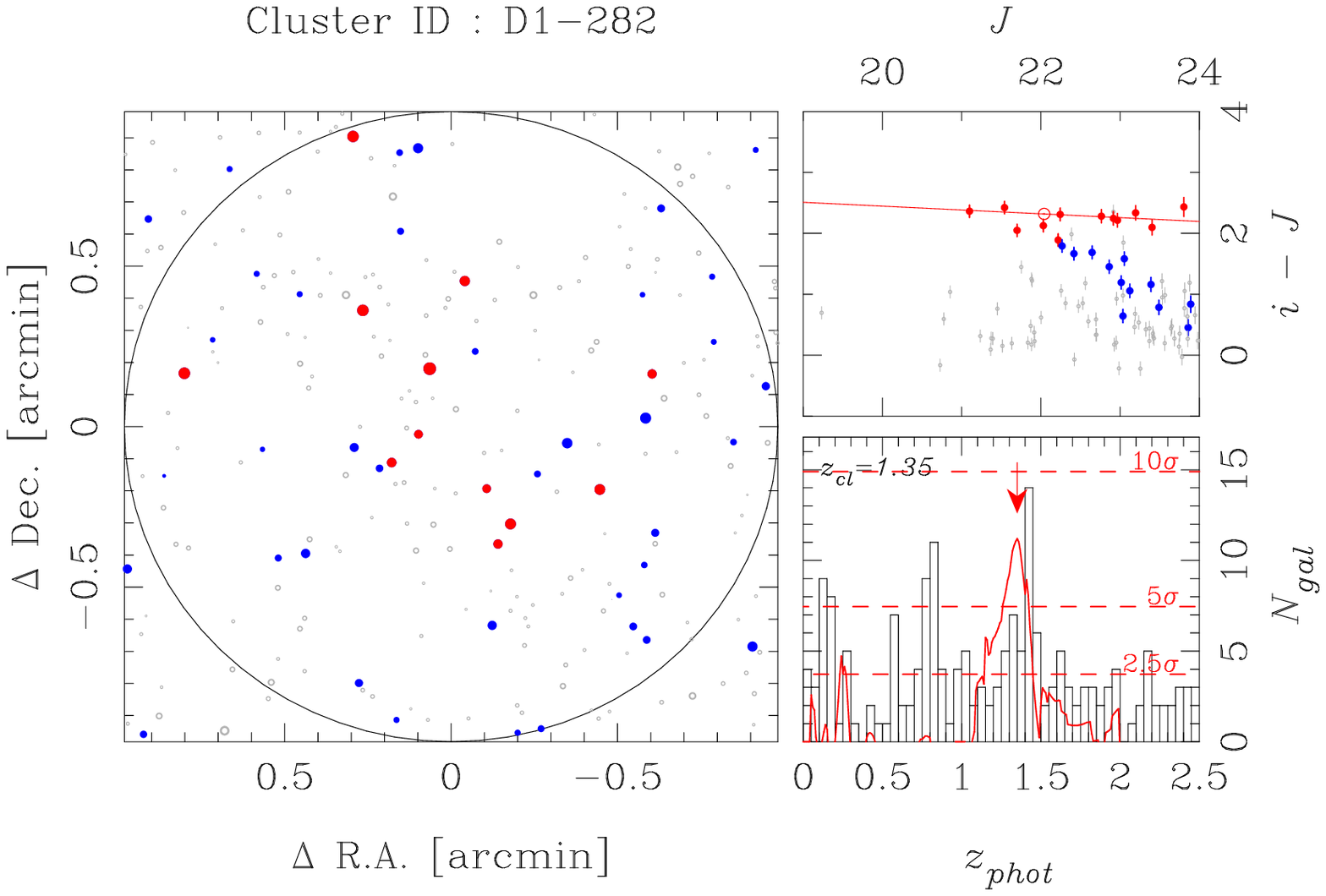}
\includegraphics[width=9.0cm]{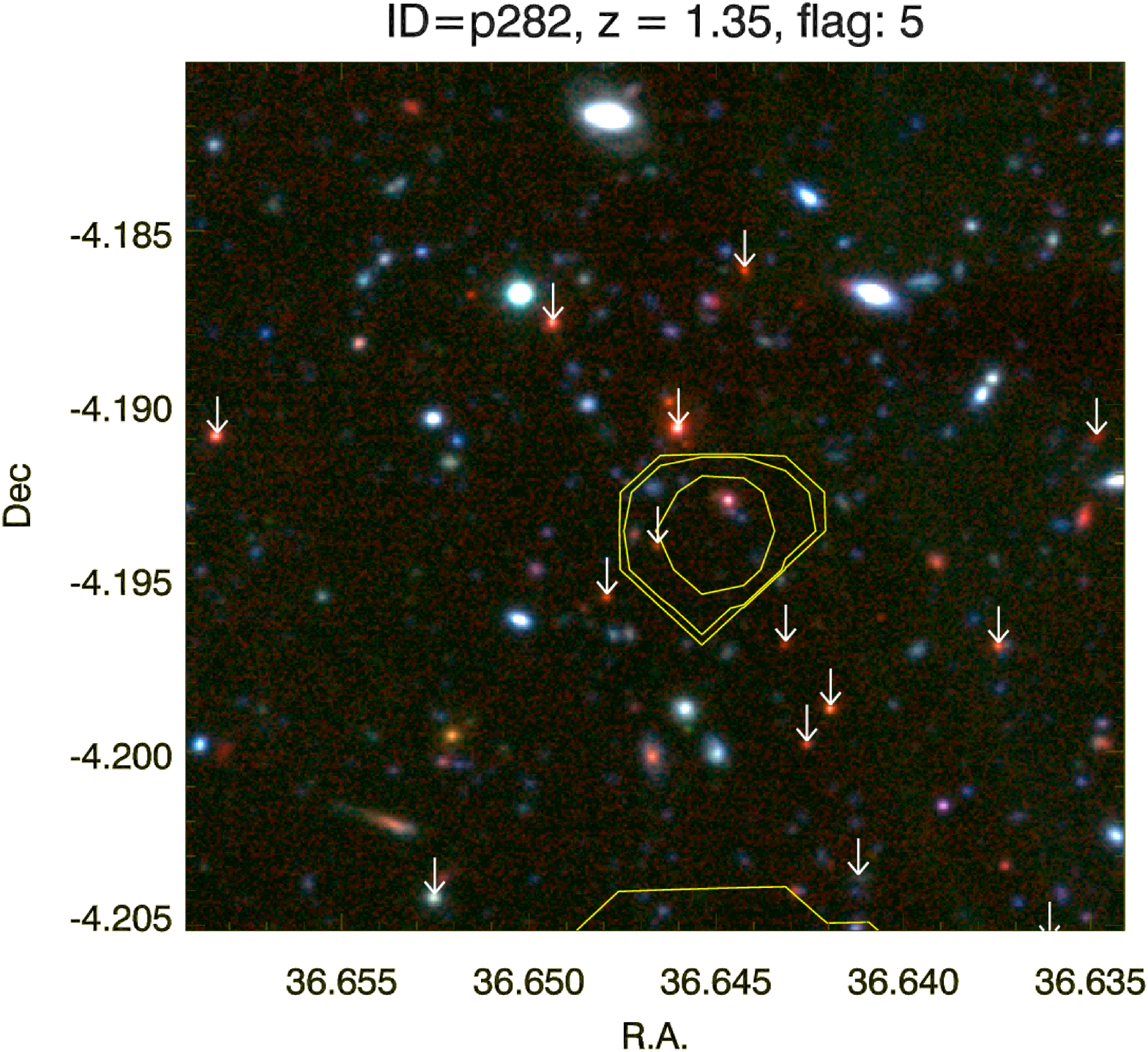}
\caption{As in Fig.~\ref{fig:colim-WIRDXC_J0224.3-0408}, but for candidate WIRDS-CS-D1-p282.}
\label{fig:colim-WIRDS-CS-D1-p282}
\end{figure}

\begin{figure}
\centering
\includegraphics[width=8.5cm]{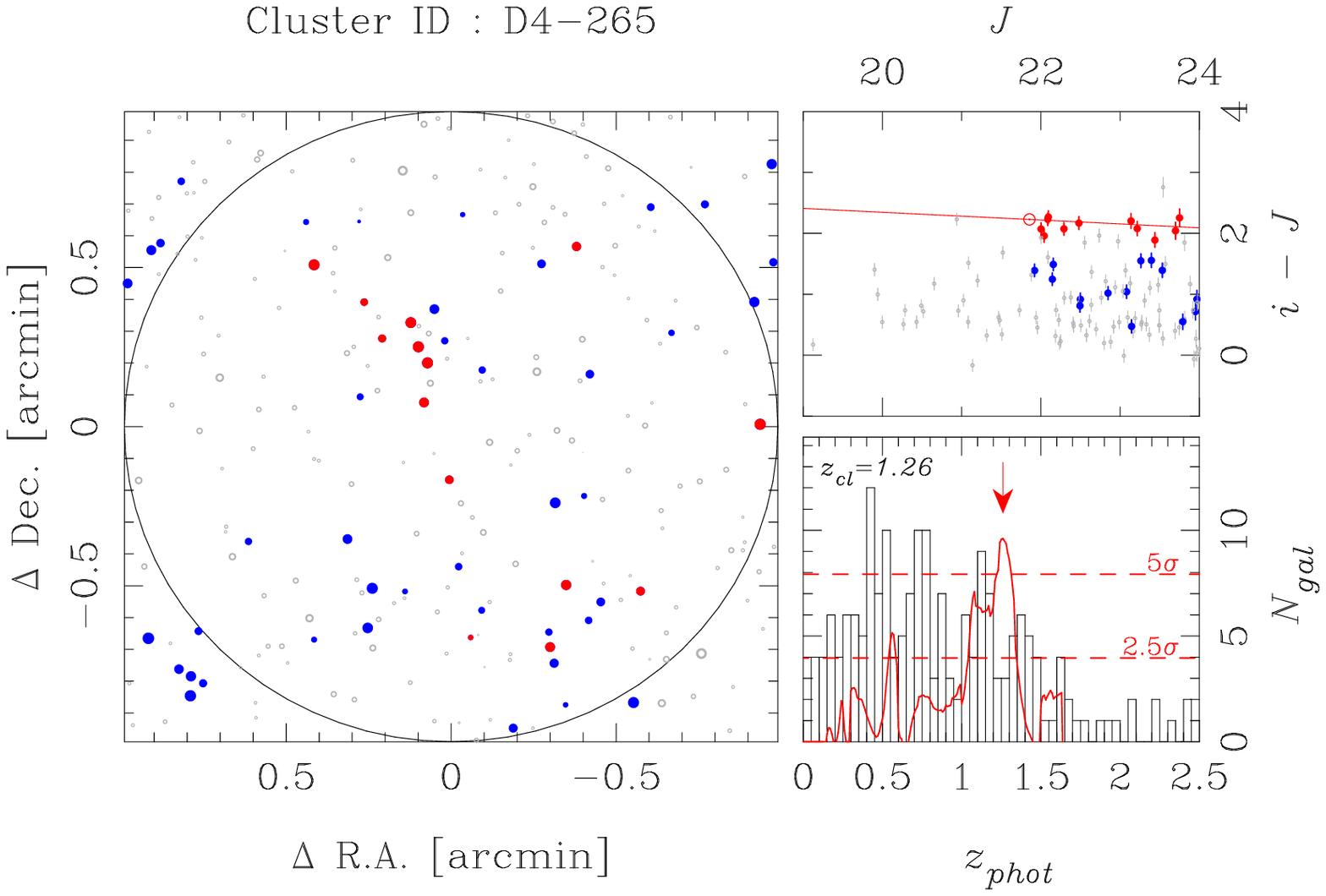}
\includegraphics[width=9.0cm]{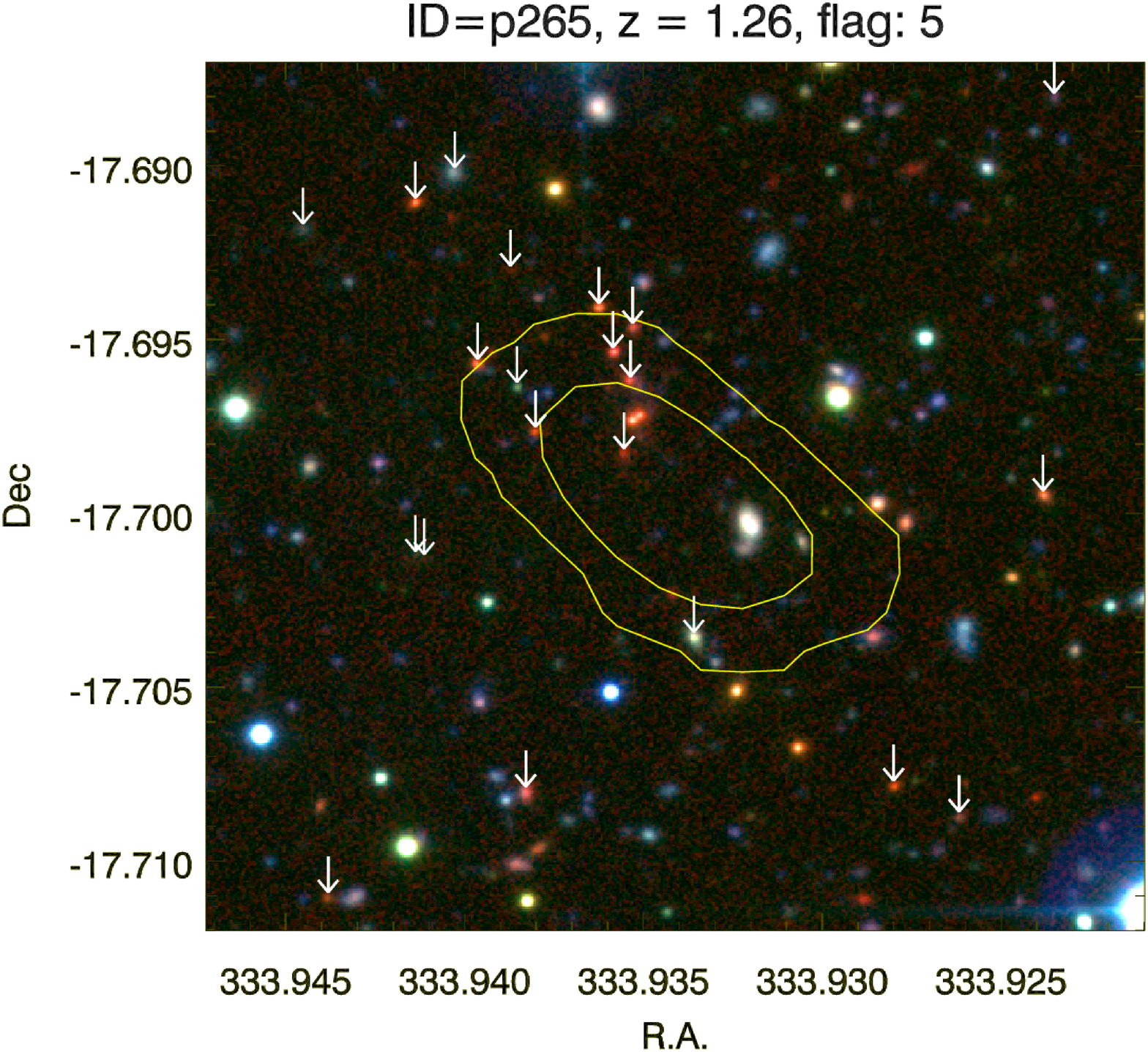}
\caption{As in Fig.~\ref{fig:colim-WIRDXC_J0224.3-0408}, but for candidate WIRDS-CS-D4-265.}
\label{fig:colim-WIRDS-CS-D4-p265}
\end{figure}

\begin{figure}
\centering
\includegraphics[width=8.5cm]{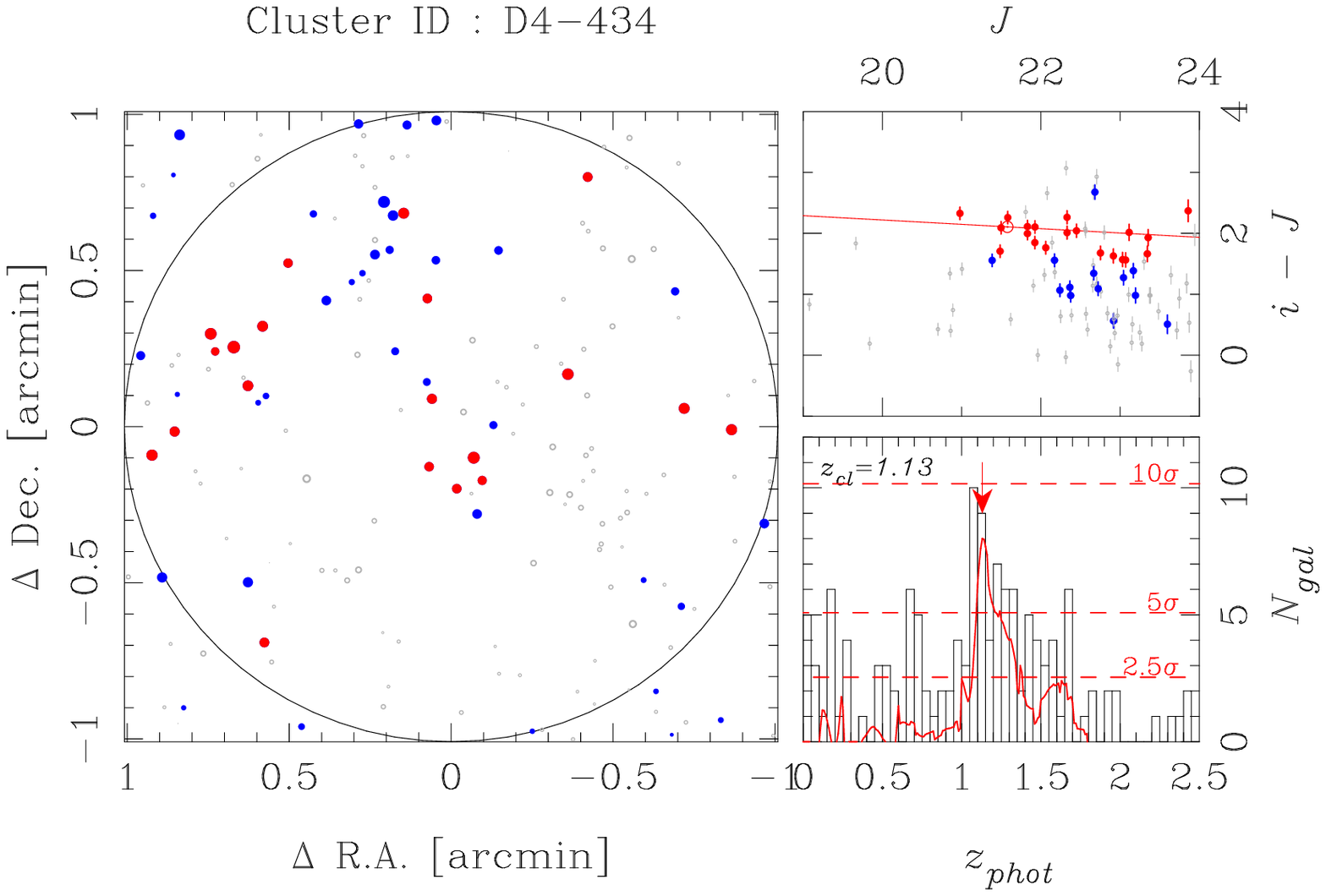}
\includegraphics[width=9.0cm]{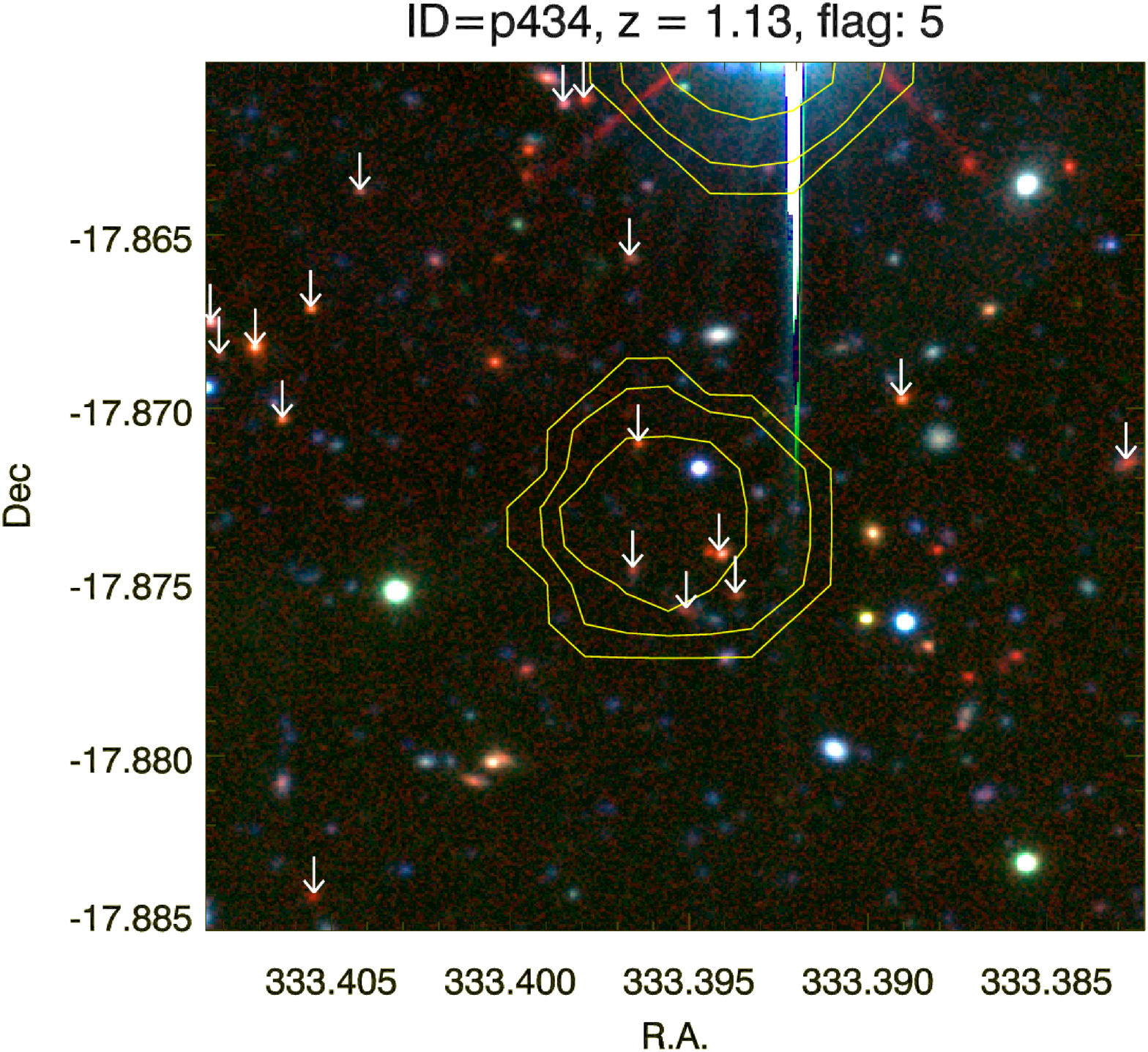}
\caption{As in Fig.~\ref{fig:colim-WIRDXC_J0224.3-0408}, but for candidate WIRDS-CS-D4-434.}
\label{fig:colim-WIRDS-CS-D4-p434}
\end{figure}

From this analysis, we find three candidates for high redshift compact clusters that are filtered out by the algorithm used to detect extended X-ray emission. The thumbnails and red-sequence results for these are shown in figures~\ref{fig:colim-WIRDS-CS-D1-p282}, \ref{fig:colim-WIRDS-CS-D4-p265} and \ref{fig:colim-WIRDS-CS-D4-p434}.

The first candidate is found in the D1 field and based on the red-sequence
analysis for this source, we estimate a redshift of $z=1.35$. This
red-sequence detection is found with a strong signal, but the red-sequence
members are distributed outside of the X-ray point source signal. We see no
obvious AGN candidates at the position of the X-ray signal and assign the
cluster candidate a flag of 5. We note that we find a small number of
objects with spectroscopic redshifts close to the X-ray emission, however
none of the red-sequence selected objects have been observed and the objects
that have been observed are not at consistent redshifts.

The second candidate found in the point-source list is in the D4 field and is shown in Fig.~\ref{fig:colim-WIRDS-CS-D4-p265}. Note that the elongated X-ray signal is due to the non-symmetric nature of the PSF across the field of view of the observations, however despite this the detection is still classed as a point-source by the X-ray detection algorithm. A relatively strong red-sequence signal is seen, dominated by the clustering of red objects to the top-left of the X-ray emission. These are found to be at a redshift of $z=1.26^{+0.08}_{-0.22}$ based on the red-sequence analysis. We note the presence of a relatively bright foreground galaxy which may hold an AGN and therefore may be the source of the X-ray emission. Again we attribute this candidate a flag of 5.

The final point-source detection with signs of a red-sequence detection is shown in Fig.~\ref{fig:colim-WIRDS-CS-D4-p434}. Again a strong red-sequence signal is observed, in this case at a redshift of $z=1.13^{+0.19}_{-0.05}$. A number of the selected galaxies are found with the PSF profile of the X-ray detection and appear well clustered. A second grouping of galaxies with the same red-sequence redshift is seen to the upper left of the X-ray emission. Again, a relatively bright foreground galaxy is observed within the PSF providing a candidate for AGN emission as the source of the detection. This candidate is also attributed with a flag of 5.

As with the primary cluster candidates, we estimate the total
flux/luminosity/mass properties of the compact cluster candidates based on
the derived redshifts. The results are given in
table~\ref{table:compact_overview}. We note that these are all low-flux
detections at the limit of our detection thresholds. As such we estimate
relatively low-masses for these objects
($\sim0.5-0.6\times10^{14}\rm{M_\odot}$). 

In summary of the check for unresolved clusters, no obvious candidate has
been found and the 3 candidates considered are likely AGNs and not
unresolved clusters. Thus, we can exclude a large influence of spatial
resolution of XMM-Newton on the reported cluster statistics.

\section{The Cluster Sample}
\label{sec:analysis}

\subsection{Cosmology from Cluster Number Counts}

\begin{figure}
\centering
\includegraphics[width=8cm]{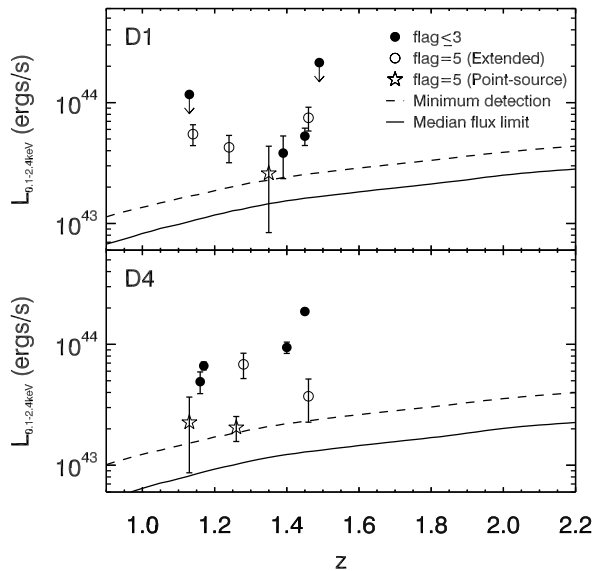}
   \caption{Cluster luminosity probed as a function of redshift in the D1 (top panel) and D4 (lower panel) fields. The flag$=$5 (open circles) and flag$\leq$3 (filled circles) extended cluster candidates are shown. Open stars show the point-source confusion cluster candidates. The dashed line shows the X-ray luminosity limit based on the minimum extended X-ray flux detection in the data, whilst the solid line shows the X-ray luminosity limit based on the median X-ray flux limit.}
      \label{fig:clustlumD1D4}
\end{figure}

In Fig.~\ref{fig:clustlumD1D4}, we show the derived X-ray luminosities of our cluster candidates in the CFHTLS D1 and D4 fields as a function of redshift. The high-redshift flag$\leq$3 cluster candidates are shown by filled circles, the flag$=$5 cluster candidates are shown by the open circles and the potentially unresolved detections are shown by open stars. For both fields, the dashed line shows the luminosity limit estimated from the minimum detected extended X-ray flux in each field, whilst the solid line gives the luminosity limit calculated from the median X-ray flux limit. The point-source detections are consistent with being faint sources in which extended signal from the cluster outskirts is below the detection threshold of the XMM data.

\begin{figure}
\centering
\includegraphics[width=8cm]{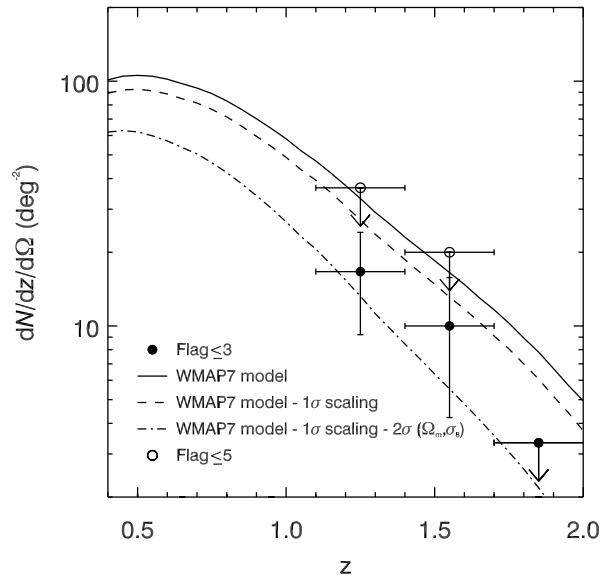}
   \caption{Cluster number counts as a function of redshift. The filled black circles show the candidate cluster counts from the flag$\leq$3 objects in the D1 and D4 fields combined. Open circles show the counts based on the flag$\leq$5 objects. The solid line shows the model predictions based on the WMAP 7-year cosmological parameters. The effect of the $1\sigma$ uncertainty in the scaling relations is illustrated by the dashed line, whilst the dash-dot line shows the effect of the $1\sigma$ uncertainty in the scaling relations, combined with a WMAP 7-yr model taking the $2\sigma$ uncertainty limits on $\Omega_m$ and $\sigma_8$.}
      \label{fig:clustnz7}
\end{figure}

We present the number counts of clusters as a function of redshift in the two fields in Fig.~\ref{fig:clustnz7}. The filled circles show the number counts based on only flag$\leq$3 candidates and the open circles show the cluster number counts for the full flag$\leq$5 candidate sample. As they include a number of tentative detections, the flag$\leq$5 counts give the upper limits to the number of detectable clusters based on the data used here. In both cases the plotted errors are statistical. We note that we do not detect any clusters at $z\gtrsim1.5$ and so the final data-point represents an estimated upper limit to the cluster number counts based on the areal coverage of our survey.

Model number counts were predicted using the cosmological code of \citet{2007MNRAS.379.1067P} with the WMAP 7-year cosmology (i.e. $\Omega_mh^2=0.1334$, $\sigma_8=0.801$, \citealt{2010arXiv1001.4538K}) and applying the areal coverage and flux limits of our survey. This is shown by the solid line in Fig.~\ref{fig:clustnz7}. We find that the model over-predicts the number counts compared to the flag$\leq$3 candidates by $\sim2\sigma$, whilst matching well with the full $z\gtrsim1.1$ upper limit flag$\leq$5 sample. There are two key elements that effect the number count measurements and models: the cluster scaling relations and the cosmological parameters used (in particular $\Omega_m$ and $\sigma_8$). Looking at the effect of changes in both of these, the dashed line shows the result of adjusting the scaling parameters by $1\sigma$, whilst the dash-dot line shows the effect of adjusting the scaling parameters again by $1\sigma$ combined with a change in the WMAP 7-year $\Omega_m$ and $\sigma_8$ values by $2\sigma$. The second of these, which takes values of $\Omega_mh^2=0.1323$ and $\sigma_8=0.741$, lowers the model number counts by a factor of $\approx2.5$ at redshifts of $z\approx1-2.0$. This illustrates the sensitivity of the cluster number counts to the cosmological parameters and is more consistent with our flag$\leq$3 candidate counts. The number counts based on the flag$\leq$3 favour a lower $\sigma_8$, which we note is in agreement with the results of \citet{2010MNRAS.403.2063F}. Based on their sample of $0<z<2$ clusters, they publish cluster number counts that prefer a $5\%$ reduction in the value of $\sigma_8$ from the WMAP 5-year value.

Significant uncertainty remains in the number count analysis. In particular the scaling relations present problems in any potential application of cluster number counts to placing constraints on cosmological parameters. The scatter in the scaling relations, particularly for galaxy groups, remains significant, the effect of which is illustrated by the dashed line in Fig.~\ref{fig:clustnz7}. In addition, we note that the redshift range over which scaling relations have been calibrated is in the main limited to $z\lesssim1$ \citep{leauthaud10}. Of course, we also note that, although we present a relatively large area for such a high-redshift survey, statistical errors due to low numbers of clusters remain high. Finally, uncertainty in a number of the cluster detections require spectroscopic confirmation of cluster members.

Returning to the lack of cluster detections at $z\gtrsim1.5$, we note that given the WMAP 7-year model we would expect $\approx3-4$ cluster detections within the survey area at such redshifts. Based on the red-sequence model, we estimate a characteristic $K_s$ band magnitude of cluster members of $m_m^*\approx21.5$ at $1.5<z<2$, which is well within our magnitude completeness limit. Increasing photometric errors will give poorer photometric redshifts (see Fig.~\ref{fig:mstar}), which will weaken the signal from a given cluster (note that the photometric redshifts are used to pre-select potential cluster members). For galaxies at this characteristic magnitude and in this redshift range, we find a median error on the photometric redshift measurements of $0.05\lesssim\sigma_z\lesssim0.1$, compared to $\approx0.03-0.05$ at $z<1.5$. Looking at the photometry for this same population, we find increases in the magnitude errors of $\sim0.05-0.08 \rm{mag}$. In order to gauge the effect of this increase in the magnitude errors, we re-run the red-sequence analysis on the same catalogues but with a 0.1 mag error added in quadrature to all 8 bands. Taking all the X-ray extended sources (including those originally classed as $z<1$), we find that 70\% of detections retain the same redshift solution given this increase in the noise of the photometry. Out of the 30\% with differing solutions given the increased photometric errors, 72\% have a red-sequence significance of $\sigma_{rs}<3$. Based on this analysis then, we may expect that the red-sequence completeness at $z>1.5$ is reduced by no more than 30\%.

Alternatively, the lack of detections may be the result of greater levels of star formation in clusters at $z>1.5$. Increased star-formation in cluster member may cause the red-sequence to be more difficult to detect, if higher fractions of bluer galaxies are present. This effect is very difficult to quantify at this point as it relies on having samples of unbiased clusters and groups at these redshifts in order to understand the properties of cluster members. This will be further complicated by the formation epoch of the red sequence varying from cluster to cluster as clusters will not all form at the same time. Ultimately, the redshifts up to which we can detect clusters via the red sequence method is something that must be learned from real data and as such the highest redshift at which the red-sequence analysis can be claimed to be effective is $z\approx1.6$ at this point \citep{2010ApJ...716L.152T}.

Finally, given the low numbers of clusters predicted to be detectable by the X-ray data in this redshift range, it is feasible that the lack of detection of any clusters at such redshifts may be the result of statistical cosmic variance (i.e. that there are no clusters above our detection thresholds at these redshifts in the volumes sampled). 

\subsection{X-ray Luminosity Function}

It is intructive to see in which way the prediction of high numbers of clusters in the WMAP7 cosmology disagrees with observed cluster characteristics, such as total mass or luminosity.  In this vein, we calculate the X-ray luminosity function in the redshift range $1.1\leq z\leq2$ based on our cluster candidate sample in the two analyzed fields. The cluster luminosity function, $\phi$, is given by:

\begin{equation}
\phi(L_X,z)=\frac{1}{\Delta L}\sum_{i=1}^{N_j}\frac{1}{V_{\rm max}(L_{X,i})}
\label{eq:Xlumfunc}
\end{equation}

\noindent where $N_j$ is the number of clusters in a given luminosity bin of width $\Delta L$ and $V_{\rm max}(L_{X,i})$ is the total comoving volume in which a cluster of luminosity $L_{X,i}$ could have been detected above the flux limits of the survey. This is given by:

\begin{equation}
V_{\rm max}(L_X)=\int_{z_{\rm min}}^{z_{\rm max}}\Omega(f_X(L_X,z))\frac{dV(z)}{dz}dz
\label{eq:vmax}
\end{equation}

\noindent where $z_{\rm min}$ and $z_{\rm max}$ are the redshift limits. The lower limit is taken as $z=1.1$, whilst we set an upper limit of $z=2.0$, assuming for the purpose of the luminosity function that the lack of detections at $1.5<z<2.0$ is due to statistical cosmic variance. $\Omega(f_X(L_X,z))$ is the sky area as a function of the X-ray flux probed by the X-ray data and $dV(z)/dz$ is the differential comoving volume element per steradian \citep{1980ApJ...235..694A,mullis04}.

\begin{figure}
\centering
\includegraphics[width=8cm]{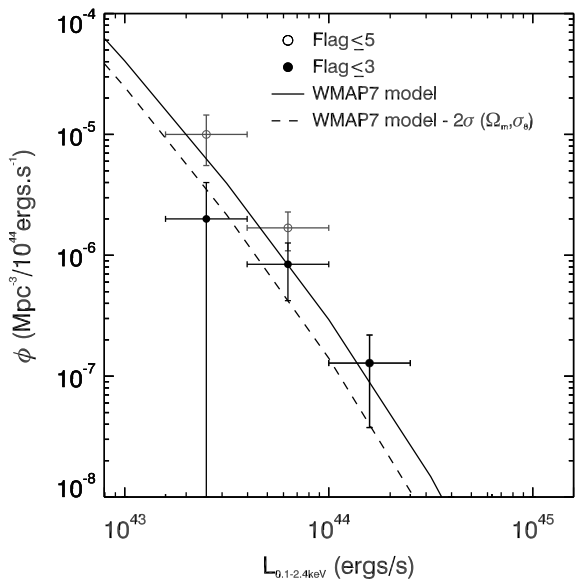}
   \caption{X-ray luminosity function calculated for $1.1\leq z\leq2.0$ candidate clusters in the CFHTLS D1 and D4 fields. The filled circles show the X-ray luminosity function for flag$\leq$3 cluster candidates only, whilst the open circles show the result for all candidates with flag$\leq$5. The solid line shows the predicted luminosity function based on our survey geometry and WMAP 7-year cosmological parameters. The dashed line shows the same but with the WMAP 7-year $\Omega_m$ and $\sigma_8$ parameters reduced by $2\sigma$.}
      \label{fig:xlumfunczge1.1}
\end{figure}

Fig.~\ref{fig:xlumfunczge1.1} shows our calculated X-ray luminosity function in the redshift range $1.1\leq z\leq2$. The luminosity function for the flag$\leq$3 sample is again shown by the filled circles, whilst the luminosity function derived from the flag$\leq$5 is shown by the open circles. Again, errors on the bin numbers are statistical (although the horizontal bars simply show the bin extent). The solid line gives the model luminosity function based on the WMAP 7-year cosmological parameters, whilst the dashed line shows the effect of reducing the WMAP 7-year $\Omega_m$ and $\sigma_8$ parameters by $2\sigma$. We see that the flag$\leq$3 luminosity function agrees well with the WMAP 7-year model at luminosities of $L_X\gtrsim4\times10^{43}{\rm ergs/s}$. Taking the single luminosity bin at $L_X\lesssim5\times10^{43}{\rm ergs/s}$, we find a deficit of clusters compared to the model luminosity function, with the cluster data point falling a factor of 3 below the model prediction (equivalent to $\approx2\sigma$ based on the error estimate). With the flag$\leq$5 luminosity function however, we find good consistency with the model over the full range of luminosities that our cluster candidates sample (although the data points prove marginally higher than the model).

The luminosity function highlights the difficulty in identifying the low-luminosity systems in the sample. Although our absolute limit for detection is $\approx10^{43}{\rm ergs/s}$ in both fields (Fig.~\ref{fig:clustlumD1D4}) at redshifts of $1.1<z<2.0$, the identification of such low luminosity groups proves difficult at $L_X\lesssim4\times10^{43}{\rm ergs/s}$. Including tentative detections appears to correct for this, however this brings greater risks of including objects falsely identified as clusters. Additionally, at fainter luminosities/fluxes the volume correction ($\propto1/\Omega(f_X)$) increases significantly making our cluster counts in the lowest luminosity bin in Fig.~\ref{fig:xlumfunczge1.1} more uncertain.

Thus a disagreement between dn/dz of clusters and the cosmological prediction comes primarily from lowest mass clusters or groups close to the detection limit and not from the massive clusters, where detection is clear and identification is more obvious. A similar picture has been seen in \citet{2010MNRAS.403.2063F} with the conclusion that, for a robust assesment of the cosmology, one needs to increase the area of such surveys to at least $50\deg2$ and use the most massive clusters in the sample.  We note, that in the SXDF field no such massive clusters have been found \citet{2010MNRAS.403.2063F}. It is also quite conceivable that the fraction of valid flag=5 identifications increases as the mass of the system decreases, which can be verified through the spectroscopic follow-up.

\section{Conclusions}
\label{sec:conclusions}

Using a combination of XMM-Newton X-ray data and optical/NIR photometry, we have performed a survey of $z\gtrsim1.1$ cluster candidates in the CFHTLS D1 and D4 fields. Crucially, NIR data is required in order to apply the red-sequence analysis at such redshifts (as the 4000\AA\ break moves out of the optical into the NIR bands). As such the WIRDS data, which provides the best combination of wide field and deep NIR data currently available, is one of the best currently available resources for the identification of high redshift clusters. Using the WIRDS data in conjunction with CFHTLS deep optical photometry, we identify a total of 15 $z>1.1$ cluster candidates in a total area of $1.0 \text{deg}^2$. All candidates were selected based on X-ray emission in deep XMM data, which provides a relatively unbiased sample.

Of our 15 cluster candidates, 7 are considered firm candidates (flag$\leq$3). These have both strong and clearly extended X-ray emission combined with strong red-sequence detections of cluster members that are clearly clustered within the X-ray detection. For a number of these, spectroscopic redshift data is available. One of these (WIRDSCS-4-25) is the previously identified cluster, XMMXCS J2215.9-1738, at $z=1.45$. Our red-sequence analysis independently determined a cluster redshift estimate of $z=1.39$, showing the success of the red-sequence method. We estimate a luminosity and mass for this cluster of $2.1\times10^{44}{\rm ergs/s}$ and $1.9\times10^{14}{\rm M_{\odot}}$, making it the most massive cluster in our sample. Additionally, WIRDSCS-1-76 has been previously identified by \citet{2009A&A...507..147A}, as a $z=1.9$ cluster (JKS041). As they note, there appear to be a number of structures in the line of sight of this X-ray source. Based on red-sequence analysis and photometric redshifts we identify four apparent structures with redshifts of $z=0.80$, $z=0.96$, $z=1.13$ and $z=1.49$. The first three of these are confirmed by VVDS Deep spectroscopic data, although the $z=0.80$ and $z=0.96$ structures appear, based on both the spectroscopic and red-sequence data, to be dominated by blue star-forming galaxies. The $z=1.13$ and $z=1.49$ both show strong signs of a clustered red-sequence correlated with the extended X-ray flux and the spectroscopic redshifts for the former again prove the effectiveness of our red-sequence analysis. Based on our analysis, we see no evidence for a cluster at $z=1.9$. Of the remaining flag$\leq$3 cluster candidates, BLOX J2215.9-1751.6 has been previously identified as a $z=1.1$ cluster based on optical photometry only \citep{dietrich07,olsen07}. With the addition of the WIRDS NIR data, we improve upon this redshift estimate, placing the cluster at $z=1.17$.

Adding to the 7 high-confidence candidates, we present 8 further candidates attributed with confidence flag$=$5, bringing our total number of candidates up to 15. These candidates have only either weak extended X-ray emission, point-source like X-ray detections or non-secure redshift estimates from the red-sequence analysis (i.e. due to small numbers of member galaxies or poorly clustered member galaxies). By including these candidates, we place upper limits on the numbers of clusters we are able to detect in the given area using the X-ray detection method on this data-set.

Comparing to other surveys incorporating $z\gtrsim1-1.1$ cluster samples, \citet{eisenhardt08} reported 106 $z>1$ cluster candidates from the IRAC Shallow Survey (based on optical and infrared photometry without any requirement for X-ray detections) with a sky density of $\approx14.6\degm2$. \citet{finoguenov07} presented a collection of 8 $z\geq1$ cluster candidates with a sky density of $\approx3.8\degm2$ and \citet{2010MNRAS.403.2063F} reported on 13 $z\geq1$ cluster candidates in the SXDF with a sky density of $\approx10.0\degm2$. Our combination of deep optical and near infrared photometry with the XMM X-ray data is therefore an important addition to the available surveys of high redshift cluster candidates. 

We show cluster number counts based on our high redshift cluster candidates
and compare these to predictions based on the WMAP 7-year cosmological
parameter estimates. The flag$\leq$3 sample shows lower number counts
compared to the WMAP 7-year model, which takes into account the survey
geometry and detection method. This tentatively favours lower values for the
cosmological parameters $\Omega_m$ and $\sigma_8$ than prescribed by the
WMAP 7-year results and we show that this is comparable to a $2\sigma$
reduction in both of these parameters. This sensitivity of cluster number
counts to $\Omega_m$ and $\sigma_8$, which is particularly strong at $z>1$,
illustrates the promise of using cluster number counts to constrain
cosmology. Based on our survey, we highlight the issues remaining in
applying this method however. As stated, modeling the number counts relies
on a clear knowledge of the detection limits of the cluster survey. We
therefore provide a focus on non-secure (flag$=$5) candidates found using
our detection methods. By including these objects, we find good agreement
between the WMAP 7-year model and the cluster number counts. Significantly,
we also present the $z>1.1$ cluster luminosity function for our sample and
find good agreement between our data and the WMAP 7-year model, except at
faint luminosities, where we only find good agreement if we include the
flag$=$5 candidates. We conclude that such non-secure candidates will
ultimately be a combination of correct and false cluster detections, which
we find introduce an uncertainty in our survey equivalent to the $2\sigma$
constraints on $\Omega_m$ and $\sigma_8$ from WMAP. Any
constraints on cosmology also rely on the cluster scaling relations, which
are required to constrain the X-ray luminosity limit of the survey. We
discuss the effect of the relations, noting that the scatter on the scaling
relations is a key issue and in addition it remains for these relations to
be calibrated beyond $z>1$.

The observations presented here suggest that there are perhaps too few clusters at $z>1$, based on the numbers of relatively secure identifications of groups and clusters, compared to predictions using WMAP-7 cosmology. However, the current level of statistical and systematic uncertainties prevent us drawing a secure conclusion. Ultimately, cluster number counts can provide a complimentary and independent method for constraining the cosmological model. In the longer term, the techniques presented here (combining X-ray cluster detection followed by red-sequence identification) present an unparallelled technique to deriving relatively unbiased group and cluster samples. With such samples we will be able to independently constrain $\Omega_m$, $\sigma_8$ and determine the level of non-gaussianity in the primordial density fluctuations.

\begin{acknowledgements}
This research was supported by the Agence Nationale de la Recherche (ANR) as part of the Deep galaxy Evolution Survey in the near InfraRed (DESIR) project (grant number ANR-07-BLAN-0228). The work was also supported in part by World Premier International Research Center Initiative (WPI Initiative), MEXT, Japan and also by the DFG cluster of excellence ÔOrigin and Structure of the UniverseÕ. We would like to thank P. Henry and M. Pierre for comments on the manuscript and J. Stott for valuable discussion. We have made use of the NASA/IPAC Extragalactic Database (NED) which is operated by the Jet Propulsion Laboratory, California Institute of Technology, under contract with the National Aeronautics and Space Administration. TERAPIX computing resources at the IAP were used for this work.
\end{acknowledgements}

\bibliographystyle{aa}
\bibliography{$HOME/Documents/rmb}

\end{document}